\documentclass[14pt]{article}
\pdfoutput=1
\usepackage{amsmath,amssymb,amsfonts,amsthm,bm,bbm,cancel,wasysym}
\usepackage{epsfig,graphics,graphicx,epstopdf}
\graphicspath{{Charts/}}
\usepackage{smartdiagram}
\usepackage{array,booktabs,colortbl,colordvi,multirow}
\usepackage{colordvi,color,xcolor}
\usepackage{hyperref}
\usepackage{rotating}

\usepackage{verbatim}
\usepackage{cite}
\usepackage{subfig}
\usepackage{setspace}
\usepackage{url}
\usepackage[percent]{overpic}
\usepackage{slashed}
\usepackage{authblk}
\usepackage{xspace}
\usepackage{fullpage}
\usepackage{hyperref}

\def\ie{{\it i.e.}}
\def\eg{{\it e.g.}}
\def\etc{{\it etc}}

\def\to{\rightarrow}

\newskip\zatskip \zatskip=0pt plus0pt minus0pt
\def\matth{\mathsurround=0pt}
\def\lsim{\mathrel{\mathpalette\atversim<}}
\def\gsim{\mathrel{\mathpalette\atversim>}}
\def\atversim#1#2{\lower0.7ex\vbox{\baselineskip\zatskip\lineskip\zatskip
  \lineskiplimit 0pt\ialign{$\matth#1\hfil##\hfil$\crcr#2\crcr\sim\crcr}}}

\begin{document}

%----------------------------------- TITLE AND AUTHORS -----------------------------------------%

%Preprint numbers
\begin{flushright}
SLAC-PUB-17572\\
\today
\end{flushright}
\vspace*{5mm}

\renewcommand{\thefootnote}{\fnsymbol{footnote}}
\setcounter{footnote}{1}

\begin{center}

{\Large {\bf SU(4) Flavorful Portal Matter}}\\
%\vspace*{0.15cm}

\vspace*{0.75cm}

{\bf George N. Wojcik$^{1,2}$ and Thomas G. Rizzo$^1$}~\footnote{gwojcik@wisc.edu,  rizzo@slac.stanford.edu}

\vspace{0.5cm}

{$^1$SLAC National Accelerator Laboratory, Menlo Park, CA, 94025 USA}\\
{$^2$Department of Physics, University of Wisconsin-Madison, Madison, WI 53703 USA}

\end{center}
\vspace{.5cm}

%--------------------------------------------- ABSTRACT ---------------------------------------------%
\begin{abstract}
\noindent In this paper, we present a model which attempts to unify a new dark sector force with a local $SU(3)$ flavor symmetry. Dark Matter (DM) and its potential interactions with the Standard Model (SM) continue to present a rich framework for model building. In the case of thermal DM of a mass between a few MeV and a few GeV, a compelling and much-explored framework is that of a dark photon/vector portal, which posits a new $U(1)$ ``dark photon" which only couples to the SM via small kinetic mixing (KM) with the SM hypercharge. This mixing can be mediated at the one-loop level by portal matter (PM) fields which are charged under both the dark $U(1)$ and the SM gauge group. In earlier work, one of the authors has noted that models with appropriate portal matter content to produce finite and calculable kinetic mixing can arise from non-minimal dark sectors, in which the dark $U(1)$ is a subgroup of a larger gauge symmetry under which SM particles might have non-trivial representations. We expand on this idea here by constructing a model in which in which the dark $U(1)$ is unified with another popular extension to the SM gauge group, a local $SU(3)$ flavor symmetry. The full dark/flavor symmetry group is $SU(4)_F \times U(1)_F$, incorporating the local $SU(3)$ flavor symmetry with PM appearing as a vector-like ``fourth generation"  to supplement the three generations of the SM. To ensure finite contributions to KM, the SM gauge group is arranged into Pati-Salam multiplets. The new extended dark gauge group presents a variety of interesting experimental signatures, including non-trivial consequences of the flavor symmetry being unified with the dark sector. 
\end{abstract}

\renewcommand{\thefootnote}{\arabic{footnote}}
\setcounter{footnote}{0}
\thispagestyle{empty}
\vfill
\newpage
\setcounter{page}{1}

%-------------------------------- DOCUMENT: INTRODUCTION ---------------------------------%

% 1 Introduction

\section{Introduction}\label{IntroSection}

In spite of composing roughly 80\% of the matter in the universe, the precise identity of dark matter (DM) remains undetermined. Models to produce the appropriate abundance of DM, however, strongly suggest that it is subject to interactions other than gravity, and null results (thus far) in searches for historically favored DM candidates such as axions \cite{Kawasaki:2013ae,Graham:2015ouw} and WIMPs \cite{Arcadi:2017kky} have led to a proliferation of new ideas on the identity, production mechanisms, and experimental signatures of DM \cite{Alexander:2016aln,Battaglieri:2017aum}. In this work, we will be exploring one of the more recently popular DM models, that of a kinetic mixing/vector portal\cite{VectorPortals,KM}. In these setups, interaction between the Standard Model (SM) fields and the DM is mediated by a massive vector dark photon $A_D$, associated with a broken dark gauge group $U(1)_D$. The DM particle is uncharged under the SM gauge group, but possesses non-zero $U(1)_D$ charge, while the SM fields are just the opposite: Charged under the SM group, but not under $U(1)_D$. Small couplings between the dark photon and the SM fields occur due to kinetic mixing (KM) between $U(1)_D$ and $U(1)_Y$, which after electroweak symmetry breaking ultimately give every SM particle a coupling to $A_D$ of $e \epsilon Q$, where $e$ is the electromagnetic coupling constant, $Q$ is the electric charge of a given particle, and $\epsilon$ is a small value which parameterizes the strength of the kinetic mixing. The simplest realizations of these models present a rich phenomenology with a remarkably simple setup: The only parameters the DM and dark photon masses (as well as potentially the mass of a scalar associated with $U(1)_D$ breaking), the $U(1)_D$ gauge coupling, and the KM parameter $\epsilon$. However, these minimal setups leave little framework for addressing a number of questions: What is the origin of the small KM parameter $\epsilon$? Is $U(1)_D$ the only dark sector gauge group, or is it embedded in some larger gauge symmetry? Could the dark sector gauge forces be connected in some way to other ongoing questions in particle physics?

In the region of parameter space where DM and the dark photon have masses ranging between $\sim$a few MeV to $\sim$a few GeV, the favored value of $\epsilon$ generally lies in the range of $\epsilon \sim 10^{-(3-5)}$ \cite{Izaguirre:2013uxa,Gherghetta:2019coi}, which can in turn suggests that the KM can arise from a simple one-loop vacuum polarization-like diagram featuring ``portal matter'' fields charged under both $U(1)_Y$ and $U(1)_D$ \cite{KM}. The kinetic mixing parameter then has the dependence  
\begin{align}\label{KMSum}
    \epsilon \propto \sum_{i} Q_{Y_i} Q_{D_i} \log \bigg( \frac{m_i^2}{\mu^2} \bigg),
\end{align}
where the sum over $i$ denotes a sum over all fermions (a nearly identical expression arising from complex scalar loops also occurs, differing only in the omitted proportionality constant in front of the sum), $Q_{Y_i}$ denotes the hypercharge of particle $i$, $Q_{D_i}$ denotes its $U(1)_D$ charge, and $\mu$ is a renormalization scale. The sum in Eq.(\ref{KMSum}) is finite and calculable when $\sum_i Q_{D_i} Q_{Y_i} = 0$. In two previous works by one of our authors \cite{Rueter:2019wdf,Rizzo:2018vlb}, henceforth referred to as \textbf{I} and \textbf{II}, respectively, the theory and phenomenology of these portal matter fields was explored. In \textbf{I}, it was argued that fermionic portal matter fields must be vector-like in order to avoid constraints due to gauge anomalies and precision electroweak measurements. Additionally, in order to ensure that the fields are unstable (and hence conform to cosmological constraints), they should only appear in the same representations under the SM gauge group as SM particles. In short, portal matter fermions should be vector-like copies of SM fields \cite{VLFs}, a possibility rarely explored in the literature \cite{PortalMatter}. The discussion in \textbf{I} limited itself to the simplest possible constructions, in which a pair of such vector-like fermions, with opposite dark charges, generate finite KM through a small mass splitting between them. Although this setup provided for interesting phenomenology, the inclusion of portal matter fields which satisfied the condition $\sum_i Q_{D_i} Q_{Y_i} = 0$ was ultimately \emph{ad hoc}: In spite of the critical importance that this cancellation had to the model, it did not happen ``naturally''. Furthermore, the $U(1)_D$ itself was still minimal -- the potentially new effects arising from a non-minimal set of dark gauge symmetries in which $U(1)_D$ could be embedded, such as the possibility that the SM fields were non-trivially charged under some part of the extended gauge group orthogonal to $U(1)_D$, were not explored. \textbf{II} began the process of addressing both of these questions. First, it was noted that the required cancellation to render Eq.(\ref{KMSum}) finite would naturally occur when the portal matter was placed in the vector-like representation $\mathbf{5}+\mathbf{\overline{5}}$ of $SU(5)$, which would then be broken down to the SM gauge group. Inspired by $E_6$ theories in which the $\mathbf{27}$ of $E_6$ breaks down to a $(\mathbf{\overline{5}},\mathbf{2})+(\mathbf{1},\mathbf{2})+(\mathbf{5},\mathbf{1})+(\mathbf{10},\mathbf{1})$ of $SU(5)\times SU(2)$ (allowing for the complete SM to be contained in a $\mathbf{\overline{5}}+\mathbf{10}$, with a vector-like set of portal fields in $\mathbf{\overline{5}}+\mathbf{5}$), \textbf{II} then developed a model based on a gauge group $SU(5)\times SU(2)_I \times U(1)_{Y_I}$. This setup exhibited a number of intriguing properties. For example, the extended dark gauge sector led to new heavy gauge bosons associated with the dark sector but coupled to the SM, as well as non-standard dark photon couplings emerging from mass mixing with SM gauge bosons (the latter a consequence of the dark group $U(1)_D$ itself only emerging as a combination of $SU(2)_I$ and $U(1)_{Y_I}$ generators). The model in \textbf{II} also exhibited potentially interesting flavor physics behavior, including flavor-changing neutral currents and flavor-dependent couplings of new gauge bosons to SM fermions. However, the setup as written had comparatively little underlying theoretical structure for the shape and magnitudes that these effects might take.

We wish now to expand on the work of \textbf{II}, and in particular the flavor physics questions it raised, by considering the following: If the extended gauge group containing $U(1)_D$ can include groups under which the SM particles are non-trivial representations, might we create a model in which the dark photon gauge symmetry $U(1)_D$ might be unified, either partially or completely, with some sort of flavor symmetry? With that goal in mind, we develop a specific model: First, we extend the SM to a Pati-Salam symmetry $SU(4)_c \times SU(2)_L \times SU(2)_R$ \cite{Pati:1974yy}, which, as a semisimple group, we shall prove in Section \ref{MatterContentSection} will guarantee that KM from contributions of the form of Eq.(\ref{KMSum}) will remain finite and calculable. For our ``dark sector'', in which the dark photon $U(1)_D$ shall be embedded, we choose the $SU(4)_F \times U(1)_F$ group. We shall see that this group is large enough to contain both a $U(1)_D$ under which the SM fields can remain uncharged, \emph{and} an embedded $SU(3)_F$ group describing quark flavor. Multiplets containing the SM fields then are placed in fundamental (and antifundamental) representations of the $SU(4)_F$ group, with the three SM generations forming triplets (and antitriplets) under $SU(3)_F$ and portal matter fields representing a ``fourth generation'' that are singlets under $SU(3)_F$. From this outline, we create a phenomenologically realistic theory in which an $SU(3)_F$ flavor symmetry  is partially unified with a vector portal for DM.

Our paper is laid out as follows: In Section \ref{ModelSetupSection}, we discuss the field content of the model and the masses of any new exotic particles we require. In addition to explicitly computing the masses and mixings of all phenomenologically relevant new fermions and gauge bosons, we comment qualitatively on several important aspects of the scalar sector, since a rigorous treatment of this sector of the model is complex enough to lie beyond the scope of the present work. In this Section, we also note that while the UV model is enormously complicated, containing a variety of new heavy fermions, the overwhelming majority of the new particles introduced will have a high enough mass scale to have very little phenomenological relevance. As such, while we make brief comments about the model at the high scale, we shall focus the majority of the paper on the comparatively few new particles which are light enough to have observable effects in the present generation of experiments. In Section \ref{CouplingsMixingsandParamsSection}, we establish several important phenomenological tools required for our later analysis: In particular, we explicitly determine the couplings of the model's fermions to both the new gauge bosons and to the SM gauge bosons and the Higgs, and determine what a ``natural'' range for our model's parameters might be. In this Section, we also explicitly compute the magnitude of the KM effect, \ie, $\epsilon$ in the model, and comment briefly on how it constrains our parameter space. In Section \ref{FCNCSection}, we compute the phenomenologically significant flavor-changing neutral currents appearing in the model, and compare them with existing experimental constraints. In Section \ref{ExoticMatterSection}, we offer a brief survey of the phenomenology of the new fields that may be produced at collider experiments, in particular focusing on how our model here differs from the results seen in \textbf{I} and \textbf{II}. Finally, Section \ref{ConclusionSection} summarizes our results and discusses possible avenues with which this work might be expounded upon.

\section{Model Setup}\label{ModelSetupSection}

In this Section, we shall outline the components of our model and demonstrate how realistic flavor structure, at least in the highly-constrained quark sector, may be effected. It should be noted that in order to attain the observed quark masses and CKM mixings, the model presented here is exceedingly complicated and requires prodigious fine-tuning. However, this finely-tuned sector of the model is associated with the new $SU(3)_F$ flavor symmetry, which generally must be broken at multi-TeV scales. At energies which may be probed experimentally, we shall find that the model is mostly agnostic to the particulars of how the $SU(3)_F$ symmetry is broken. In the interest of completeness, then, we present a construction of our model in which the $SU(3)_F$ symmetry is broken in a manner consistent with the observed quark masses and CKM parameters, but we note that the behavior of the model is in most cases insensitive to the specifics of the model in the multi-TeV regime. When we discuss the phenomenology of the model in later Sections, we shall explicitly note when the results may be sensitive to modifications of the multi-TeV sector of the theory.

\subsection{Fermion Content}\label{MatterContentSection}

To begin our construction, it is useful to make several assumptions. First, to ensure that our fermions will always produce a finite and calculable kinetic mixing between SM hypercharge and any new $U(1)$'s appearing in our model, we will restrict our efforts to theories where the gauge group can be written $\mathcal{G} \times \mathcal{G}_P$, where $\mathcal{G}_P$ is an arbitrary gauge group containing a dark $U(1)$ symmetry, and $\mathcal{G}$ is some \emph{semisimple} group containing the SM. Our inclusion of the condition that $\mathcal{G}$ is semisimple now guarantees that once $\mathcal{G}$ is broken down to a group which includes $U(1)$ factors, any kinetic mixing between these $U(1)$ factors and any $U(1)$ contained in $\mathcal{G}_P$ generated by one-loop contributions of the form of Eq.(\ref{KMSum}) will be finite. 

We can prove this claim of finiteness straightforwardly: To start, we consider a field in the representation $\mathcal{R}_1 \times \mathcal{R}_2$ of $\mathcal{G} \times \mathcal{G}_P$. Then, consider an arbitrary $U(1)$ embedded in $\mathcal{G}$, the generator for which we'll call $T_Y$, and an arbitrary $U(1)$ embedded in $\mathcal{G}_P$, the generator for which we'll call $T_P$. We can demonstrate that for \emph{any} representation $\mathcal{R}_1 \times \mathcal{R}_2$ of $\mathcal{G} \times \mathcal{G}_P$, the loop-level contribution to the kinetic mixing between the two $U(1)$ groups is finite and calculable. We know from Eq.(\ref{KMSum})that this contribution to the KM coefficient $\epsilon$ will be finite and calculable as long as the trace $Tr[T_Y T_P]=0$, where the generators $T_Y$ and $T_P$ are given for the representation $\mathcal{R}_1 \times \mathcal{R}_2$. However, since this representation is a direct product, this trace is simply $Tr[T_Y]\times Tr[T_P]$. For a semisimple group, the trace $Tr[T_Y]$ is always zero for any $U(1)$ generator and any representation. As as a result, $Tr[T_Y T_P]=0\times Tr[T_P]=0$, and the contribution of any representation under $\mathcal{G}\times \mathcal{G}_P$ to KM will always be finite and calculable for semisimple $\mathcal{G}$.\footnote{There are two caveats worth mentioning here: First, this works equally well assuming $\mathcal{G}_P$, rather than $\mathcal{G}$, is semisimple, however, given the large number of extensions of the SM gauge group to a semisimple one (Pati-Salam, trinification, and all GUT models), we have elected here to restrict our attention to the scenario where $\mathcal{G}$ is semisimple. Second, we note that in the high-energy theory, the absence of an independent $U(1)$ factor in $\mathcal{G}$ (since $U(1)$ is not simple) means that KM arising from vacuum polarization diagrams as described in the Introduction are forbidden by gauge invariance. However, we note that these terms still arise via effective operators featuring the insertion of scalar vev's when these gauge symmetries are broken-- we shall argue in Section \ref{KMSection} that this does not vitiate our KM calculation there, at least for the purposes of evaluating the magnitude of the mixing.}

For our purposes, we shall select (as mentioned in the Introduction) the Pati-Salam gauge group $SU(4)_c \times SU(2)_L \times SU(2)_R$ as our semisimple group $\mathcal{G}$ \cite{Pati:1974yy}.
To ensure that any new exotic fermions occur in representations that will quickly decay into SM fields, we then assume that all fermions in our model occur in the traditional Pati-Salam multiplets $(\mathbf{4},\mathbf{2},\mathbf{1})$ and $(\mathbf{\overline{4}},\mathbf{1},\mathbf{2})$, or their conjugate representations.\footnote{This is not strictly necessary, since there exist other Pati-Salam representations that contain only fields with SM-like quantum numbers. However, we shall stick to the familiar canonical choices for simplicity.} As our work above for a general $\mathcal{G}$ would suggest, it is straightforward to see that each of these multiplets has $Tr[Y_{SM}]=0$, and hence, if we only introduce new matter in these representations, then regardless of its transformation properties under $\mathcal{G}_P$, the new fermions will invariably produce a finite and calculable kinetic mixing between the SM hypercharge and any new $U(1)\subset \mathcal{G}_P$. Dark matter, if we wish it to be fermionic, can then be introduced as a Pati-Salam singlet (or collection of such singlets) or as the SM singlet component of a $(\mathbf{4},\mathbf{1},\mathbf{2})$ multiplet, with some non-trivial charge under $U(1)_D$. For the sake of brevity, we shall employ some further simplifying notation: We shall denote the groups $SU(N)_A \times SU(M)_B \times ... \times U(1)_C$ by $N_A M_B ... 1_C$, so that, for example, $\mathcal{G}_{PS} = 4_c 2_L 2_R$, and the SM gauge group $SU(3)_c \times SU(2)_L \times U(1)_Y$ is simply written $3_c 2_L 1_Y$.

Our choice of $\mathcal{G}_P$ is then further restricted by our requirement that the model incorporate three generations of the usual SM fermion fields, all of which must be uncharged under $U(1)_D$. A simple method of accomplishing this is to assume that $\mathcal{G}_P$ contains an $SU(3)$ flavor symmetry, $SU(3)_F$, the generators of which are orthogonal to the generators of the $U(1)_D$ symmetry. That is, we can decompose $\mathcal{G}_P \rightarrow SU(3)_F \times \mathcal{G'}_P$, with $\mathcal{G'}_P \supset U(1)_D$. Then, if we can find a model which gives the complete set of SM fermions in fundamental or antifundamental representations of $SU(3)_F$ with zero charge under $U(1)_D$, we'll have recreated all three generations of the SM. The sole remaining requirement is that all other additional fermions in the model be vector-like once $\mathcal{G'}_P$ is broken down to $U(1)_D$, leaving the SM as the only `light' chiral matter content.

Precisely this sort of construction can be arranged if we take, \eg, $\mathcal{G}_P = SU(4)_F \times U(1)_F= 4_F 1_F$, and write the matter content of our theory in terms of left-handed multiplets (not to be confused with the $SU(2)_L$ gauge group) of the form $4_c 2_L 2_R 4_F 1_F$ as
\begin{align}\label{originalMatterContent}
    (\mathbf{4},\mathbf{2},\mathbf{1},\mathbf{4},-1/4)+(\mathbf{\overline{4}},\mathbf{1},\mathbf{2},\mathbf{\overline{4}},+1/4)+(\mathbf{\overline{4}},\mathbf{2},\mathbf{1},\mathbf{1},+1)+(\mathbf{4},\mathbf{1},\mathbf{2},\mathbf{1},-1).
\end{align}
By breaking $4_F 1_F \rightarrow 3_F 1_{F'} 1_F \rightarrow 3_F 1_D$, we can then decompose this field content into
\begin{align}\label{brokenOriginalMatterContent}
    (\mathbf{4},\mathbf{2},\mathbf{1},\mathbf{3},0)+(\mathbf{\overline{4}},\mathbf{1},\mathbf{2},\mathbf{\overline{3}},0)+(\mathbf{4},\mathbf{2},\mathbf{1},\mathbf{1},-1)+(\mathbf{\overline{4}},\mathbf{2},\mathbf{1},\mathbf{1},+1)+(\mathbf{4},\mathbf{1},\mathbf{2},\mathbf{1},-1)+(\mathbf{\overline{4}},\mathbf{1},\mathbf{2},\mathbf{1},+1),
\end{align}
where multiplets are now labelled according to $4_c 2_L 2_R 3_F 1_D$. As is readily apparent from Eq.(\ref{brokenOriginalMatterContent}), all of our criteria are satisfied: The SM is contained in a chiral $SU(3)_F$ triplet and an antitriplet, which only gain mass at the scale of electroweak symmetry breaking, while all additional fields, which might serve as portal matter to contribute to kinetic mixing, are vector-like and therefore able to obtain far larger masses. Inspection of Eq.(\ref{originalMatterContent}) also shows this model to be anomaly-free.

In order to effect realistic flavor mixing within this setup, we finally introduce additional vector-like multiplets to act as seesaw partners to the SM fermions, analogous to the treatment of the charged leptons and quarks in the model construction of \cite{Bao:2015pva}. Specifically, we find that adding a vector-like multiplet in the $(\mathbf{4},\mathbf{1},\mathbf{2},\mathbf{4},-1/4)+(\mathbf{\overline{4}},\mathbf{1},\mathbf{2},\mathbf{\overline{4}},+1/4)$ representation to the matter content of Eq.(\ref{originalMatterContent}) allows us, with appropriate selections for the model's scalar content, to exhibit the observed mass hierarchy for quarks and charged leptons, as well as the SM CKM matrix. We also note that in order to ensure the existence of light, SM-like fermions, certain mass terms will have to be forbidden by imposing a discrete $Z_2$ symmetry on our model, discussed in detail in Section \ref{FermionSpectrumSection}. Our final matter content can then be written as 
\begin{align}\label{finalGUTMatterContent}
    (\mathbf{4},\mathbf{2},\mathbf{1},\mathbf{4},-1/4)+(\mathbf{\overline{4}},\mathbf{1},\mathbf{2},\mathbf{\overline{4}},+1/4) &+(\mathbf{\overline{4}},\mathbf{2},\mathbf{1},\mathbf{1},+1) +(\mathbf{4},\mathbf{1},\mathbf{2},\mathbf{1},-1)\\
    &+(\mathbf{4},\mathbf{1},\mathbf{2},\mathbf{4},-1/4)+(\mathbf{\overline{4}},\mathbf{1},\mathbf{2},\mathbf{\overline{4}},+1/4). \nonumber
\end{align}
For convenience, we have listed this matter content, labelled by Pati-Salam and SM multiplets, in Table \ref{table:fermions}. We have also labelled the various multiplets in Table \ref{table:fermions} based on their SM quantum numbers and their role in the theory: $q_L$, $d^c_R$, and $u^c_R$ represent (up to small mass mixing with vector-like fermions) $SU(2)_L$ doublet, down-like $SU(2)_L$ singlet, and up-like $SU(2)_L$ singlet quarks, respectively, while $l_L$, $e^c_R$, and $\nu^c_R$ represent the same for $SU(2)_L$ doublet leptons, charged $SU(2)_L$ singlet leptons, and sterile neutrinos respectively. The multiplets labelled by capitalized letters represent (again up to small mass mixing) the new vector-like particles in the theory, which act as portal matter and serve as seesaw partners to the SM states.

The addition of the new vector-like states in Eq.(\ref{finalGUTMatterContent}), however, raises two issues that warrant some discussion. First, we note that these extra states vitiate any discrete left-right symmetry that previously existed in the model, and which is frequently assumed in Pati-Salam constructions. Our selection here is ultimately one of convenience: The additional vector-like fields represent the \emph{minimal} additions we need to effect the observed SM flavor structure, at least via the method outlined in Section \ref{FermionSpectrumSection}. In principle, nothing prevents us from adding a corresponding vector-like $SU(2)_L$ multiplet to restore the left-right discrete symmetry. In practice, however, the addition of these states significantly complicates the numerical structure of the model, so for our purposes here we simply include the extra $SU(2)_R$ vector-like multiplet. If we wish to maintain a left-right discrete symmetry, we can assume that the particles of the corresponding $SU(2)_L$ vector-like multiplet are significantly more massive than those of the $SU(2)_R$ multiplet, and therefore do not have a significant numerical effect on the theory at low energy. We also note that even if we were to include an $SU(2)_L$ multiplet with similar particle masses to those of its $SU(2)_R$ counterpart, the effect on low-energy phenomenology would be minimal; we shall see in Section \ref{FermionSpectrumSection} that the natural masses of the particles in the extra vector-like multiplets are so high that only one of these vector-like particles, acting as a seesaw partner to the top quark, will be accessible at the LHC or any likely future planned colliders. We therefore can determine that the effect of adding an extra vector-like multiplet to the theory to restore left-right symmetry would, even if the masses of the particles in the multiplet were comparable, likely only result in the introduction of a single additional particle, a vector-like $SU(2)_L$ doublet quark, at accessible energies. Given the substantial increase in numerical complexity coupled with relatively limited phenomenological impact, we leave an exploration of the strict enforcement of discrete left-right symmetry in this model to future work.

We also note that the fermion content of Eq.(\ref{finalGUTMatterContent}) pushes us over the constraint for asymptotically free QCD: If the Pati-Salam $SU(4)_c$ symmetry is broken down to the SM QCD at a higher scale than any of these fermions acquire mass, then the theory has 18 flavors, while QCD only remains asymptotically free for 16 or fewer flavors. While we wish to focus our efforts on the low-energy implications of this model, rather than the complexities associated with its full UV completion, the abandonment of asymptotic freedom does merit some justification. We might note that as long as we don't introduce any further colored fermions (as we would need to in order to, for example, enforce a discrete left-right symmetry), the matter content of Eq.\ref{finalGUTMatterContent} should result in a confined $SU(4)_c$ theory (even accounting for the existence of an $SU(4)_c$ adjoint scalar to break this symmetry).
In fact, using the rough order-of-magnitude estimates for the various fermionic masses in the theory given in Section \ref{FermionSpectrumSection}, we find that $\alpha_s$ remains comfortably perturbative even if $SU(4)_c$ is broken as high as the Planck scale, due to the fact that a number of fermion fields must have masses in the high multi-TeV range in order to recreate the observed fermion mass hierarchy. Given that the Pati-Salam symmetry is generally broken at significantly sub-Planckian energies \cite{Saad:2017pqj,Chamseddine:2015ata,Hartmann:2014fya}, we can safely assume that for this construction, our abandonment of asymptotic freedom does not present a severe difficulty.
% However, given the fact that we shall assume the Pati-Salam symmetry, including $SU(4)_c$, will be broken at an exceedingly high scale $\sim O(10^{13})$ TeV (following setups such as \cite{Saad:2017pqj,Chamseddine:2015ata,Hartmann:2014fya}), it is quite likely irresponsible to simply assume that the $SU(3)_c$ theory won't reach its Landau pole before $SU(4)_c$ symmetry is restored and asymptotic freedom regained.
For more general cases, we also note that it has been conjectured that it's not unreasonable to abandon asymptotic freedom at some intermediate scales, such as the TeV-scale, in exchange for asymptotic \emph{safety}, where a UV interacting fixed point is reinstated by physics at some high energy before the Planck scale \cite{Sannino:2015sel}. As our discussion in this paper is limited to physics well below the Planck scale, we won't conjecture as to how this asymptotic freedom is restored, and only note that such a restoration may be possible, even in setups with an even greater number of excess fermions than the minimal construction we employ here.

\begin{table}[h!]
\centering
\begin{tabular}{| c | l | c | c | c | c | c|} 
 \hline
 Fermion & $4_c 2_L 2_R 4_F 1_F$ Multiplet & SM Label & $SU(3)_c$ & $Y/2$ & $SU(3)_F$ & $U(1)_D$\\ [0.5ex] 
 \hline
 \multirow{4}{4em}{$\Psi_a$} & \multirow{4}{4em}{$(\mathbf{4},\mathbf{2},\mathbf{1},\mathbf{4},-1/4)$} & $q_L$ & $\mathbf{3}$ & $+1/6$ & $\mathbf{3}$ & 0\\
 & & $l_L$ & $\mathbf{1}$ & $-1/2$ & $\mathbf{3}$ & 0 \\
 & & $Q_L$ & $\mathbf{3}$ & $+1/6$ & $\mathbf{1}$ & -1 \\
 & & $L_L$ & $\mathbf{1}$ & $-1/2$ & $\mathbf{1}$ & -1 \\
 \hline
 \multirow{2}{4em}{$\Psi_b$} & \multirow{2}{4em}{$(\mathbf{\overline{4}},\mathbf{2},\mathbf{1},\mathbf{1},+1)$} & $Q_R^c$ & $\mathbf{\overline{3}}$ & $-1/6$ & $\mathbf{1}$ & +1 \\
 & & $L_R^c$ & $\mathbf{1}$ & $+1/2$ & $\mathbf{1}$ & +1 \\
 \hline
 \multirow{8}{4em}{$\Psi_c$} & \multirow{8}{4em}{$(\mathbf{\overline{4}},\mathbf{1}, \mathbf{2}, \mathbf{\overline{4}},+1/4)$} & $d_R^c$ & $\mathbf{\overline{3}}$ & +1/3 & $\mathbf{\overline{3}}$ & 0 \\
 & & $u_R^c$ & $\mathbf{\overline{3}}$ & -2/3 & $\mathbf{\overline{3}}$ & 0 \\
 & & $e_R^c$ & $\mathbf{1}$ & +1 & $\mathbf{\overline{3}}$ & 0 \\
 & & $\nu_R^c$ & $\mathbf{1}$ & 0 & $\mathbf{\overline{3}}$ & 0 \\
 & & $(D_1)_R^c$ & $\mathbf{\overline{3}}$ & +1/3 & $\mathbf{1}$ & +1 \\
 & & $(U_1)_R^c$ & $\mathbf{\overline{3}}$ & -2/3 & $\mathbf{1}$ & +1 \\
 & & $(E_1)_R^c$ & $\mathbf{1}$ & +1 & $\mathbf{1}$ & +1 \\
 & & $(N_1)_R^c$ & $\mathbf{1}$ & 0 & $\mathbf{1}$ & +1 \\
 \hline
 \multirow{4}{4em}{$\Psi_d$} & \multirow{4}{4em}{$(\mathbf{4},\mathbf{1},\mathbf{2},\mathbf{1},-1)$} & $(D_1)_L$ & $\mathbf{3}$ & -1/3 & $\mathbf{1}$ & -1 \\
 & & $(U_1)_L$ & $\mathbf{3}$ & +2/3 & $\mathbf{1}$ & -1 \\
 & & $(E_1)_L$ & $\mathbf{1}$ & -1 & $\mathbf{1}$ & -1 \\
 & & $(N_1)_L$ & $\mathbf{1}$ & 0 & $\mathbf{1}$ & -1 \\
 \hline
 \multirow{8}{4em}{$\Psi_e$} & \multirow{8}{4em}{$(\mathbf{4},\mathbf{1}, \mathbf{2}, \mathbf{4},-1/4)$} & $(D_2)_L$ & $\mathbf{3}$ & -1/3 & $\mathbf{3}$ & 0 \\
 & & $(U_2)_L$ & $\mathbf{3}$ & +2/3 & $\mathbf{3}$ & 0 \\
 & & $(E_2)_L$ & $\mathbf{1}$ & -1 & $\mathbf{3}$ & 0 \\
 & & $(N_2)_L$ & $\mathbf{1}$ & 0 & $\mathbf{3}$ & 0 \\
 & & $(D_3)_L$ & $\mathbf{3}$ & -1/3 & $\mathbf{1}$ & -1 \\
 & & $(U_3)_L$ & $\mathbf{3}$ & +2/3 & $\mathbf{1}$ & -1 \\
 & & $(E_3)_L$ & $\mathbf{1}$ & -1 & $\mathbf{1}$ & -1 \\
 & & $(N_3)_L$ & $\mathbf{1}$ & 0 & $\mathbf{1}$ & -1 \\
 \hline
 \multirow{8}{4em}{$\Psi_f$} & \multirow{8}{4em}{$(\mathbf{\overline{4}},\mathbf{1}, \mathbf{2}, \mathbf{\overline{4}},+1/4)$} & $(D_2)_R^c$ & $\mathbf{\overline{3}}$ & +1/3 & $\mathbf{\overline{3}}$ & 0 \\
 & & $(U_2)_R^c$ & $\mathbf{\overline{3}}$ & -2/3 & $\mathbf{\overline{3}}$ & 0 \\
 & & $(E_2)_R^c$ & $\mathbf{1}$ & +1 & $\mathbf{\overline{3}}$ & 0 \\
 & & $(N_2)_R^c$ & $\mathbf{1}$ & 0 & $\mathbf{\overline{3}}$ & 0 \\
 & & $(D_3)_R^c$ & $\mathbf{\overline{3}}$ & +1/3 & $\mathbf{1}$ & +1 \\
 & & $(U_3)_R^c$ & $\mathbf{\overline{3}}$ & -2/3 & $\mathbf{1}$ & +1 \\
 & & $(E_3)_R^c$ & $\mathbf{1}$ & +1 & $\mathbf{1}$ & +1 \\
 & & $(N_3)_R^c$ & $\mathbf{1}$ & 0 & $\mathbf{1}$ & +1 \\
 \hline
\end{tabular}
\caption{The fermion content of the model, grouped by their representation under $4_c 2_L 2_R 4_F 1_F$}
\label{table:fermions}
\end{table}

\subsection{Scalar Content}\label{ScalarContentSection}
In order to break the gauge symmetry in our model from its original GUT-scale $4_c 2_L 2_R 4_F 1_F$ to the SM gauge group, and then down to $3_c 1_{\textrm{em}}$, as well as generate the appropriate spectrum for SM fermions, we must posit the existence of a significant number of scalar fields. For simplicity, we shall assume that the Pati-Salam symmetries $SU(4)_c$ and $SU(2)_R$ are broken at some exceedingly high scale (or scales). We shall see that this is not in fact an unreasonable assumption -- if we were to assume a discrete left-right symmetry, for example, renormalization group running of the $SU(2)_L$ and $SU(2)_R$ in a minimal Pati-Salam model suggests symmetry breaking around $\sim 10^{13-14} \; \textrm{GeV}$ \cite{Hartmann:2014fya,Saad:2017pqj,Chamseddine:2015ata}, while the highest scales we shall require for breaking the $SU(4)_F \times U(1)_F$ symmetry and reproducing the fermion spectra and mixings will not exceed $O(10^{8-9} \; \textrm{GeV})$. Assuming the Pati-Salam symmetry is broken, then, we are now left with the need for scalars which break the $SU(4)_F \times U(1)_F$ symmetry down to a dark $U(1)$, which we shall call $U(1)_D$, which is then broken entirely at some low scale at roughly $O(0.1-1 \; \textrm{GeV})$, as well as some scalar content to perform the role of the SM Higgs, breaking $2_L 1_Y \rightarrow 1_{\textrm{em}}$. To accomplish these tasks, we posit 5 Higgs scalars, given in the $4_c 2_L 2_R 4_F 1_F$ representations as
\begin{align}\label{ScalarContent}
    &\Phi_A \sim (\mathbf{1}, \mathbf{1}, \mathbf{1}, \mathbf{15}, 0), \nonumber \\
    &\Phi_B \sim (\mathbf{1}, \mathbf{1}, \mathbf{3}, \mathbf{15}, 0), \nonumber \\
    &\Phi_S \sim (\mathbf{1}, \mathbf{1}, \mathbf{1}, \mathbf{1}, 0), \\
    &\Phi_P \sim (\mathbf{1}, \mathbf{1}, \mathbf{1}, \mathbf{4}, +3/4), \nonumber \\
    &H \sim (\mathbf{1}, \mathbf{2}, \mathbf{2}, \mathbf{1}, 0). \nonumber
\end{align}
For convenience, we have listed these scalars, along with rough orders of magnitude for their vacuum expectation values necessary to achieve the observed SM quark mass spectrum (which we shall derive in Section \ref{FermionSpectrumSection}), in Table \ref{table:scalars}. The scalar content that we have selected here can be easily deduced from phenomenological considerations: The scalars $\Phi_A$, $\Phi_B$, $\Phi_S$, and $H$ are all immediately required for a treatment of the quark masses analogous to that of \cite{Bao:2015pva}, generalized to treat our larger model. In \cite{Bao:2015pva}, the masses and mixings of the quarks and leptons are reproduced with the aid of a traditional SM Higgs doublet, two $SU(3)_F$ adjoint scalars, and a singlet. To adapt the construction of that work to the present scenario, we require that the $SU(3)_F$ adjoints in that work are promoted to $SU(4)_F$ adjoints, that one of the two $SU(4)_F$ adjoints also be a triplet of $SU(2)_R$ in order to effect the large discrepancy between up-like and down-like quark masses (and the non-trivial CKM matrix) without relying on renormalization group evolution, and that the scalar $H$ containing the SM Higgs field is promoted to a bidoublet under $SU(2)_L \times SU(2)_R$. It is useful to note that the scalar content of Eq.(\ref{ScalarContent}) does not contain any scalars that are charged under both $U(1)_Y$ and $U(1)_F$. Loop-level kinetic mixing between these two $U(1)$ symmetries, then, such as that mediated by the fermion content in our model, does NOT occur in the scalar sector. In the scalar content of Table \ref{table:scalars}, the addition of the singlet $\Phi_S$ may seem superfluous, and so merits some further clarification. As we shall see in Section \ref{FermionSpectrumSection}, the most general set of Yukawa couplings that can be written with the contents of Tables \ref{table:fermions} and \ref{table:scalars} does \emph{not} reproduce the fermion spectrum accurately. In order to forbid undesirable Yukawa terms, the authors of \cite{Bao:2015pva} impose a $Z_2$ parity under which certain fermion fields are odd and others are even. However, in both this work and \cite{Bao:2015pva}, this $Z_2$ parity forbids one mass term that is still necessary for the recreation of realistic fermion masses, and so this term is reintroduced with the $Z_2$-odd scalar $\Phi_S$.

\begin{table}[h!]
\centering
\begin{tabular}{|c | c | c | c | c | c|} 
 \hline
 Scalar & $SU(2)_L$ & $SU(2)_R$ & $SU(4)_F$ & $U(1)_F$ & $\langle \Phi \rangle$ \\ [0.5ex] 
 \hline
 $\Phi_A$ & $\mathbf{1}$ & $\mathbf{1}$ & $\mathbf{15}$ & $0$ & $\begin{pmatrix}
 \langle A \rangle & \vec{\gamma}_A \\
 \vec{\gamma}_A^\dagger & -Tr[\langle A \rangle]
 \end{pmatrix} \sim \begin{pmatrix}
 10^3-10^9 \; \textrm{GeV} & 0.1-1 \; \textrm{GeV} \\
 0.1-1 \; \textrm{GeV} & 10^9 \; \textrm{GeV}
 \end{pmatrix}$\\ 
 \hline
 $\Phi_B$ & $\mathbf{1}$ & $\mathbf{3}$ & $\mathbf{15}$ & 0 & $\frac{\sigma_3}{2} \otimes \begin{pmatrix}
 \langle B \rangle & \vec{\gamma}_B \\
 \vec{\gamma}_B^\dagger & -Tr[\langle B \rangle]
 \end{pmatrix} \sim \begin{pmatrix}
 10^3-10^9 \; \textrm{GeV} & 0.1-1 \; \textrm{GeV} \\
 0.1-1 \; \textrm{GeV} & 10^9 \; \textrm{GeV}
 \end{pmatrix}$\\
 \hline
 $\Phi_P$ & $\mathbf{1}$ & $\mathbf{1}$ & $\mathbf{4}$ & +3/4 & $(\vec{\gamma}_P, v_P) \sim (0.1-1 \; \textrm{GeV}, \; 10^3-10^4 \; \textrm{GeV})$  \\
 \hline
 $\Phi_S$ & $\mathbf{1}$ & $\mathbf{1}$ & $\mathbf{1}$ & 0  & $v_S \sim 10^3-10^4 \; \textrm{GeV}$\\
 \hline
 $H$ & $\mathbf{2}$ & $\mathbf{2}$ & $\mathbf{1}$ & 0 & $\frac{v}{\sqrt{2}}\begin{pmatrix}
 \cos \beta & 0 \\
 0 & \sin \beta
 \end{pmatrix}$ \\ [1ex] 
 \hline
 
\end{tabular}
\caption{The scalars introduced to break the $4_F 1_F$ flavor symmetry and provide masses to the fermions. Here, $\langle A \rangle$ and $\langle B \rangle$ are $3 \times 3$ Hermitian matrices, $\vec{\gamma}_{A,B,P}$ are 3-component complex vectors, and $v_P$, $v_S$, and $v_{EW}$ are simply real vevs, with $v \approx 246 \; \textrm{GeV}$ being the SM Higgs vev. In the definition of $\Phi_B$'s vev, $\sigma_3$ denotes the third Pauli matrix.}
\label{table:scalars}
\end{table}

It is helpful here to use the labelling of Table \ref{table:scalars} to illustrate the rough pattern of symmetry breaking from the scale where the Pati-Salam group is broken down to the SM group $3_c 2_L 1_Y$ down to the scale where the dark photon gauge symmetry $U(1)_D$ is broken -- or in other words, how the $SU(4)_F \times U(1)_F$ symmetry is broken.
\begin{align}\label{FlavorSymmetryBreaking}
    SU(4)_F \times U(1)_F \xrightarrow{\langle A,B \rangle} U(1)_{F}' \times U(1)_F \xrightarrow{v_P} U(1)_D \xrightarrow{\vec{\gamma}_{A,B,P}} \textrm{Nothing}.
\end{align}
In Eq.(\ref{FlavorSymmetryBreaking}), we have simply listed how the various vev terms from Table \ref{table:scalars} break the $SU(4)_F \times U(1)_F$ symmetry down, arranged by the rough scale at which each breaking occurs. Consulting Table \ref{table:scalars}, we can see that $SU(4)_F$ is broken by $\langle A \rangle$ and $\langle B \rangle$ down to $U(1)_F'$ at a series of scales spanning between $\sim 10^9 \; \textrm{GeV}$ and $10^3 \; \textrm{GeV}$ (in practice, we shall that this symmetry breaking usually completes at a substantially higher scale than $10^3 \; \textrm{GeV}$), while $v_P$ then breaks $U(1)_F' \times U(1)_F$ down to $U(1)_D$ at a scale of roughly $10^{3-4}$ GeV, and finally the small vev's $\vec{\gamma_{A,B,P}}$ break $U(1)_D$ entirely at a scale of $\sim 0.1-1 \; \textrm{GeV}$. The contents of Table \ref{table:scalars} highlights two notable characteristics of our vev arrangements. First, there exist substantial hierarchies between the various vev scales in the model, the values of which span nine orders of magnitude. As we'll see in Section \ref{FermionSpectrumSection}, these hierarchies are necessary to recreate the observed hierarchy in SM quark masses, as well as ensure that the dark photon remains at roughly the scale of $\sim 1 \; \textrm{GeV}$. Due to the significant complexity of the scalar sector here, with numerous possible potential terms, a detailed exploration on the nauralness of these hierarchies is beyond the scope of this paper, however it is clear that any potential which might produce these vev's is likely quite finely tuned.

The second notable characteristic of our vev's evinced in Table \ref{table:scalars} is that the vacuum expectation values of certain components of the scalar fields are exceedingly large (we remind the reader that even the largest scales here, $\sim 10^9 \; \textrm{GeV}$, are still well below what we can anticipate for the breaking of the symmetries $SU(4)_c$ and $SU(2)_R$, which as noted before we assume to be $\sim 10^{13-14} \; \textrm{GeV}$). As a result, we can anticipate (and, in Sections \ref{FermionSpectrumSection} and \ref{GaugeSpectrumSection}, prove) that much of the new physics which shall arise in this model appears at scales far in excess of what can be directly probed in the foreseeable future. The task left to us is therefore identifying the elements of our model construction that yield new physics at scales that we do expect to be probed in the near future. As with the problem of quantitatively evaluating the vev arrangement's tuning, performing this task in detail is far beyond the scope of the present work, given the large number of multiplets and the resulting high degree of complexity for the scalar potential.

We can, however, briefly comment on some of the scalars that will likely be relevant at accessible energies. First, we note that if the left-right symmetry breaking scale is quite high (as we have already assumed), the SM Higgs field is well-approximated as the sole element of the bidoublet $H$ that achieves a non-zero vev, so that
\begin{align}\label{LightHiggsState}
    H = \begin{pmatrix}
    \phi_1^0 & \phi_2^+ \\
    \phi_1^- & -\phi_2^{0*}
    \end{pmatrix},
    \;\;\langle H \rangle = \frac{v}{\sqrt{2}}\begin{pmatrix}
    \cos \beta & 0 \\
    0 & \sin \beta
    \end{pmatrix}, \;\; h_{SM} \approx \phi_1^0 \cos \beta + \phi_2^0 \sin \beta,
\end{align}
where $h_{SM}$ denotes the real scalar identified with the SM Higgs boson, and $v \approx 246 \; \textrm{GeV}$ is the SM Higgs vev. In immediate analogy with conventional left-right symmetric models, additional physical scalars associated with the bidoublet $H$ will all have masses at approximately the scale of $SU(2)_R$ breaking \cite{Maiezza:2016ybz,Maiezza:2010ic,Gunion:1989in}, and hence given our assumptions be unobservable at present or immediately foreseeable experiments. Finally, we note that to ensure perturbativity for the Yukawa couplings of the heavy scalars arising from the bidoublet, again in analogy with well-known left-right symmetric model building principles, we must also require that $\tan \beta \lsim 0.8$ \cite{Maiezza:2010ic}.\footnote{This specifically stems from the fact that the heavy doublet's Yukawa coupling to SM quarks is enhanced by $\sec(2 \beta)$, so having $\tan \beta$ overly close to 1 will result in an unacceptably large magnitude of heavy scalar couplings to the third-generation quarks. In principle, one might accomodate this bound by having $\tan \beta \gsim 1.2$, but this is identical to simply choosing an appropriate $\tan \beta \lsim 0.8$ and changing the Yukawa coupling parameters, in particular the angle $\alpha$, in the Yukawa action given in Eq.(\ref{GUTYukawa}).}

Apart from the SM Higgs, there is only one other scalar that we shall consider here. Specifically, given the exceptionally low scale at which the symmetry $U(1)_D$ is broken in comparison to the other scales of the theory, we anticipate that some combination of the scalars described in this section will emerge as a physical ``dark Higgs'' boson of mass $\sim O(|\vec{\gamma}_{P,A,B}|)$, as the physical counterpart to the combination of Goldstone bosons which become the longitudinal component of $A_D$-- such a scalar appears in, for example, the discussions of \textbf{I} and \textbf{II}. Because the precise structure of the scalar sector is too complex to be able to probe in detail in this work, we instead shall simply deduce the contribution of this scalar to relevant processes (in particular, in Sections \ref{PortalMatterSection} and \ref{TopPartnerSection} when discussing the decay processes of heavy exotic fermions) via the Equivalence Theorem \cite{GoldstoneEquivalence} where applicable, and comment briefly upon the potential effect of these scalars where such a quantitative treatment fails.

\subsection{Fermion Spectrum}\label{FermionSpectrumSection}

In order to determine the structure of the model's scalar vevs, we turn to the CKM mixing matrix and the spectrum of SM quark masses for guidance. In order to recreate the observed flavor structure of the SM, we present a construction based heavily on that of \cite{Bao:2015pva}. In the unbroken theory, we shall write the Yukawa sector of the model as
\begin{align}\label{GUTYukawa}
    \mathcal{L}_{Y} &=-y_H \cos \alpha \Psi_f^T i \sigma_2 H \Psi_a - y_H \sin \alpha \Psi_f^T i \sigma_2 \Tilde{H} \Psi_a - y_{P1} \Psi_b^T i \sigma_2 \Phi_P^\dagger \Psi_a - y_{P2} \Psi_c^T i \sigma_2 \Phi_P \Psi_d\\
    &- y_A \Psi_f^T i \sigma_2 \Phi_A \Psi_e - y_B \Psi_f^T i \sigma_2 \Phi_B \Psi_e - M \Psi_f^T i \sigma_2 \Psi_e - y_S \Psi_c^T i \sigma_2 \Phi_S \Psi_e \; + h.c. \nonumber
\end{align}
Here, $\Tilde{H} \equiv \sigma_2 H^* \sigma_2$, $\sigma_2$ is the second Pauli matrix, $\alpha$ is an arbitrary real angle between $-\pi$ and $\pi$, and $M$ is an arbitrary mass term not forbidden by any symmetries. For simplicity, we assume that all Yukawa coupling parameters here are real, as would be the case in a model with no explicit CP violation. The expression in Eq.(\ref{GUTYukawa}) is not, a priori, the most general set of Yukawa couplings that can be written with the scalar and fermion content of Tables \ref{table:fermions} and \ref{table:scalars}. However as mentioned in Section \ref{ScalarContentSection}, any additional terms can be easily eliminated by imposing a discrete $Z_2$ symmetry. Specifically, we outline our parity assignments for the fermions and scalars in Table \ref{table:Z2}.

\begin{table}[h!]
\centering
\begin{tabular}{|c | c | c|} 
 \hline
 $Z_2$ Parity & Fermions & Scalars \\
 \hline
 + & $\Psi_a, \, \Psi_b, \, \Psi_e, \, \Psi_f$ & $H, \, \Phi_P, \, \Phi_A, \, \Phi_B$ \\
 \hline
 - & $\Psi_c, \, \Psi_d$ & $\Phi_S$\\
 \hline
\end{tabular}
\caption{The parity of the model's fermion and scalar fields under $Z_2$, in order to ensure that Eq.(\ref{GUTYukawa}) is the most general set of Yukawa couplings that we can write.}
\label{table:Z2}
\end{table}

Following the parity assignments of Table \ref{table:Z2}, Eq.(\ref{GUTYukawa}) becomes the most general set of Yukawa couplings that we can construct. We can now write the up- and down-like quark mass matrices explicitly by referencing Tables \ref{table:fermions} and \ref{table:scalars}. The mass terms for the up- and down-like quarks take the form
\begin{align}
    \mathbf{\overline{U}}_R \mathcal{M}_u \mathbf{U}_L, \\
    \mathbf{\overline{D}}_R \mathcal{M}_d \mathbf{D}_L, \nonumber
\end{align}
where
\begin{align}\label{FermionFlavorBasis}
    &\mathbf{\overline{U}}_R \equiv (\overline{u}_R,(\overline{U}_2)_R, \overline{Q}_R^u, (\overline{U}_1)_R, (\overline{U}_3)_R), \nonumber\\
    &\mathbf{U}_L \equiv (u_L, (U_2)_L, Q_L^u, (U_1)_L, (U_3)_L), \\
    &\mathbf{\overline{D}_R} \equiv (\overline{d}_R, (\overline{D}_2)_R, \overline{Q}_R^d, (\overline{D}_1)_R, (\overline{D}_3)_R), \nonumber \\
    &\mathbf{D}_L \equiv (d_L, (D_2)_L, Q_L^d, (D_1)_L, (D_3)_L). \nonumber
\end{align}
Here, $Q_{L,R}^{u(d)}$ denotes the up(down)-like component of the $SU(2)_L$ doublet $Q_{L,R}$. We also remind the reader that $u_{L,R}$ and $(U_{2})_{L,R}$ are triplets of $SU(3)_F$, and therefore have three components in flavor space, in contrast to $Q_{L,R}^{u,d}$, $(U_1)_{L,R}$, and $(U_3)_{L,R}$, which each have only one. So, $\mathbf{U}_{L,R}$ and $\mathbf{D}_{L,R}$ are both 9-component vectors in flavor space, where we have grouped the $SU(3)_F$ triplets $u_{L,R}$ and $(U_2)_{L,R}$ together in the first 6 components, leaving the last 3 components for the $SU(3)_F$ singlets. Notably, the $SU(3)_F$ triplets are all uncharged under $U(1)_D$, while the $SU(3)_F$ singlets all possess a common non-zero charge under this symmetry. The mass matrices $\mathcal{M}_{u,d}$ are then given by
\begin{align}\label{FermionMassMats}
    &\mathcal{M}_u = \begin{pmatrix}
    \mathbf{0}_{3\times3} & y_S v_S \mathbf{1}_{3\times3} & \mathbf{0}_{3\times1} & y_{P2} \vec{\gamma}_P & \mathbf{0}_{3\times1} \\
    \frac{y_u v}{\sqrt{2}} \mathbf{1}_{3\times3} & \mathbf{M}_u + M \mathbf{1}_{3\times3} & \mathbf{0}_{3\times1} & \mathbf{0}_{3\times1} & \vec{\gamma}_u \\
    y_{P1} \vec{\gamma}_P^\dagger & \mathbf{0}_{1\times3} & y_{P1} v_P  & 0 & 0 \\
    \mathbf{0}_{1\times3} & \mathbf{0}_{1\times3} & 0 & y_{P2} v_P & y_S v_S \\
    \mathbf{0}_{1\times3} & \vec{\gamma}_u^\dagger & \frac{y_u v}{\sqrt{2}} & 0 & -Tr[\mathbf{M}_u]+M
    \end{pmatrix}, \nonumber\\
    \\
    &\mathcal{M}_d = \begin{pmatrix}
    \mathbf{0}_{3\times3} & y_S v_S \mathbf{1}_{3\times3} & \mathbf{0}_{3\times1} & y_{P2} \vec{\gamma}_P & \mathbf{0}_{3\times1} \\
    \frac{y_d v}{\sqrt{2}} \mathbf{1}_{3\times3} & \mathbf{M}_d + M \mathbf{1}_{3\times3} & \mathbf{0}_{3\times1} & \mathbf{0}_{3\times1} & \vec{\gamma}_d \\
    y_{P1} \vec{\gamma}_P^\dagger & \mathbf{0}_{1\times3} & y_{P1} v_P  & 0 & 0 \\
    \mathbf{0}_{1\times3} & \mathbf{0}_{1\times3} & 0 & y_{P2} v_P & y_S v_S \\
    \mathbf{0}_{1\times3} & \vec{\gamma}_d^\dagger & \frac{y_d v}{\sqrt{2}} & 0 & -Tr[\mathbf{M}_d]+M
    \end{pmatrix}, \nonumber\\
    \nonumber\\
    &\mathbf{M}_u \equiv y_A \langle A \rangle + y_B \langle B \rangle, \; \; \mathbf{M}_d \equiv y_A \langle A \rangle - y_B \langle B \rangle, \nonumber\\
    &\vec{\gamma}_u \equiv y_A \vec{\gamma}_A + y_B \vec{\gamma}_B, \; \; \vec{\gamma}_d \equiv y_A \vec{\gamma}_A - y_B \vec{\gamma}_B. \nonumber\\
    &y_u \equiv y_H \cos(\alpha-\beta), \; \; \; y_d \equiv y_H \sin(\alpha+\beta) \nonumber,
\end{align}
where we remind the reader that $v$, $\beta$, $v_S$, $v_P$, $\vec{\gamma}_{P,A,B}$, $\langle A \rangle$, and $\langle B \rangle$ are defined in Table \ref{table:scalars}, while all remaining parameters here appear in Eq.(\ref{GUTYukawa}). In general, we can exploit $SU(4)_F$ gauge freedom to render $\mathbf{M}_u$ or $\mathbf{M}_d$ diagonal, and eliminate either $\vec{\gamma}_u$ (if we diagonalize $\mathbf{M}_u$) or $\vec{\gamma}_d$ (if we diagonalize $\mathbf{M}_d$). This corresponds to moving to a basis in which either the combination of $\Phi_A$ and $\Phi_B$ vev's which couple to the up-like or down-like quarks are diagonal. Although we shall find it most convenient to work in a basis in which $\mathbf{M}_u$ is diagonal and $\vec{\gamma}_u=0$, for now we will make no such choice, so that our analytical results for the up-like quarks will apply straightforwardly to the down-like sector as well. Continuing our analysis, the matrices $\mathcal{M}_u$ and $\mathcal{M}_d$ can be bi-diagonalized as usual to produce mass eigenstates. That is, there exist unitary matrices $\mathcal{U}_{L,R}^{u,d}$ such that
\begin{align}\label{calUDefs}
    (\mathcal{U}_{L}^u)^\dagger \mathcal{M}_u^\dagger \mathcal{M}_u \mathcal{U}_L^u = (\mathcal{U}_{R}^u)^\dagger \mathcal{M}_u \mathcal{M}_u^\dagger \mathcal{U}_R^u = diag(m_{u1}^2, m_{u2}^2, m_{u3}^2, M_{u1}^2, M_{u2}^2, M_{u3}^2, (m_{P1}^u)^2, (m_{P2}^u)^2, (m_{P3}^u)^2),
\end{align}
where the $m_{ui}$ denotes the mass of the $i^{th}$-generation up-like SM quark, $M_{ui}$ is the mass of the (also uncharged under $U(1)_D$) heavy vector-like partner to this quark, and $m_{P1,2,3}^u$ denote the masses of the three portal matter fields which are charged under $U(1)_D$. An analogous expression holds for $\mathcal{M}_d$. As an example, we shall now determine the matrices $\mathcal{U}_{L,R}^u$, with a completely analogous treatment holding for the determination of $\mathcal{U}_{L,R}^d$.

We start by determining $\mathcal{U}_{L}^u$. To begin, we find it useful to define
\begin{align}\label{ULdefs}
    \mathbf{u} \equiv \mathbf{M}_u +M \mathbf{1}_{3\times3}, \;\;\; \mathbf{u}_D = diag(u_1, u_2, u_3) = \mathbf{W}_u^\dagger \mathbf{u} \mathbf{W}_u, \;\;\;
    X_u \equiv -Tr[\mathbf{M}_u]+M,
\end{align}
where $\mathbf{u}_D$ is the diagonalized form of $\mathbf{u}$ and $\mathbf{W}_u$ is a $3\times3$ unitary matrix. Then, we may write (dropping terms of $O(\vec{\gamma}_{P,u}^2)$, which shall be significantly smaller than the other mass scales occurring in the matrix) 
\begin{align}\label{MMatuL}
    \mathcal{M}_u^\dagger \mathcal{M}_u = \begin{pmatrix}
    \frac{y_u^2 v_u^2}{2} \mathbf{1}_{3\times3} & \frac{y_u v}{\sqrt{2}} \mathbf{u} & y_{P1}^2 v_P \vec{\gamma}_P & \mathbf{0}_{3\times1} & \frac{y_u v}{\sqrt{2}}\vec{\gamma}_u \\
    \frac{y_u v}{\sqrt{2}}\mathbf{u} & \mathbf{u}^2+y_S^2 v_S^2 \mathbf{1}_{3\times3} & \frac{y_u v}{\sqrt{2}}\vec{\gamma_u} & y_{P2} y_S v_S \vec{\gamma}_P & (\mathbf{u}+X_u \mathbf{1}_{3\times3})\cdot \vec{\gamma}_u\\
    y_{P1}^2 v_P \vec{\gamma}_P^\dagger & \frac{y_u v}{\sqrt{2}} \vec{\gamma}_u^\dagger & y_{P1}^2 v_P^2+\frac{y_u^2 v^2}{2} & 0 & \frac{y_u v}{\sqrt{2}} X_u\\
    \mathbf{0}_{1\times3} & y_{P2} y_S v_S \vec{\gamma}_P^\dagger & 0 & y_{P2}^2 v_P^2 & y_{P2}v_P y_S v_S \\
    \frac{y_u v}{\sqrt{2}} \vec{\gamma}_u^\dagger & \vec{\gamma}_u^\dagger \cdot (\mathbf{u}+X_u \mathbf{1}_{3\times3}) & \frac{y_u v}{\sqrt{2}} X_u & y_{P2}v_P y_S v_S & y_S^2 v_S^2 + X_u^2
    \end{pmatrix}.
\end{align}
To derive expressions for the mass eigenvectors here, it is easiest to split the problem into two basic parts. First, we shall determine the unitary matrix $(\mathcal{U}_L^u)_0$ which diagonalizes the matrix of Eq.(\ref{MMatuL}) in the limit where all components of $\vec{\gamma}_{P,u}$ are zero, that is, in the absence of all mixing between the fields with no charge under $U(1)_D$ (namely, $u_{L}$, and $(U_2)_L$, the first six rows/columns of Eq.(\ref{MMatuL})) and those with non-vanishing $U(1)_D$ charge ($Q_L^u$,$(U_1)_L$, and $(U_3)_L$, the last three rows/columns). With this diagonalization done, we can then determine the effects of $\vec{\gamma}_{P,u}$ as small perturbations, in a rotation matrix $(\mathcal{U}_L^u)_\gamma$. The matrix which diagonalizes Eq.(\ref{MMatuL}) is then simply given as
\begin{align}\label{UuLDef}
    \mathcal{U}_L^u \approx (\mathcal{U}_L^u)_0 (\mathcal{U}_L^u)_\gamma.
\end{align}
We now begin with our determination of $(\mathcal{U}_L^u)_0$. Since there's no mixing between the $U(1)_D$-charged and $U(1)_D$-uncharged fermions in this limit, we shall treat the two separately. We start with the mass matrix for the fermions with no $U(1)_D$ charge, which takes the form,

\begin{align}\label{MMatuLQD0}
    \begin{pmatrix}
    \frac{y_u^2 v^2}{2} \mathbf{1}_{3\times3} & \frac{y_u v}{\sqrt{2}} \mathbf{u} \\
    \frac{y_u v}{\sqrt{2}} \mathbf{u} & \mathbf{u}^2 + y_S^2 v_S^2
    \end{pmatrix}.
\end{align}
We can break this matrix down into three $2\times2$ matrices via rotation by the unitary block matrix
\begin{align}\label{MMatuLCKMRot}
    \begin{pmatrix}
    \mathbf{1}_{3\times3} & \mathbf{0}_{3\times3} \\
    \mathbf{0}_{3\times3} & \mathbf{W}_u
    \end{pmatrix},
\end{align}
where we recall that $\mathbf{W}_u^\dagger \mathbf{u} \mathbf{W}_u = diag (u_1, u_2, u_3)$. Then, each $2\times2$ matrix can be easily diagonalized, yielding (assuming that $v \ll v_S, u_i$)
\begin{align}\label{uLSMMat}
    &diag(m_{ui}^2, M_{ui}^2)=\begin{pmatrix}
    \cos(\rho^u_i) & \sin(\rho^u_i) \\
    -\sin(\rho^u_i) & \cos(\rho^u_i)
    \end{pmatrix} \begin{pmatrix}
    \frac{y_u^2 v^2}{2} & \frac{y_u v}{\sqrt{2}} u_i \\
    \frac{y_u v}{\sqrt{2}} u_i & u_i^2+y_S^2 v_S^2
    \end{pmatrix} \begin{pmatrix}
    \cos(\rho^u_i) & -\sin(\rho^u_i) \\
    \sin(\rho^u_i) & \cos(\rho^u_i)
    \end{pmatrix} \nonumber\\
    &\cos(\rho^u_i) \approx 1-\frac{1}{2}\frac{m_{ui}^2}{M_{ui}^2} \bigg( \frac{M_{ui}^2-y_S^2 v_S^2}{y_S^2 v_S^2} \bigg), \; \; \sin(\rho^u_i) \approx -sgn(y_u u_i) \frac{m_{ui}}{M_{ui}} \bigg( \frac{M_{ui}^2-y_S^2 v_S^2}{y_S^2 v_S^2} \bigg)^{1/2}, \\
    &m_{ui}^2 \approx \frac{y_u^2 v^2}{2}\frac{y_S^2 v_S^2}{(u_i^2+y_S^2 v_S^2)}, \; \; M_{ui}^2 \approx u_i^2+y_S^2 v_S^2 \nonumber
\end{align}

Having diagonalized the mass matrix for the left-handed quarks with no charge under $U(1)_D$, we next turn our attention to the left-handed fields which are charged under $U(1)_D$. In the case of the up-like quark sector, these are $Q_L^u$, $(U_1)_L$, and $(U_3)_L$. From Eq.(\ref{MMatuL}), we see that the mass matrix for these fields is
\begin{align}\label{MMatuLPortals}
    \begin{pmatrix}
    y_{P1}^2 v_P^2 + \frac{y_u^2 v^2}{2} & 0 & \frac{y_u v}{\sqrt{2}}X_u \\
    0 & y_{P2}^2 v_P^2 & (y_S v_S)(y_{P2} v_P) \\
    \frac{y_u v}{\sqrt{2}}X_u & (y_S v_S)(y_{P2} v_P) & y_S^2 v_S^2+X_u^2
    \end{pmatrix}.
\end{align}
In the absence of extreme fine-tuning, we shall see that natural Yukawa couplings yield $X_u$ at roughly the scale $X_u \sim 10^5 y_S v_S$, so we can assume that $v_S,v_P \ll X_u$. Any mixing between the three fields of Eq.(\ref{MMatuLPortals}), then, is exceedingly small and may be treated perturbatively. Dropping terms of $O(v_P^2/X_u^2, v_S^2/X_u^2, v^2/X_u^2)$, the mass matrix of Eq.(\ref{MMatuLPortals}) can be diagonalized by a rotation by the matrix
\begin{align}\label{MMatuLPortalsRot}
    \mathbf{W}^P_L \equiv \begin{pmatrix}
    1 & \alpha_{12}^L & \alpha_{13}^L \\
    -\alpha_{12}^L & 1 & \alpha_{23}^L \\
    -\alpha_{13}^L & -\alpha_{23}^L & 1
    \end{pmatrix},
\end{align}
where
\begin{align}\label{alphaLs}
    \alpha_{12}^L \equiv \frac{(y_u v)(y_S v_S)(y_{P2}v_P)}{\sqrt{2}X_u(y_{P1}^2-y_{P2}^2)v_P^2}, \; \; \; \alpha_{13}^L \equiv \frac{y_u v}{\sqrt{2}X_u}, \; \; \; \alpha_{23}^L \equiv \frac{(y_S v_S)(y_{P2}v_P)}{X_u^2}.
\end{align}
In the end, we arrive at
\begin{align}\label{uLPortalMat}
    &diag((m_{P1}^u)^2,(m_{P2}^u)^2,(m_{P3}^u)^2) = \mathbf{W}^{P \dagger}_L \begin{pmatrix}
    y_{P1}^2 v_P^2 + \frac{y_u^2 v^2}{2} & 0 & \frac{y_u v}{\sqrt{2}}X_u \\
    0 & y_{P2}^2 v_P^2 & (y_S v_S)(y_{P2} v_P) \\
    \frac{y_u v}{\sqrt{2}}X_u & (y_S v_S)(y_{P2} v_P) & y_S^2 v_S^2+X_u^2
    \end{pmatrix} \mathbf{W}^{P}_L, \\
    &(m_{P1}^u)^2 \equiv y_{P1}^2 v_P^2, \; \; \; (m_{P2}^u)^2 \equiv y_{P2}^2 v_P^2, \;\;\; (m_{P3}^u)^2 \equiv X_u^2. \nonumber
\end{align}
We can now apply the combined rotations of Eqs.(\ref{MMatuLCKMRot}), (\ref{uLSMMat}), and (\ref{MMatuLPortalsRot}) in order to arrive at the full rotation matrix $(\mathcal{U}^u_L)_0$. The combined rotation matrix is
\begin{align}\label{UuL0}
    &(\mathcal{U}_L^u)_0 \equiv \begin{pmatrix}
    \mathbf{W}_u \mathbf{c}_\rho^u & -\mathbf{W}_u \mathbf{s}_\rho^u & \mathbf{0}_{3\times1} & \mathbf{0}_{3\times1} & \mathbf{0}_{3\times1} \\
    \mathbf{W}_u \mathbf{s}_\rho^u & \mathbf{W}_u \mathbf{c}_\rho^u & \mathbf{0}_{3\times1} & \mathbf{0}_{3\times1} & \mathbf{0}_{3\times1} \\
    \mathbf{0}_{1\times3} & \mathbf{0}_{1\times3} & 1 & \alpha_{12}^L & \alpha_{13}^L \\
    \mathbf{0}_{1\times3} & \mathbf{0}_{1\times3} & -\alpha_{12}^L & 1 & \alpha_{23}^L \\
    \mathbf{0}_{1\times3} & \mathbf{0}_{1\times3} & -\alpha_{13}^L & -\alpha_{23}^L & 1
    \end{pmatrix},\\
    &\mathbf{c}_\rho^u \equiv diag(\cos(\rho_1^u),\cos(\rho_2^u),\cos(\rho_3^u)), \; \; \; \mathbf{s}_\rho^u \equiv diag(\sin(\rho_1^u),\sin(\rho_2^u),\sin(\rho_3^u)), \nonumber
\end{align}
where we remind the reader that $\mathbf{W}_u$ is defined in Eq.(\ref{ULdefs}), $\rho^u_1$, $\rho^u_2$, and $\rho^u_3$ are given in Eq.(\ref{uLSMMat}), and the variables $\alpha_{ij}^L$ are defined in Eq.(\ref{alphaLs}). The mass matrix of Eq.(\ref{MMatuL}) then takes the form
\begin{align}\label{MMatuLRot}
    (\mathcal{U}^u_L)_0^\dagger \mathcal{M}_u^\dagger \mathcal{M}_u (\mathcal{U}^u_L)_0 = diag(m_u^2, m_c^2, m_t^2, M_u^2, M_c^2, M_t^2, (m_{P1}^u)^2, (m_{P2}^u)^2, (m_{P3}^u)^2) + O(\vec{\gamma}_u, \vec{\gamma}_P).
\end{align}
It is now straightforward to determine the perturbed eigenvectors of the matrix in Eq.(\ref{MMatuLRot}) up to $O(\vec{\gamma}_u, \vec{\gamma}_P)$. Once the dust settles, we have
\begin{align}\label{UuLgamma}
    (\mathcal{U}^u_L)_\gamma \approx \begin{pmatrix}
    \mathbf{1}_{3\times3} & \mathbf{0}_{3\times3} & \vec{\Delta}_{P1L}^u & \vec{\Delta}_{P2L}^u & \vec{\Delta}_{P3L}^u \\
    \mathbf{0}_{3\times3} & \mathbf{1}_{3\times3} & \vec{\Delta}_{P1L}^U & \vec{\Delta}_{P2L}^U & \vec{\Delta}_{P3L}^U \\
    -\vec{\Delta}_{P1L}^{u \dagger} & -\vec{\Delta}_{P1L}^{U \dagger} & 1 & 0 & 0\\
    -\vec{\Delta}_{P2L}^{u \dagger} & -\vec{\Delta}_{P2L}^{U \dagger} & 0 & 1 & 0\\
    -\vec{\Delta}_{P3L}^{u \dagger} & -\vec{\Delta}_{P3L}^{U \dagger} & 0 & 0 & 1
    \end{pmatrix},
\end{align}
where
\begin{align}\label{LongDeltaExpressions}
    &(\vec{\Delta}_{P1L}^u)_i = \frac{((y_{P1}^2 v_P \mathbf{c}_{\rho}^u - \alpha_{12}^L y_{P2} y_S v_S \mathbf{s}_{\rho}^u)\mathbf{W}_u^\dagger \vec{\gamma}_P+ (\frac{y_u v}{\sqrt{2}} \mathbf{s}_\rho^u -\alpha_{13}^L (\frac{y_u v}{\sqrt{2}}\mathbf{c}_\rho^u + \mathbf{u}_D\mathbf{s}_{\rho}^u+ X_u \mathbf{s}_\rho^u))\mathbf{W}_u^\dagger \vec{\gamma}_u)_i}{m_{P1}^2-m_{ui}^2}, \nonumber\\
    &(\vec{\Delta}_{P2L}^u)_i = \frac{((\alpha_{12}^L y_{P1}^2 v_P \mathbf{c}_\rho^u + y_{P2} y_S v_S \mathbf{s}_\rho^u)\mathbf{W}_u^\dagger \vec{\gamma}_P +(\alpha_{12}^L \frac{y_u v}{\sqrt{2}}\mathbf{s}_\rho^u -\alpha_{23}^L (\frac{y_u v}{\sqrt{2}}\mathbf{c}_\rho^u + \mathbf{u}_D \mathbf{s}_\rho^u + X_u \mathbf{s}_{\rho}^u))\mathbf{W}_u^\dagger \vec{\gamma}_u)_i}{m_{P2}^2-m_{ui}^2} \nonumber\\
    &(\vec{\Delta}_{P3L}^u)_i = \frac{((\alpha_{13}^L y_{P1}^2 v_P \mathbf{c}_\rho^u+\alpha_{23}^L y_{P2} y_S v_S \mathbf{s}_\rho^u) \mathbf{W}_u^\dagger \vec{\gamma}_P + (\alpha_{13}^L \frac{y_u v}{\sqrt{2}} \mathbf{s}_\rho^u +\frac{y_u v}{\sqrt{2}} \mathbf{c}_\rho^u + \mathbf{u}_D \mathbf{s}_\rho^u+X_u \mathbf{s}_\rho^u)\mathbf{W}^\dagger_u \vec{\gamma}_u)_i}{m_{P3}^2-m_{ui}^2}\\
    &(\vec{\Delta}_{P1L}^U)_i = \frac{((-y_{P1}^2 v_P \mathbf{s}_\rho^u-\alpha_{12}^L y_{P2} y_S v_S \mathbf{c}_\rho^u)\mathbf{W}_u^\dagger \vec{\gamma}_P+(\frac{y_u v}{\sqrt{2}} \mathbf{c}_\rho^u-\alpha_{13}^L(\mathbf{u}_D \mathbf{c}_\rho^u+X_u \mathbf{c}_\rho^u-\frac{y_u v}{\sqrt{2}}\mathbf{s}_\rho^u))\mathbf{W}_u^\dagger \vec{\gamma}_u)_i}{(m_{P1}^u)^2-M_{ui}^2} \nonumber\\
    &(\vec{\Delta}_{P2L}^U)_i = \frac{((-\alpha_{12}^L y_{P1}^2 v_P \mathbf{s}_\rho^u + y_{P2}y_S v_S \mathbf{c}_\rho^u)\mathbf{W}_u^\dagger \vec{\gamma}_P+(\alpha_{12}^L\frac{y_u v}{\sqrt{2}} \mathbf{c}_\rho^u -\alpha_{23}^L(\mathbf{u}_D\mathbf{c}_\rho^u + X_u \mathbf{c}_\rho^u-\frac{y_u v}{\sqrt{2}}\mathbf{s}_\rho^u))\mathbf{W}_u^\dagger \vec{\gamma}_u)_i}{(m_{P2}^u)^2-M_{ui}^2}, \nonumber\\
    &(\vec{\Delta}_{P3L}^u)_i = \frac{((-\alpha_{13}^L y_{P1}^2 v_P \mathbf{s}_\rho^u+\alpha_{23}^L y_{P2} y_S v_S \mathbf{c}_\rho^u)\mathbf{W}_u^\dagger \vec{\gamma}_P+(\alpha_{13}^L \frac{y_u v}{\sqrt{2}} \mathbf{c}_\rho^u + \mathbf{u}_D \mathbf{c}_\rho^u + X_u \mathbf{c}_\rho^u-\frac{y_u v}{\sqrt{2}} \mathbf{s}_\rho^u)\mathbf{W}_u^\dagger \vec{\gamma}_u)_i}{(m_{P3}^u)^2-M_{ui}^2}, \nonumber
\end{align}
where $\mathbf{u}_D$ is the diagonalized form of $\mathbf{u}$ discussed in Eq.(\ref{ULdefs}).

The lengthy expressions of Eq.(\ref{LongDeltaExpressions}), while extremely cumbersome in their full form, can be dramatically simplified by examining the numerical hierarchies between various parameters, and dropping all but the numerically dominant contributions to each expression. To get a better sense of the numerics of the system, we begin by exploring the $SU(3)_F$ triplet sector, containing the SM quarks and their heavy partners. Examining Eq.(\ref{uLSMMat}), we note that the expressions for $m_{ui}$ and $M_{ui}$ only hold here up to $O(v^2/v_S^2)$. To maintain numerical accuracy, we then assume that $y_S v_S \gsim 1 \; \textrm{TeV}$-- given that $v \sim 246 \; \textrm{GeV}$, this will ensure that the expressions given here are accurate to within a few percent. This also ensures that there is a significant hierarchy between $m_{ui}$ and $M_{ui}$, as can be seen from the relation $m_{ui}^2/M_{u_i}^2 = 2 m_{ui}^4/(y_S^2 v_S^2 y_u^2 v^2)$ (which holds exactly in the limit that $\vec{\gamma}_{u,P}\rightarrow 0$). If $y_S v_S$ were much smaller than 1 TeV, the heavy counterpart of the top quark would have roughly the same mass as the SM top, and be subject to extremely severe constraints from direct production searches at the LHC and from modifications to top quark couplings; the former limit the mass of such a vector-like quark to $\gsim 1.3-1.5 \; \textrm{TeV}$, due to null results in searches for pair production \cite{Aaboud:2018pii}. We can also use the relation $M_{ui}^2 = (y_S v_S)^2 (y_u v)^2/(2 m_{ui}^2)$, and its trivially analogous expression for down-like quarks, to derive approximate estimates of the masses of the various heavy quark partners. 

Assuming that we want to reproduce the SM quark masses at the scale of $\sim 1 \; \textrm{TeV}$, roughly the scale at which the lightest heavy partner quark, the top partner, will be integrated out\footnote{A technically more correct procedure (albeit one that still ignores the running of the CKM matrix) would be for each SM quark's mass to be determined near the scale of the mass of its specific partner, so the model would reproduce the mass $m_u$ run up to $\sim 10^9 \; \textrm{GeV}$, $m_c$ near $\sim 10^6 \; \textrm{GeV}$, $m_t$ near $1 \; \textrm{TeV}$, \etc. However, due to the additional quarks in the model, this computation would be substantially more complicated than our treatment here, and the numerical changes to the results would be minimal, especially given the fact that the most pronounced discrepancies with our calculation would arise in the least certain quark masses, $m_u$ and $m_d$.}, we arrive at
\begin{align}\label{ApproxQuarkMs}
    &m_u(1 \; \textrm{TeV}) = 1.07 \; \textrm{MeV}, &M_u \sim (2 \times 10^5)y_S v_S, \nonumber \\
    &m_c(1 \; \textrm{TeV}) = 532 \; \textrm{MeV}, &M_c \sim 300 y_S v_S, \nonumber\\
    &m_t(1 \; \textrm{TeV}) = 144 \; \textrm{GeV}, &M_t \sim y_S v_S, \\
    &m_d(1 \; \textrm{TeV}) = 2.28 \; \textrm{MeV}, &M_d \sim (8\times 10^4)y_S v_S, \nonumber\\
    &m_s(1 \; \textrm{TeV}) = 46.5 \; \textrm{MeV}, &M_s \sim (4 \times 10^4) y_S v_S, \nonumber\\
    &m_b(1 \; \textrm{TeV}) = 2.40 \; \textrm{GeV}, &M_b \sim 70 y_S v_S. \nonumber
\end{align}
Given these estimates, and assuming $y_S v_S \gsim 1 \; \textrm{TeV}$, only the heavy partner to the top quark, with a mass $M_t$, can reasonably be expected to be experimentally directly observable in the foreseeable future. We do, however, note that Eq.(\ref{ApproxQuarkMs}) assumes that the up-like and down-like Yukawa couplings to the SM Higgs, that is, $y_H \cos (\alpha-\beta)$ and $y_H \sin(\alpha+\beta)$ are both roughly of $O(1)$. While this may be expected from naturalness, it is far from the only feasible scenario. If, for example, $y_H \sin(\alpha+\beta) \sim O(10^{-2})$, equivalent to a value arising in a number of multi-Higgs doublet scenarios which seek to explain the mass hierarchy between the top and $b$ quark, we might expect that $M_b \sim M_t$, and hence the heavy partner of the $b$ may also play a significant phenomenological role. In this case, we would expect significant additional constraints arising from both direct production of the $b$ partner at the LHC and possibly significant modifications to the SM $Z \overline{b} b$ coupling. For the sake of simplicity, however, we leave a detailed exploration of the effects of a lighter $b$ partner within this model to future work.

Meanwhile, in the mass matrix of Eq.(\ref{MMatuLPortals}), we note that $X_u = -Tr[\mathbf{M}_u]+M$ should, in the absence of extreme fine tuning, be roughly the same order of magnitude as the largest eigenvalue of the matrix $\mathbf{u}$. We have established, however, that this eigenvalue is approximately $\sim 10^5 \times y_S v_S$. The mass $m_{P3}^u \approx |X_u|$ will naturally be extremely large, on the order of $\sim 10^8 \; \textrm{GeV}$.

Our expressions in Eq.(\ref{LongDeltaExpressions}) can then be dramatically simplified by merely assuming that the extremely heavy particles (the fermion with mass $\approx |X_u|$ and the heavy partners of the up and charm quarks) are virtually entirely decoupled from the lower mass fields (the SM quarks, the fermions with masses $\sim v_P$, and the heavy partner to the top quark) -- in practice, this involves taking the limits where $|u_{1,2}|,|X_u|,M_{u1,u2} \rightarrow \infty$ in the expressions of Eq.(\ref{LongDeltaExpressions}). This permits us to write
\begin{align}\label{ShortDeltaExpressionsL}
    &\vec{\Delta}_{P1L}^u \approx \frac{\mathbf{W}_u^\dagger \vec{\gamma}_P}{v_P}, \nonumber\\
    &(\vec{\Delta}_{P2L}^u)_3 \approx -sgn(y_u y_s y_{P2} u_3)\frac{m_t}{M_t}\frac{(M_t^2-y_S^2 v_S^2)^{1/2}}{m_{P2}^u} \frac{(\mathbf{W}_u^\dagger \vec{\gamma}_P)_3}{v_P} \\
    &(\vec{\Delta}_{P1L}^U)_3 \approx sgn(y_u u_3) \frac{m_t}{M_t} \bigg(\frac{M_t^2-y_S^2 v_S^2}{y_S^2 v_S^2}\bigg)^{1/2}\frac{(m_{P1}^u)^2}{((m_{P1})^2-M_t^2)} \frac{(\mathbf{W}_u^\dagger \vec{\gamma}_P)_3}{v_P} \nonumber\\
    &(\vec{\Delta}_{P2L}^U)_3 \approx sgn(y_{P2}) \frac{y_S v_S m_{P2}^u}{(m_{P2}^u)^2-M_t^2} \frac{(\mathbf{W}_u^\dagger \vec{\gamma}_P)_3}{v_P}, \nonumber
\end{align}
where the rest of the $\vec{\Delta}$ terms are either negligibly small or parameterize mixing between the extremely heavy states. The expressions in Eq.(\ref{ShortDeltaExpressionsL}) already provide some useful information: In particular, in the limit where the super-heavy states decouple the mixing of the phenomenologically accessible states is independent of $\vec{\gamma}_u$, the parameter corresponding to the $U(1)_D$-breaking components of the scalar vev's $\langle \Phi_A \rangle$ and $\langle \Phi_B \rangle$.  Additionally, there exists a suppression factor of $m_t/M_t$ in front of $(\vec{\Delta}_{P2L}^u)_3$ and $(\vec{\Delta}_{P1L}^U)_3$, corresponding to the mixing between the top quark and the $m_{P2}^u$ state, and the mixing between the heavy top partner and the $m_{P1}^u$ state, respectively -- these couplings between $SU(3)_F$ triplet and singlet states only arise due to mixing between the top quark and its heavy partner, and therefore vanish in the limit where the top partner decouples from the SM.

Armed with Eq.(\ref{ShortDeltaExpressionsL}), it is not difficult to extract a complete expression for $\mathcal{U}_L^u$,  or at least the components of this rotation matrix that are relevant for the phenomenologically accessible particles. Taking the same decoupling limit in the expression for the diagonalization matrix $(\mathcal{U}_L^u)_0$ in Eq.(\ref{UuL0}) (which here corresponds to letting $\rho_{1,2}^u, \alpha_{ij}^L \rightarrow 0$), we can use Eqs.(\ref{UuLDef}), (\ref{UuL0}), (\ref{UuLgamma}), and (\ref{ShortDeltaExpressionsL}) in order to extract the approximate mass eigenvectors for each left-handed up-like (and, through trivial generalization, down-like) quark with a phenomenologically accessible mass. Truncating our 9-dimensional flavor-space quark vectors to omit the super-heavy particles, we can write these vectors as
\begin{align}\label{ShortVecs}
    (\overline{\mathbf{U}}_R)_{\textrm{dec}} &\equiv (\overline{u}_R, \overline{T}_R, \overline{Q}^u_R, (\overline{U}_1)_R ), \nonumber\\
    (\mathbf{U}_L)_{\textrm{dec}} &\equiv (u_L, T_L, Q^u_L, (U_1)_L ), \\
    (\overline{\mathbf{D}}_R)_{\textrm{dec}} &\equiv (\overline{d}_R, \overline{Q}^d_R, (\overline{D}_1)_R ), \nonumber\\
    (\mathbf{D}_L)_{\textrm{dec}} &\equiv (d_L, Q^d_L, (D_1)_L ). \nonumber
\end{align}
Notably, the above truncation in the up-like sector, in which the top quark's heavy partner is retained, is much easier in a basis in which $\mathbf{W}_u = \mathbf{1}_{3\times3}$. Otherwise, as we can see from Eq.(\ref{FermionFlavorBasis}), the top partner will be represented by a non-trivial combination of the elements of the $SU(3)_F$ triplets $(U_2)_L$ and $(U_2)_R$, which would require retaining rows and columns of $\mathbf{U}^u_{L,R}$ that correspond to heavy states until a rotation into a basis in which $\mathbf{W}_u=\mathbf{1}_{3\times3}$ could be performed. Therefore, for the following expressions, we have assumed that $\mathbf{W}_u = \mathbf{1}_{3\times3}$; as we mentioned at the beginning of this section, it is always possible to use $SU(4)_F$ freedom to work in such a basis from the beginning of our analysis. Using Eqs.(\ref{UuL0}) and (\ref{UuLgamma}), we can write the truncated up-like quark rotation matrix as
\begin{align}\label{UuLTrunc}
    &\mathcal{U}^u_L \approx \begin{pmatrix}
    1 & 0 & 0 & 0 & \frac{(\vec{\gamma}_P)_1}{v_P} & 0 \\
    0 & 1 & 0 & 0 & \frac{(\vec{\gamma}_P)_2}{v_P} & 0\\
    0 & 0 & 1 & r\frac{m_t}{M_t} & \frac{(\vec{\gamma}_{P})_3}{v_P} &  r q \bigg(\frac{z_{P2}^3}{1-z_{P2}^2}\bigg)\frac{m_t}{M_t}\frac{(\vec{\gamma}_P)_3}{v_P}\\
    0 & 0 & -r \frac{m_t}{M_t} & 1 & r\bigg( \frac{z_{P1}^2}{1-z_{P1}^2}\bigg)\frac{m_t}{M_t}\frac{(\vec{\gamma}_P)_3}{v_P} & q \bigg( \frac{z_{P2}}{1-z_{P2}^2} \bigg) \frac{(\vec{\gamma}_P)_3}{v_P}\\
    -\frac{(\vec{\gamma}_P)_1^*}{v_P} & -\frac{(\vec{\gamma}_P)_2^*}{v_P} & -\frac{(\vec{\gamma}_P)_3^*}{v_P} & -r \bigg(\frac{1}{1-z_{P1}^2}\bigg)\frac{m_t}{M_t} \frac{(\vec{\gamma}_P)^*_3}{v_P} & 1 & 0\\
    0 & 0 & r q z_{P2} \frac{m_t}{M_t} \frac{(\vec{\gamma}_P)_3^*}{v_P} & -\frac{q z_{P2}}{1-z_{P2}^2}\frac{(\vec{\gamma}_P)_3^*}{v_P} & 0 & 1
    \end{pmatrix}\\
    &r \equiv sgn(y_u u_3)\bigg( \frac{M_t^2-y_S^2 v_S^2}{y_S^2 v_S^2} \bigg)^{1/2}, \; \; q \equiv \frac{y_S v_S}{M_t}, \; \; z_{P1} \equiv sgn(y_{P2}) \frac{M_t}{m_{P1}^u}, \; \; z_{P2} \equiv sgn(y_{P2}) \frac{M_t}{m_{P2}^u}, \nonumber
\end{align}
where we have dropped terms of $O(m_t^2/M_t^2)$ and higher. Note that each of the variables $r$, $q$, $z_{P1}$, and $z_{P2}$ are all $O(1)$ parameters, if we assume that the portal scale $v_P$ is close to the heavy top partner mass scale $v_S$. The result for the down-like quarks is dramatically simpler, since we assume that the heavy $b$ quark partner, unlike the top partner, is too massive to be phenomenologically relevant. In this case, we simply have
\begin{align}\label{UdLTrunc}
    \mathcal{U}^d_L \approx \begin{pmatrix}
    \mathbf{W}_d & \frac{\vec{\gamma}_P}{v_P} & \mathbf{0}_{3\times1}\\
    -\frac{\vec{\gamma}_P^\dagger \mathbf{W}_d}{v_P} & 1 & 0\\
    \mathbf{0}_{1\times3} & 0 & 1
    \end{pmatrix}.
\end{align}

Having determined the relevant components of the left-handed rotation  matrix $\mathcal{U}_L^u$, we can follow an analogous procedure for the right-handed rotation matrix $\mathcal{U}_R^u$ identified in Eq.(\ref{calUDefs}). For the sake of brevity, we shall simply summarize the results here. The right-handed fermion mass matrix is given (up to $O(\vec{\gamma}_{P,u})$) by
\begin{align}\label{MMatuR}
    \mathcal{M}_u \mathcal{M}_u^\dagger = \begin{pmatrix}
    y_S^2 v_S^2 \mathbf{1}_{3\times3} & y_S v_S \mathbf{u} & \mathbf{0}_{3\times1} & y_{P2}^2 v_P \vec{\gamma}_P & y_S v_S \vec{\gamma}_u \\
    y_S v_S \mathbf{u} & \frac{y_u^2 v^2}{2}+\mathbf{u}^2 & \frac{y_u v}{\sqrt{2}} y_{P1} \vec{\gamma}_P & y_S v_S \vec{\gamma}_u & (\mathbf{u}+X_u \mathbf{1}_{3\times3})\vec{\gamma}_u \\
    \mathbf{0}_{1\times3} & \frac{y_u v}{\sqrt{2}}y_{P1} \vec{\gamma}_P^\dagger & y_{P1}^2 v_P^2 & 0 & \frac{y_u v}{\sqrt{2}}y_{P1} v_P \\
    y_{P2}^2 v_P \vec{\gamma}_P^\dagger & y_S v_S \vec{\gamma}_u^\dagger & 0 & y_{P2}^2 v_P^2+y_S^2 v_S^2 & X_u y_S v_S \\
    y_S v_S \vec{\gamma}_u^\dagger & \vec{\gamma}_u^\dagger (\mathbf{u}+X_u\mathbf{1}_{3\times3}) & \frac{y_u v}{\sqrt{2}}y_{P1} v_P & X_u y_S v_S & \frac{y_u^2 v^2}{2} +X_u^2
    \end{pmatrix}.
\end{align}
In the same framework as our treatment of $\mathcal{U}_L^u$, we shall split the result into a matrix $(\mathcal{U}_R^u)_0$ which diagonalizes $\mathcal{M}_u \mathcal{M}_u^\dagger$ in the limit where $\vec{\gamma}_{P,u} \rightarrow 0$, and a matrix $(\mathcal{U}_R^u)_\gamma$ which provides the leading-order corrections due to the $\vec{\gamma}_{P,u}$ terms. For $(\mathcal{U}_R^u)_0$, we arrive at
\begin{align}\label{UuR0}
    &(\mathcal{U}_{R}^u)_0 = \begin{pmatrix}
    -\mathbf{W}_u \mathbf{c}^u_\eta sgn(y_S y_u \mathbf{u}_D) & -\mathbf{W}_u \mathbf{s}^u_\eta sgn(\mathbf{u}_D) & \mathbf{0}_{3\times1} & \mathbf{0}_{3\times1} & \mathbf{0}_{3\times1}\\
    -\mathbf{W}_u \mathbf{s}^u_\eta sgn(y_S y_u \mathbf{u}_D) & \mathbf{W}_u \mathbf{c}^u_\eta sgn(\mathbf{u}_D) & \mathbf{0}_{3\times1} & \mathbf{0}_{3\times1} & \mathbf{0}_{3\times1}\\
    \mathbf{0}_{1\times3} & \mathbf{0}_{1\times3} & sgn(y_{P1}) & \alpha_{12}^R sgn(y_{P2}) & \alpha_{13}^R sgn(X_u) \\
    \mathbf{0}_{1\times3} & \mathbf{0}_{1\times3} & -\alpha_{12}^R sgn(y_{P1}) & sgn(y_{P2}) & \alpha_{23}^R sgn(X_u) \\
    \mathbf{0}_{1\times3} & \mathbf{0}_{1\times3} & -\alpha_{13}^R sgn(y_{P1}) & -\alpha_{23}^R sgn(y_{P2}) & sgn(X_u)
    \end{pmatrix},\\
    &\mathbf{c}_\eta^u \equiv diag(\cos(\eta_1^u),\cos(\eta_2^u),\cos(\eta_3^u)), \; \; \; \mathbf{s}_\eta^u \equiv diag(\sin(\eta_1^u),\sin(\eta_2^u),\sin(\eta_3^u)), \nonumber \\
    &\cos(\eta_i^u) = \bigg(\frac{M_{ui}^2-y_S^2 v_S^2}{M_{ui}^2} \bigg)^{1/2}, \; \; \; \sin(\eta_i^u) = -\frac{y_S v_S}{M_{ui}}sgn(u_i), \nonumber\\
    &\alpha_{12}^R \equiv \frac{y_u v}{\sqrt{2}X_u}\frac{(y_S v_S)(y_{P1} v_P)}{(y_{P1}^2-y_{P2}^2)v_P^2}, \; \; \; \alpha_{13}^R \equiv \frac{y_u v}{\sqrt{2} X_u} \frac{y_{P1} v_P}{X_u} \bigg(1-\frac{y_S^2 v_S^2}{(y_{P1}^2-y_{P2}^2)v_P^2} \bigg), \;\;\; \alpha_23^R \equiv \frac{y_S v_S}{X_u}.\nonumber
\end{align}
The sole non-trivial difference in the result of Eq.(\ref{UuR0}) and that of Eq.(\ref{UuL0}) is the introduction of sign flips (or rather, phase rotations) of the right-handed quarks in order to ensure that the mass terms in the final action correspond to real positive fermion masses-- these rotations could have just as well been done in the left-handed sector, without altering the physical results of the model. The leading perturbations to $(\mathcal{U}_R^u)_0$ arising from the $\vec{\gamma}_{P,u}$ terms of the mass matrix of Eq.(\ref{MMatuR}) are then given by
\begin{align}\label{UuRGamma}
    (\mathcal{U}_R^u)_\gamma = \begin{pmatrix}
    \mathbf{1}_{3\times3} & \mathbf{0}_{3\times3} & \vec{\Delta}^u_{P1R} & \vec{\Delta}^u_{P2R} & \vec{\Delta}^u_{P3R} \\
     \mathbf{0}_{3\times3} & \mathbf{1}_{3\times3} & \vec{\Delta}^U_{P1R} & \vec{\Delta}^U_{P2R} & \vec{\Delta}^U_{P3R} \\
     -\vec{\Delta}^{u\dagger}_{P1R} & -\vec{\Delta}^{U\dagger}_{P1R} & 1 & 0 & 0 \\
     -\vec{\Delta}^{u\dagger}_{P2R} & -\vec{\Delta}^{U\dagger}_{P2R} & 0 & 1 & 0 \\
     -\vec{\Delta}^{u\dagger}_{P3R} & -\vec{\Delta}^{U\dagger}_{P3R} & 0 & 0 & 1
    \end{pmatrix},
\end{align}
where
\begin{align}\label{LongDeltaRExpressions}
    &(\vec{\Delta}^u_{P1R})_i = \frac{-\bigg( (\frac{y_u v}{\sqrt{2}} y_{P1} \mathbf{s}^u_\eta -\alpha_{12}^R y_{P2}^2 v_P \mathbf{c}^u_\eta)\mathbf{W}_u^\dagger \vec{\gamma}_P - (\alpha_{12}^R y_S v_S \mathbf{s}^u_\eta + \alpha_{13}^R(y_S v_S \mathbf{c}^u_\eta+\mathbf{u}_D \mathbf{s}^u_\eta+X_u \mathbf{s}^u_\eta))\mathbf{W}_u^\dagger \vec{\gamma}_u\bigg)_i}{((m_{P1}^u)^2-m_{ui}^2)sgn(y_{P1}y_S y_u u_i)} \nonumber\\
    &(\vec{\Delta}^u_{P2R})_i = \frac{-\bigg( (\alpha_{12}^R \frac{y_u v}{\sqrt{2}}y_{P1} \mathbf{s}^u_\eta + y_{P2}^2 v_P \mathbf{c}^u_\eta)\mathbf{W}_u^\dagger \vec{\gamma}_P +(y_S v_S \mathbf{s}^u_\eta- \alpha_{23}^R(y_S v_S \mathbf{c}^u_\eta + \mathbf{u}_D \mathbf{s}^u_\eta+X_u \mathbf{s}^u_\eta))\mathbf{W}_u^\dagger \vec{\gamma}_u\bigg)_i}{((m_{P2}^u)^2-(m_{ui}^2))sgn(y_{P2}y_S y_u u_i)} \nonumber\\
    &(\vec{\Delta}_{P3R}^u)_i = \frac{-\bigg( (\alpha_{13}^R \frac{y_u v}{\sqrt{2}}y_{P1} \mathbf{s}^u_\eta + \alpha_{23}^R y_{P2}^2 v_P \mathbf{c}^u_\eta)\mathbf{W}_u^\dagger \vec{\gamma}_u + (\alpha_{23}^R y_S v_S \mathbf{s}^u_\eta +y_S v_S \mathbf{c}^u_\eta+\mathbf{u}_D \mathbf{s}^u_\eta + X_u \mathbf{s}^u_\eta)\mathbf{W}_u^\dagger \vec{\gamma}_u\bigg)_i}{((m_{P2}^u)^2-m_{ui}^2)sgn(X_u y_S y_u u_i)}\\
    &(\vec{\Delta}^U_{P1R})_i = \frac{\bigg( (\frac{y_u v}{\sqrt{2}} y_{P1} \mathbf{c}^u_\eta +\alpha_{12}^R y_{P2}^2 v_P \mathbf{s}^u_\eta)\mathbf{W}_u^\dagger \vec{\gamma}_P -(\alpha_{12}^R y_S v_S \mathbf{c}^u_\eta + \alpha_{13}^R(-y_S v_S \mathbf{s}^u_\eta+\mathbf{u}_D\mathbf{c}^u_\eta + X_u \mathbf{c}^u_\eta))\mathbf{W}_u^\dagger \vec{\gamma}_u \bigg)_i}{((m_{P1}^u)^2-M_{ui}^2)sgn(y_{P1}u_i)}, \nonumber\\
    &(\vec{\Delta}^U_{P2R})_i = \frac{\bigg( (\alpha_{12}^R \frac{y_u v}{\sqrt{2}}y_{P1} \mathbf{c}^u_\eta - y_{P2}^2 v_P \mathbf{s}^u_\eta)\mathbf{W}_u^\dagger \vec{\gamma}_P + (y_S v_S \mathbf{c}^u_\eta -\alpha_{23}^R(-y_S v_S \mathbf{s}^u_\eta+ \mathbf{u}_D \mathbf{c}^u_\eta+X_u \mathbf{c}^u_\eta))\mathbf{W}_u^\dagger \vec{\gamma}_u\bigg)_i}{((m_{P2}^u)^2-M_{ui}^2)sgn(y_{P2}u_i)}, \nonumber\\
    &(\vec{\Delta}^U_{P3R})_i = \frac{\bigg( (\alpha_{13}^R \frac{y_u v}{\sqrt{2}} y_{P1} \mathbf{c}^u_\eta-\alpha_{23}^R y_{P2}^2 v_P \mathbf{s}^u_\eta)\mathbf{W}_u^\dagger \vec{\gamma}_P +(\alpha_{23}^R y_S v_S \mathbf{c}^u_\eta-y_S v_S \mathbf{s}^u_\eta+\mathbf{u}_D \mathbf{c}^u_\eta+ X_u \mathbf{c}^u_\eta)\mathbf{W}_u^\dagger \vec{\gamma}_u\bigg)_i}{((m_{P3}^u)^2-M_{ui}^2)sgn(X_u u_i)}. \nonumber
\end{align}
In the same manner as the left-handed eigenvectors, the expressions in Eq.(\ref{LongDeltaRExpressions}) simplify dramatically in the limit where the $m_{P3}^u$ quark and the heavy up and charm partners decouple from the theory. In this limit, we arrive at
\begin{align}\label{ShortDeltaExpressionsR}
    &(\vec{\Delta}_{P1R}^u)_3 \approx \frac{m_t}{m_{P1}^u} \frac{(\mathbf{W}_u^\dagger \vec{\gamma}_P)_3}{v_P}, \nonumber \\
    &(\vec{\Delta}_{P2R}^u)_i \approx -sgn(y_u y_S u_i y_{P2})\bigg( \frac{M_{ui}^2-y_S^2 v_S^2}{M_{ui}^2}\bigg)^{1/2} \frac{(\mathbf{W}_u^\dagger \vec{\gamma}_P)_i}{v_P}, \\
    &(\vec{\Delta}_{P1R}^U)_3 \approx sgn(y_u u_3)\frac{m_{P1}^u m_t}{(m_{P1}^u)^2-M_t^2}\bigg( \frac{M_t^2-y_S^2 v_S^2}{y_S^2 v_S^2}\bigg)^{1/2} \frac{(\mathbf{W}_u^\dagger \vec{\gamma}_P)_3}{v_P}, \nonumber \\
    &(\vec{\Delta}_{P2R}^U)_3 \approx sgn(y_{P2}) \frac{(m_{P2}^u)^2}{(m_{P2}^u)^2-M_t^2}\bigg( \frac{y_S v_S}{M_t}\bigg) \frac{(\mathbf{W}_u^\dagger \vec{\gamma}_P)_3}{v_P}, \nonumber
\end{align}
with all other $\vec{\Delta}$ terms for the right-handed quarks being either numerically negligible or parameterizing mixing between the extremely heavy fermions. We can then insert the results of Eq.(\ref{ShortDeltaExpressionsR}) into Eq.(\ref{UuRGamma}) in order to derive the rotation matrix $(\mathcal{U}^u_R)_\gamma$, which when combined with $(\mathcal{U}^u_R)_0$ given in Eq.(\ref{UuR0}) will give us the mass eigenstates for the right-handed up-like quarks in the model. As in the case of the left-handed diagonalization matrices, we shall restrict our attentions to the limit in which only the SM quarks, the heavy top partner, and the two lighter portal matter fields mix among one another, with the other much more massive additional fermions decoupled from the low-energy theory. Using the truncated flavor-space vectors of Eq.(\ref{ShortVecs}), we can write
\begin{align}\label{UuRTrunc}
    &\mathcal{U}^u_R \approx \mathbf{N}_1 \begin{pmatrix}
    1 & 0 & 0 & 0 & 0 & \frac{(\vec{\gamma}_P)_1}{v_P}\\
    0 & 1 & 0 & 0 & 0 & \frac{(\vec{\gamma}_P)_2}{v_P}\\
    0 & 0 & q r & q & r q \bigg( \frac{z_{P1}^3}{1-z_{P1}^2}\bigg) \frac{m_t}{M_t}\frac{(\vec{\gamma}_P)_3}{v_P} & \bigg( \frac{1-q^2 r^2 z_{P2}^2}{1-z_{P2}^2}\bigg)\frac{(\vec{\gamma}_P)_3}{v_P}\\
    0 & 0 & -q & q r & \frac{z_{P1}}{q}\bigg( \frac{1-q^2 z_{P1}^2}{1-z_{P1}^2}\bigg)\frac{m_t}{M_t} \frac{(\vec{\gamma}_P)_3}{v_P} & r \frac{q^2 z_{P2}^2}{1-z_{P2}^2}\frac{(\vec{\gamma}_P)_3}{v_P}\\
    0 & 0 & z_{P1} \frac{m_t}{M_t} \frac{(\vec{\gamma}_P)_3^*}{v_P} & -r \bigg( \frac{z_{P1}}{1-z_{P1}^2} \bigg)\frac{m_t}{M_t}\frac{(\vec{\gamma}_P)_3^*}{v_P} & 1 & 0\\
    -\frac{(\vec{\gamma_P})_1^*}{v_P} & -\frac{(\vec{\gamma_P})_2^*}{v_P} & -q r \frac{(\vec{\gamma_P})_3^*}{v_P} & -\frac{q}{1-z_{P2}^2}\frac{(\vec{\gamma}_P)_3^*}{v_P} & 0 & 1
    \end{pmatrix} \mathbf{N}_2,\\
    &\mathbf{N_1} \equiv diag(1,1,1,sgn(y_u y_S), 1, 1), \; \; \mathbf{N}_2 \equiv diag(-sgn(u_1 y_S y_u), -sgn(u_2 y_S y_u), -1, 1, sgn(y_{P1}),sgn(y_{P2})), \nonumber
\end{align}
up to $O(m_t^2/M_t^2)$ corrections, where $r$, $q$, $z_{P1}$, and $z_{P2}$ are all defined as in Eq.(\ref{UuLTrunc}). Considering the case of the down-like quarks, where just as we did for $\mathcal{U}^d_L$ we have assumed that the heavy $b$ partner has decoupled from the theory, we arrive at
\begin{align}\label{UdRTrunc}
    \mathcal{U}^d_R \approx \begin{pmatrix}
    \mathbf{W}_d & \mathbf{0}_{3\times1} & \frac{\vec{\gamma}_P}{v_P}\\
    \mathbf{0}_{1\times3} & 1 & 0\\
    -\frac{\vec{\gamma}_P^\dagger \mathbf{W}_d}{v_P} & 0 & 1
    \end{pmatrix}\begin{pmatrix}
    -sgn(y_S y_d \mathbf{d}_D) & \mathbf{0}_{3\times1} & \mathbf{0}_{3\times1}\\
    \mathbf{0}_{1\times3} & sgn(y_{P1}) & 0\\
    \mathbf{0}_{1\times3} & 0 & sgn(y_{P2})
    \end{pmatrix}.
\end{align}

At this point, we have determined the rotation matrices $\mathcal{U}^u_L$ and $\mathcal{U}^u_R$ necessary to diagonalize the up-like quark mass matrix, and by trivial generalization, these results will also give us the rotation matrices $\mathcal{U}^d_{L,R}$ needed to diagonalize the down-like quark mass matrices. With this knowledge, we can now comment briefly on how to ensure that this construction reproduces the observed CKM matrix. Considering Eq.(\ref{UuL0}) and its down-like equivalent, we can see that in the limit where $\vec{\gamma}_{P,u,d} \rightarrow 0$, the CKM matrix (as defined by the coupling matrix to the SM $W$ boson, which couples to the left-handed SM quarks but not their heavy partners) is approximately given by
\begin{align}
    V_{CKM} = \mathbf{c}^u_\rho \mathbf{W}_u^\dagger \mathbf{W}_d \mathbf{c}^d_\rho,
\end{align}
where we remind the reader that $\mathbf{W}_u$ and $\mathbf{W}_d$ are the unitary matrices which diagonalize the Hermitian matrices $\mathbf{u}$ and $\mathbf{d}$ (as defined in Eq.(\ref{ULdefs}) for $\mathbf{u}$, with an analogous expression for $\mathbf{d}$) respectively. Since $\mathbf{c}^u_\rho \approx \mathbf{c}^d_\rho \approx 1$ up to $O(m_{q}^2/M_q^2)$ corrections (where $q=u,d,s,c,b,t$), the CKM matrix can be well-approximated as $\mathbf{W}_u^\dagger \mathbf{W}_d$, or in other words, by specifying $\mathbf{u}$ and $\mathbf{d}$ such that the clash of their diagonalization matrices matches the observed CKM.

Our recipe for reconstructing the quark sector in this construction is therefore compete, and it is useful at this point to take stock of the parameters which we may freely specify while still producing SM quark masses and mixings consistent with observation. First, we recall that there are two $SU(4)_F$ adjoint scalars in the model, $\Phi_A$ and $\Phi_B$, with vev's that we can write $\langle \Phi_A \rangle$ and $\sigma_2/2 \otimes \langle \Phi_B \rangle$. Then, we can use $SU(4)$ freedom to work in a basis in which the combination $y_A \langle \Phi_A \rangle + y_B \langle \Phi_B \rangle$ is diagonal, which we can see from the action in Eq.(\ref{GUTYukawa}) results in $\gamma_{u} =0$ and $\mathbf{u}=\mathbf{u}_D = diag(u_1,u_2,u_3)$. Then, we have $V_{CKM} \approx \mathbf{W}_d$, that is, the rotation matrix to diagonalize $\mathbf{d}$ is uniquely determined by the need to recreate the CKM matrix ($\mathbf{W}_u$ is, of course, simply the identity matrix in this basis). Turning to quark masses, we note that if we specify $y_u \equiv y_H \cos(\alpha-\beta) \sim O(1)$, $y_d \equiv y_H \sin(\alpha+\beta) \sim O(1)$, and $y_S v_S \sim 1\; \textrm{TeV}$, Eqs.(\ref{uLSMMat}) and (\ref{UuR0}) uniquely determine the eigenvalues of $\mathbf{u}$ and $\mathbf{d}$ up to a sign, and therefore also give us the masses of the heavy partner quarks and the mixing angles $\rho^{u,d}_{1,2,3}$ and $\eta^{u,d}_{1,2,3}$. If we additionally specify $y_{P1} \sim 1$, $y_{P2} \sim 1$, $v_P \gsim 1 \; \textrm{TeV}$, and the mass term $M$ (as seen in Eqs.(\ref{GUTYukawa}), (\ref{FermionMassMats}), and (\ref{ULdefs})), we can then specify the masses and mixings of the portal matter fermions (that is, those that consist primarily of fields with non-zero charge under $U(1)_D$). Notably, in our present construction, the masses of the some of the portal matter fermions are degenerate between the up- and down-quark sectors (specifically, $m_{P1}^u = m_{P1}^d$ and $m_{P2}^u = m_{P2}^d$) up to radiative corrections. In practice, we might expect such a relation to not hold precisely, due to differing renormalization group evolution of the Yukawa couplings $y_{P1,2}$ to the up- and down-like sectors, but for the purposes of our simple numerical study we shall take this relation at face value. The sole remaining parameters we may specify are the $U(1)_D$-breaking terms, in form of the complex vectors $\vec{\gamma}_{P,d} \lsim 1 \; \textrm{GeV}$ (recalling that in our basis, $\vec{\gamma}_u=0$), which in turn give us the effects of mixing between states of different $U(1)_D$ charge, as parameterized in Eqs.(\ref{LongDeltaExpressions}) and (\ref{LongDeltaRExpressions}).

In summary, then, we can generate a point in the model parameter space that produces the observed quark masses and mixings by specifying the $O(1)$ real parameters $y_H$, $y_{P1}$, $y_{P2}$, as well as $y_S v_S$, the angles $\alpha$ and $\beta$, the real vev's $v_P, v_S \gsim 1 \; \textrm{TeV}$, and the complex $U(1)_D$-breaking vev's $\vec{\gamma}_P$ and $\vec{\gamma}_d$. Altogether, these selections generate a unique point in parameter space up to the signs of the eigenvalues of the matrices $\mathbf{u}$ and $\mathbf{d}$.

Having addressed the quark sector, we move on to discussing the lepton sector of the theory. Notably, because none of the scalars introduced thus far break the group $SU(4)_c$, the Yukawa couplings in Eq.(\ref{GUTYukawa}) do not account for any discrepancy between down-like quark and charged lepton masses at tree level. However, because we've already assumed that the scale for $SU(4)_c$ breaking is incredibly high (and indeed must be, due to constraints on processes mediated by new $SU(4)_c$ gauge bosons), we can assume that discrepancies between the charged lepton and down-like quark masses are due to significant renormalization group effects. Specifically, we can posit differing running of the couplings $y_A$, $y_B$, and $y_H$, as well as the mass term $M$, for color triplets versus singlets. In practice, the task of generating a specific numerical realization of the leptonic Yukawa couplings necessary to reproduce the observed charged lepton spectrum (consistent with the vev structure necessary to reproduce the SM quark masses and CKM matrix) is analytically onerous and not terribly enlightening. However, we can expect that the effect of the specific realization of the charged lepton sector is largely phenomenologically irrelevant: Because the spectrum of charged leptons resembles that of the down-like quarks (at least in order of magnitude), it suffices here to point out that such a realization is almost certainly achievable with choices of leptonic $y_A$, $y_B$, $y_H$, and $M$ of roughly the same order of magnitude as those for the down-like quark sector, as would be consistent with RG evolution. Phenomenologically, then, we would expect that the $SU(3)_F$ heavy partner leptons here are too heavy to be realistically producible in existing and currently planned collider experiments, like their counterparts in the down-like quark sector. Therefore, the only potentially accessible new fermions arising from the charged leptons would be the $SU(3)_F$ singlet portal matter fields, specifically those which should have masses of $O(\textrm{TeV})$. However, due to their lack of color charge, experimental constraints on their production at the LHC are somewhat weaker than those of the quark portal matter fields with comparable masses, so we shall not address their phenomenology in detail here.

Meanwhile, the neutrino sector within this model presents substantial additional challenges. In particular, the tiny neutrino masses $\sim 0.1 \; \textrm{eV}$ and near-maximal neutrino flavor mixings are in general inconsistent with any sort of high-energy degeneracy with the up-like quark sector broken by renormalization group running, as the Yukawa couplings in Eq.(\ref{GUTYukawa}) suggest. However, there do exist extensions of the model that can account for this discrepancy. For example, including scalars in the representations $(\mathbf{15},\mathbf{1},\mathbf{1},\mathbf{15},0)$, $(\mathbf{15},\mathbf{1},\mathbf{3},\mathbf{15},0)$, $(\mathbf{15},\mathbf{1},\mathbf{1},\mathbf{1},0)$ and/or $(\mathbf{15},\mathbf{1},\mathbf{3},\mathbf{1},0)$ could allow for dramatically different Yukawa couplings and mixings between the quark and lepton sectors, at the expense of a yet larger scalar sector and extreme fine-tuning to reproduce the tiny neutrino masses. Alternatively, the introduction of a scalar in the representation $(\mathbf{10},\mathbf{1},\mathbf{3},\mathbf{10},-1/2)$ with the appropriate vev would permit the inclusion of a large Majorana mass for the right-handed $SU(3)_F$ triplet neutrinos, which would in turn suppress the SM neutrino masses via a seesaw mechanism. In both of these scenarios, the extension of the scalar sector would also result in substantial additional contributions to the masses of the new $SU(3)_F$ gauge bosons-- given the fact that the SM neutrino masses are so small, it's even probable that given these setups the vev's of some of these scalars would dominate the mass terms of the gauge bosons. Given the model-building ambiguity here and the substantial complexity that such considerations would introduce, however, we determine that a full exploration of potential ways to realize the neutrino mass spectrum and mixing within this framework is beyond the scope of our present work. In our discussion of the gauge boson masses in Section \ref{GaugeSpectrumSection}, however, we shall note that even in the absence of any additional mass terms which might arise from added scalar content to the model, the new gauge bosons associated with the $SU(3)_F$ symmetry will be extremely heavy, with only one of these bosons approaching a low enough mass scale to have a potentially observable effect even in highly constrained measurements of flavor-changing neutral currents. As a result, we can expect that any modifications to the model setup that must be made to accommodate the neutrino masses and mixings should have a minimal effect on the experimentally observable new physics in the model considered here.

\subsection{Gauge Boson Spectrum}\label{GaugeSpectrumSection}

Having selected our scalar vacuum expectation values such that the SM quark masses and mixings can be faithfully recreated, we now turn to how these selections will generate masses for the new gauge bosons in the theory. As noted in Section \ref{ScalarContentSection}, we shall assume for the sake of simplicity that the Pati-Salam symmetry group, $SU(4)_c \times SU(2)_L \times SU(2)_R$, is broken down to the SM gauge group $SU(3)_c \times SU(2)_L \times U(1)_Y$ at a significantly higher energy scale than any of the symmetry breakings of the group $SU(4)_F \times U(1)_F$. While this is certainly feasible, we do also note that even if that assumption is dropped, the $SU(2)_R$ symmetry must still be broken at a scale much higher than is presently observable, due to the vev of the field $\Phi_B \sim (\mathbf{1},\mathbf{1},\mathbf{3}, \mathbf{15},0)$, which we can estimate from our results in Section \ref{FermionSpectrumSection} will break this symmetry at a scale of roughly $\sim 10^8 \; \textrm{GeV}$. Assuming that the new gauge bosons associated with the Pati-Salam extension to the SM are too heavy to be relevant, we turn to the mass matrices of the gauge bosons corresponding to the $SU(4)_F \times U(1)_F$ generators. In this case, the squared mass matrix of the 16 gauge bosons here takes the form,
\begin{align}\label{GMassMat}
    &\mathcal{M}_G^2 =\begin{pmatrix}
    g_4^2 \mathbf{M}_{SU(4)} & \frac{3}{2} g_4 g_1 \vec{M}_{SU(4)\times U(1)} \\
    \frac{3}{2} g_4 g_1 \vec{M}^T_{SU(4)\times U(1)} & \frac{9}{8} g_1^2 |\langle \Phi_P \rangle |^2
    \end{pmatrix},\\
    &(\mathbf{M}_{SU(4)})_{ij} \equiv -2(Tr[[t^i, \langle \Phi_A \rangle][t^j, \langle \Phi_A \rangle]]+Tr[[t^i, \langle \Phi_B \rangle][t^j, \langle \Phi_B \rangle]])+ \langle \Phi_P\rangle^\dagger \{t^i, t^j\} \langle \Phi_P \rangle, \nonumber \\
    &(\vec{M}_{SU(4)\times U(1)})_i \equiv \langle \Phi_P \rangle^\dagger t^i \langle \Phi_P \rangle \nonumber,
\end{align}
where $g_4$ and $g_1$ are the $SU(4)_F$ and $U(1)_F$ coupling constants respectively, $[A,B]$ denotes the commutator of $A$ and $B$, $\{A,B\}$ denotes the anticommutator, and $t^i$ denotes the $i^{th}$ generator matrix for the fundamental representation of the group $SU(4)$, in the basis described in Appendix \ref{appendix:SU4Generators}. Note that here, $\mathbf{M}_{SU(4)}$ is a $15 \times 15$ real symmetric matrix, while $\vec{M}_{SU(4) \times U(1)}$ is a 15-component real vector, and both of these terms have dimensions of mass squared.

Referring to Eq.(\ref{GMassMat}), our first task to determine the mass eigenstates of  $SU(4)\times U(1)$ gauge bosons is to determine what the vev's $\langle \Phi_A \rangle$ and $\langle \Phi_B \rangle$ are, given our requirements that the SM quark masses and mixing matrices are properly reproduced. To that end, we consider the combinations of $\Phi_A$ and $\Phi_B$ that are coupled to the up- and down-like quarks, referring to Eq.(\ref{FermionMassMats}) and several useful variable definitions in Eq.(\ref{ULdefs}). Comparing these to the vev's of $\Phi_A$ and $\Phi_B$ as given in Table \ref{table:scalars}, we arrive at
\begin{align}\label{PhiMats}
    &\begin{pmatrix}
    \mathbf{u}_D & \mathbf{0}_{3\times1} \\
    \mathbf{0}_{1\times3} & X_u
    \end{pmatrix} = y_A \begin{pmatrix}
    \langle A \rangle & \vec{\gamma}_A \\
    \vec{\gamma}_A^\dagger & -Tr[\langle A \rangle]
    \end{pmatrix}+y_B \begin{pmatrix}
    \langle B \rangle & \vec{\gamma}_B \\
    \vec{\gamma}_B^\dagger & -Tr[\langle B \rangle]
    \end{pmatrix}+M \mathbf{1}_{4\times4}, \\
    &\begin{pmatrix}
    V_{CKM}^\dagger \mathbf{d}_D V_{CKM} & \vec{\gamma}_d \\
    \vec{\gamma}_d^\dagger & X_d
    \end{pmatrix} = y_A \begin{pmatrix}
    \langle A \rangle & \vec{\gamma}_A \\
    \vec{\gamma}_A^\dagger & -Tr[\langle A \rangle]
    \end{pmatrix}-y_B \begin{pmatrix}
    \langle B \rangle & \vec{\gamma}_B \\
    \vec{\gamma}_B^\dagger & -Tr[\langle B \rangle]
    \end{pmatrix}+M \mathbf{1}_{4\times4}, \nonumber
\end{align}
where we remind the reader that we can use $SU(4)$ gauge freedom to work in a basis where $\mathbf{u}$ is diagonal and $\vec{\gamma}_u=0$, as we have done in Section \ref{FermionSpectrumSection}, and that $V_{CKM}$ is the CKM matrix. We recall that in that Section, the procedure we developed to produce a point in parameter space that recreates the SM quark masses and mixings already leaves us with $\mathbf{u}_D$, $\mathbf{d}_D$, and $M$ specified. Then, simply selecting the Yukawa couplings $y_A$ and $y_B$ here will allow us to uniquely determine the matrices $\langle A \rangle$ and $\langle B \rangle$, as well as the vectors $\vec{\gamma}_{A}$ and $\vec{\gamma}_B$. Specifically, we have
\begin{align}\label{PhiIdentities}
    &\langle A \rangle = \frac{1}{2 y_A} (\mathbf{u}_D+V_{CKM}^\dagger \mathbf{d}_D V_{CKM}-2 M), \; \; \langle B \rangle = \frac{1}{2 y_B} (\mathbf{u}_D-V_{CKM}^\dagger \mathbf{d}_D V_{CKM}), \\
    &\vec{\gamma}_A = \frac{1}{2 y_A} \vec{\gamma}_d, \; \; \vec{\gamma}_B = -\frac{1}{2 y_B} \vec{\gamma}_d. \nonumber
\end{align}
Hence, given a construction which generates the SM quark masses and mixings, we now can readily determine the necessary vev's $\langle \Phi_A \rangle$ and $\langle \Phi_B \rangle$ by only specifying two additional parameters, the Yukawa couplings $y_A$ and $y_B$. With this information, we can now discuss the spectrum of the gauge bosons which results. In general, the eigenvalues of the mass matrix in Eq.(\ref{GMassMat}) are highly complicated expressions and difficult to present in a compact closed form. However, we can avail ourselves of the hierarchies present between eigenvalues of $\mathbf{u}_D$ and $\mathbf{d}$, as well as the nearly-diagonal form of the CKM matrix, in order to dramatically simplify matters. Specifically, we will employ the Wolfenstein parameterization of the CKM matrix,
\begin{align}\label{CKMWolf}
    &\begin{pmatrix}
    1-\frac{\lambda^2}{2}-\frac{\lambda^4}{8} & \lambda & A \lambda^3(\rho - i \eta) \\
    -\lambda & 1-\frac{\lambda^2}{2}-(1+4 A^2) \frac{\lambda^4}{8} & A \lambda^2 \\
    -A(\rho+i \eta -1) \lambda^3 & -A(\lambda^2+\frac{\lambda^4}{2}(2(\rho+i \eta) -1)) & 1-\frac{A^2 \lambda^4}{2}
    \end{pmatrix}, \\
    \nonumber\\
    &\lambda = 0.22453, \; \; A = 0.836, \; \; \rho = 0.122, \; \; \eta = 0.355, \nonumber
\end{align}
and note that, numerically, the eigenvalues of $\mathbf{u}_D$ and $\mathbf{d}$ ($u_{1-3}$ and $d_{1-3}$, respectively) satisfy $u_2 \sim \lambda^4 u_1$, $u_3 \sim \lambda^8 u_1$, $d_2 \sim \lambda^2 d_1$, and $d_3 \sim \lambda^4 d_1$ for values of these parameters that reproduce the SM quark masses. By rewriting Eq.(\ref{GMassMat}) using the identities in Eq.(\ref{PhiIdentities}), and making the substitutions $u_2 \rightarrow \lambda^4 \upsilon_2$, $u_3 \rightarrow \lambda^8 \upsilon_3$, $d_2 \rightarrow \lambda^2 \delta_2$, and $d_3 \rightarrow \lambda^4 \delta_3$, where $u_1 \sim d_1 \sim \upsilon_2 \sim \delta_2 \sim \upsilon_3 \sim \delta_3$, we can determine an approximate hierarchy between gauge boson mass eigenvalues by expanding in the parameter $\lambda$. To start, we note that 13 of the gauge bosons acquire masses that are generally in excess of any currently observable energy scale. In particular, we find that 10 gauge bosons possess masses that are of $O(u_1,d_1) \sim 10^8 \; \textrm{GeV}$, while three more have masses of $O(\lambda^2 u_1, \lambda^2 d_1) \sim 10^7 \; \textrm{GeV}$. Even more conveniently, the gauge bosons which acquire masses at these scales correspond to the operators facilitating the most constrained flavor-changing effects, those involving light quarks: These gauge bosons in fact must develop high masses, because the scales of the scalar vev's which contribute to these operators are ultimately set by the masses of the light quarks' seesaw partners, which dictate how we constructed our scalar vev's. As we have seen in Section \ref{FermionSpectrumSection}, these masses are enormous, and therefore in our construction the masses of the gauge bosons which couple to the light quarks must be similarly large.

For particles at the scale of these heavy bosons (around $10^{7-8} \; \textrm{GeV})$, the only even remotely observable effects stem from the imaginary part of effective 4-quark operators that facilitate $K^0-\overline{K}^0$ mixing \cite{Alpigiani:2017lpj,Blanke:2017ohr}. However, from the $\lambda$ expansions outlined above, it can be seen that effects which give a non-trivial phase to the coefficients of these 4-quark operators only appear at the $O(\lambda^3)$ level or higher. Since the masses of the potentially offending gauge bosons are already near the limit of the sensitivity of this tree-level probe, this essentially rules out any observable effect from the heavy bosons. Given that any observable signatures of these bosons are likely well outside the sensitivity of any experiment, we can shift our attention to the three remaining gauge bosons that acquire mass at a far lower scale.

The first of these, which we shall refer to as $Z_F$ (the ``flavor $Z$''), consists almost entirely of a combination of the generators of the $SU(3)_F$ group embedded in $SU(4)_F$ (corresponding to the first eight $t^i$ generators, and the first eight rows/columns of the mass matrix in Eq.(\ref{GMassMat})), up to numerically negligible corrections due to $\vec{\gamma}_{d,P}$ terms. There does not exist a compact numerically accurate approximation for the mass of $Z_F$. However, expanding in the Wolfenstein parameter $\lambda$ as above yields the result that $m_{Z_F} \sim O(\lambda^4 d_1, \lambda^4 u_1) \sim 10^5 \; \textrm{GeV}$. While this initially seems quite large, we note that a boson of this mass may still mediate observable tree-level flavor-changing neutral currents \cite{Charles:2020dfl,Bao:2015pva,Blanke:2017ohr}. As such, it behooves us to determine how $Z_F$ couples to the quarks, and crucially what flavor-changing interactions it can mediate. Expanding Eq.(\ref{GMassMat}) in the Wolfenstein parameter $\lambda$ offers an appealing approximate expression for the relevant combinations of generators corresponding to this state. Specifically, up to $O(\lambda^2)$, the $SU(4)_F$ generator combination corresponding to $Z_F$ becomes
\begin{align}\label{ApproxZFGenerator}
    O(\lambda^7) \; t^1+O(\lambda^7) \; t^2+O(\lambda^4) \; t^3+ O(\lambda^5) \; t^4 + O(\lambda^5) \; t^5+\sqrt{3} A \lambda^2 (\rho-1) \; t^6 + \sqrt{3} A \lambda^2 \eta\; t^7 + t^8.
\end{align}
Consulting the explicit form of the generator matrices $t^i$ in Appendix \ref{appendix:SU4Generators}, we see that $Z_F$ only mediates significant flavor-changing neutral currents between the second and third generations -- in both the up- and down-like sectors, flavor-changing neutral currents featuring the first generation of quarks only appear at the $O(\lambda^5)\sim 10^{-4}$ level, at most. Our chief phenomenological concern regarding $Z_F$, then, is the constraint arising from tree-level flavor-changing neutral currents that it mediates between the second and third generation of quarks. We shall discuss phenomenological constraints arising from this type of interaction in Section \ref{ZFMesonSection}.

The next two light gauge bosons we consider will arise from the generators that are left unbroken by $\langle \Phi_A \rangle$ and $\langle \Phi_B \rangle$, at least in the approximation where $\vec{\gamma}_d \rightarrow 0$. In particular, we see that in the limit where $\vec{\gamma}_{d,P}\rightarrow 0$, $\langle \Phi_A \rangle$ and $\langle \Phi_B \rangle$ break $SU(4)_F \times U(1)_F$ down to $U(1)_F' \times U(1)_F$, where the generator for $U(1)_F'$ is given by the matrix $t^{15}$. In the limit where $\vec{\gamma}_P \rightarrow 0$, the scalar $\langle \Phi_P \rangle = (\vec{\gamma}_P, v_P)$ breaks $U(1)_F'\times U(1)_F$ down to $U(1)_D$ at the scale $v_P \sim 1-10 \; \textrm{TeV}$. The $\vec{\gamma}_d$ and $\vec{\gamma}_P$ terms then finally break the $U(1)_D$ symmetry at $\sim 0.1-1 \; \textrm{GeV}$. The two mass eigenstates we consider here, then, will be well-approximated as combinations of the $U(1)_F$ and $U(1)_F'$ bosons, with one achieving a mass of $O(v_P) \sim 1-10 \; \textrm{TeV}$, the scale at which $U(1)_F'\times U(1)_F$ breaks down to $U(1)_D$, and the other achieving mass only at the scale where $U(1)_D$ is broken, at approximately $0.1-1 \; \textrm{GeV}$. Referring to Eq.(\ref{GMassMat}), we see that in the limit where $\vec{\gamma}_{d,P}\rightarrow 0$, the squared mass matrix for the eventual mass eigenstates, $Z_P$ and $A_D$ (\ie, the dark photon),  becomes
\begin{align}\label{GMassMatZP}
    \begin{pmatrix}
    B' & B
    \end{pmatrix}\begin{pmatrix}
    \frac{3 g_4^2 v_P^2}{4} & -\sqrt{\frac{3}{2}}\frac{3 g_4 g_1 v_P^2}{4} \\
    -\sqrt{\frac{3}{2}}\frac{3 g_4 g_1 v_P^2}{4} & \frac{9 g_1^2 v_P^2}{8}
    \end{pmatrix} \begin{pmatrix}
    B' \\
    B
    \end{pmatrix},
\end{align}
where $B$ and $B'$ refer to the $U(1)_F$ and $U(1)_F'$ gauge bosons, respectively. It is convenient to define an angle $\theta_P$ such that $g_1 = \sqrt{2/3} g_4 \tan \theta_P$. Then, we may write the mass squared matrix as
\begin{align}\label{GMassMatZPTheta}
    \frac{3 g_4^2 v_P^2}{4}\begin{pmatrix}
    B' & B
    \end{pmatrix}\begin{pmatrix}
    1 & - \tan \theta_P \\
    - \tan \theta_P & \tan^2 \theta_P
    \end{pmatrix}\begin{pmatrix}
    B' \\
    B
    \end{pmatrix}.
\end{align}
The mass eigenstates for these gauge bosons can then be easily determined, up to $O(\vec{\gamma}_{P,d}^2/v_P^2) \sim 10^{-4}$ corrections. Given the definition of $\theta_P$ we've provided, we see that the eigenstates are simply
\begin{align}\label{AdZPDefs}
    A_D \approx B' s_P + B c_P, \; \; Z_P \approx B' c_P - B s_P,
\end{align}
where $A_D$ refers to the dark photon, with a mass of $\sim 0.1-1 \; \textrm{GeV}$, while $Z_P$ refers to the heavier gauge boson, a dark ``portal Z'', with $O(v_P)$ mass. Here we have also used the abbreviations $c_P \equiv \cos \theta_P$ and $s_P \equiv \sin \theta_P$, which for simplicity we shall also employ going forward.

The masses of $A_D$ and $Z_P$ can also be straightforwardly determined. In the case of $Z_P$, we can directly extract the mass from the mass matrix of Eq.(\ref{GMassMatZPTheta}), arriving at
\begin{align}\label{PortalZMass}
    m_{Z_P}^2 \approx \frac{3 g_4^2 v_P^2}{4 c^2_P},
\end{align}
which is accurate up to numerically negligible $O(\vec{\gamma}_{P,d}^2/v_P^2)$ corrections. Meanwhile, the mass of the dark photon can be given up to $O(\lambda^2) \sim O(10^{-2})$ corrections by
\begin{align}\label{DarkPhotonMass}
    m_{A_D}^2 &\approx \frac{2 g_4^2 s^2_P}{3} \bigg(\frac{7}{8}(\vec{\gamma}_P^* \cdot \vec{\gamma_P}) +\mu_1^{A_D}(\vec{\gamma}_d^* \cdot \vec{\gamma}_d) + \mu_2^{A_D}|(\vec{\gamma}_d)_1|^2 \bigg),\\
    \mu_1^{A_D} &\equiv \frac{4(4M-u_1)^2}{y_A^2 (d_1-u_1)^2 + y_B^2(d_1+u_1-8 M)^2}, \nonumber\\
    \mu_2^{A_D} &\equiv \frac{16 M (d_1-u_1)(y_A^2(d_1-u_1)(-3 M +u_1)+y_B^2((-4M+u_1)^2-M u_1 + d_1(-3M+u_1)))}{(y_A^2(d_1-u_1)^2+y_B^2(d_1+u_1-8M)(d_1+u_1-4M))^2+16 y_A^2 y_B^2 M^2 (d_1-u_1)^2}, \nonumber
\end{align}
where to arrive at this expression, we have exploited the fact that $d_1,u_1 \gg d_{2,3},u_{2,3}, v_P$.
At this point, we have obtained the masses and eigenvectors for the three new gauge bosons that may be phenomenologically relevant -- namely, the $SU(3)_F$ boson $Z_F$ with a mass of $O(10^5 \; \textrm{GeV})$, the ``portal $Z$'' field $Z_P$ with a mass roughly of $O(1 \; \textrm{TeV})$, and the dark photon field $A_D$, with a mass of $O(0.1-1 \; \textrm{GeV})$.

\section{Couplings, Kinetic Mixing, and Low-Energy Parameters}\label{CouplingsMixingsandParamsSection}

Before computing the observable experimental signatures of this setup, it is useful to explicitly determine the couplings between various phenomenologically relevant particles in the model. In particular, in light of our discussion of the fermion spectrum in Section \ref{FermionSpectrumSection}, we shall focus on the SM and new physics gauge boson and scalar couplings of the SM fermions, the portal matter fermions, and the heavy top partner.

\subsection{$SU(4)_F \times U(1)_F$ Gauge Bosons}\label{FlavorGaugeCouplingsSection}

We begin our discussion by deriving the couplings of the new gauge bosons that arise in our setup due to our $SU(4)_F \times U(1)_F$ extension of the SM gauge group. In Section \ref{GaugeSpectrumSection}, we have determined that only three of the new gauge bosons are sufficiently light so as to have observable phenomenological effects: A (comparatively) light $SU(3)_F$ boson that we have dubbed $Z_F$, with a mass of roughly $O(10^5 \; \textrm{GeV})$, a ``portal $Z$'' field with a mass of $O(1-10 \; \textrm{TeV})$, which we denote as $Z_P$, and a dark photon, $A_D$, with a mass of approximately $O(0.1-1 \; \textrm{GeV})$.

To start, we consider the couplings of the dark photon $A_D$ to the fermions of the theory {\it before} KM takes place. In the limit where $\vec{\gamma}_{P,d} \rightarrow 0$, the dark photon is given as a simple combination of two $U(1)$ bosons, described in Eq.(\ref{AdZPDefs}). In fact, the results of Eq.(\ref{AdZPDefs}) are exceedingly numerically accurate, even once contributions from $\vec{\gamma}_{P,d}$ terms are included: Mixing between $A_D$, the gauge boson $Z_P$, and the gauge bosons corresponding to the $SU(3)_F$ embedded in $SU(4)_F$ only occurs at second order in the quantities $\vec{\gamma}_{P,d}$, and is hence numerically negligible, while mixing between $A_D$ and the gauge bosons corresponding to the other $SU(4)_F$ generators is suppressed by the latter's enormous masses -- in Section \ref{GaugeSpectrumSection}, we found these to be of $O(10^{7-8}) \; \textrm{GeV}$. Thus, we can use the combinations given in Eq.(\ref{AdZPDefs}) to derive the couplings of $A_D$ and $Z_P$ without concerning ourselves with any additional complicating effects. Noting that the $B'$ boson corresponds to the $15^{\textrm{th}}$ generator of $SU(4)_F$ ($t^{15}$, as listed in Appendix \ref{appendix:SU4Generators}), and $B$ is simply the $U(1)_F'$ boson, we can straightforwardly find a coupling matrix by writing the fermions as 9-dimensional vectors in flavor space, as in Eq.(\ref{FermionFlavorBasis}). We find that for up-like and down-like quarks, charged leptons, and neutrinos (at least ignoring any extra structure the model may need to accommodate neutrino flavor phenomenology), the dark photon couplings are given by 
\begin{align}\label{DarkPhotonCoupling}
    g_4 \sqrt{\frac{2}{3}}s_P (\mathbf{C}^{u,d,e,\nu}_{A_D})_{L,R} = -g_4 \sqrt{\frac{2}{3}}s_P \mathcal{U}^{u,d,e,\nu \dagger}_{L,R} \begin{pmatrix}
    \mathbf{0}_{3\times3} & \mathbf{0}_{3\times3} & \mathbf{0}_{3\times3}\\
    \mathbf{0}_{3\times3} & \mathbf{0}_{3\times3} & \mathbf{0}_{3\times3}\\
    \mathbf{0}_{3\times3} & \mathbf{0}_{3\times3} & \mathbf{1}_{3\times3}
    \end{pmatrix} \mathcal{U}^{u,d,e,\nu}_{L,R},
\end{align}
where we have used the fact that the $U(1)_F$ coupling constant, $g_1$, is given in terms of the $SU(4)_F$ coupling constant by the relation $g_1 = \sqrt{2/3} g_4 \tan \theta_P$. The overall negative sign in this expression is simply an artifact of the sign conventions we have selected for the definitions of $U(1)_F$ and $U(1)_D$. At this point, we have omitted the additional couplings arising from kinetic mixing with the SM hypercharge field as noted earlier -- these will introduce shifts of $e \epsilon Q \mathbf{1}_{9\times9}$ to the above expression, where $e$ is the proton charge, $\epsilon$ is the kinetic mixing parameter and $Q$ is the electric charge of the given field. While these shifts are obviously significant in dark matter phenomenology, since they encapsulate the means by which, for example, a dark matter field might annihilate to form SM particles, they are of less importance to us when considering other constraints here: Because the $e \epsilon Q$ shift is always proportional to the identity matrix for a given fermion electric charge, it does not facilitate flavor-changing neutral currents or portal matter decay to SM particles.
Restricting our attention to only fermions which are light enough to be phenomenologically relevant, namely the SM quarks, the heavy top partner, and the lightest two portal matter fields in each sector, we can rewrite the coupling matrices using the truncated flavor vectors of Eq.(\ref{ShortVecs}). In the up-like quark sector, we arrive at the truncated coupling matrices
\begin{align}\label{ADuCouplingTrunc}
    &(\mathbf{C}_{A_D}^u)_L \approx \begin{pmatrix}
    & & & 0 & -\frac{(\vec{\gamma}_P)_1}{v_P} & 0\\
    & (\mathbf{C}_{A_D}^u)_L^{SM} & & 0 & -\frac{(\vec{\gamma}_P)_2}{v_P} & 0\\
    & & & 0 & -\frac{(\vec{\gamma}_P)_3}{v_P} & q r z_{P2} \frac{m_t}{M_t} \frac{(\vec{\gamma}_P)_3}{v_P}\\ 
    0 & 0 & 0 & 0 & -\frac{r}{(1-z_{P1}^2)} \frac{m_t}{M_t} \frac{(\vec{\gamma}_P)_3}{v_P} & - \frac{q z_{P2}}{1-z_{P2}^2} \frac{(\vec{\gamma}_P)_3}{v_P}\\
    -\frac{(\vec{\gamma}_P)_1^*}{v_P} & -\frac{(\vec{\gamma}_P)_2^*}{v_P} & -\frac{(\vec{\gamma}_P)_3^*}{v_P} & -\frac{r}{(1-z_{P1}^2)} \frac{m_t}{M_t}\frac{(\vec{\gamma}_P)_3^*}{v_P} & 1 & 0\\
    0 & 0 & q r z_{P2} \frac{m_t}{M_t} \frac{(\vec{\gamma}_P)_3^*}{v_P} & -\frac{q z_{P2}}{1-z_{P2}^2}\frac{(\vec{\gamma}_P)_3^*}{v_P} & 0 & 1
    \end{pmatrix},\\
    &(\mathbf{C}_{A_D}^u)_R \approx \mathbf{N}_2 \begin{pmatrix}
    & & & 0 & 0 & \frac{(\vec{\gamma}_P)_1}{v_P}\\
    & (\mathbf{C}_{A_D}^u)_R^{SM} & & 0 & 0 & \frac{(\vec{\gamma}_P)_2}{v_P}\\
    & & & 0 & -z_{P1} \frac{m_t}{M_t} \frac{(\vec{\gamma}_P)_3}{v_P} & qr \frac{(\vec{\gamma}_P)_3}{v_P}\\ 
    0 & 0 & 0 & 0 & -\frac{r z_{P1}}{(1-z_{P1}^2)} \frac{m_t}{M_t} \frac{(\vec{\gamma}_P)_3}{v_P} & - \frac{q}{1-z_{P2}^2} \frac{(\vec{\gamma}_P)_3}{v_P}\\
    0 & 0 & -z_{P1}\frac{m_t}{M_t}\frac{(\vec{\gamma}_P)_3^*}{v_P} & -\frac{r z_{P1}}{(1-z_{P1}^2)} \frac{m_t}{M_t}\frac{(\vec{\gamma}_P)_3^*}{v_P} & 1 & 0\\
    \frac{(\vec{\gamma}_P)_1^*}{v_P} & \frac{(\vec{\gamma}_P)_2^*}{v_P} & q r \frac{(\vec{\gamma}_P)_3^*}{v_P} & -\frac{q}{1-z_{P2}^2}\frac{(\vec{\gamma}_P)_3^*}{v_P} & 0 & 1
    \end{pmatrix} \mathbf{N}_2, \nonumber\\
    &(\mathbf{C}_{A_D}^u)_L^{SM} \equiv \frac{(\vec{\gamma}_P \otimes \vec{\gamma}_P^\dagger)}{v_P^2}, \; \; \; (\mathbf{C}_{A_D}^u)_R^{SM} \equiv \begin{pmatrix}
    \frac{|(\vec{\gamma}_P)_1|^2}{v_P^2} & \frac{(\vec{\gamma}_P)_1(\vec{\gamma}_P)_2^*}{v_P^2} & q r \frac{(\vec{\gamma}_P)_1(\vec{\gamma}_P)_3^*}{v_P^2}\\
    \frac{(\vec{\gamma}_P)_2(\vec{\gamma}_P)_1^*}{v_P^2} & \frac{|(\vec{\gamma}_P)_2|^2}{v_P^2} & q r \frac{(\vec{\gamma}_P)_2(\vec{\gamma}_P)_3^*}{v_P^2}\\
    q r\frac{(\vec{\gamma}_P)_3(\vec{\gamma}_P)_1^*}{v_P^2} & q r\frac{(\vec{\gamma}_P)_3(\vec{\gamma}_P)_2^*}{v_P^2} & q^2 r^2 \frac{|(\vec{\gamma}_P)_3|^2}{v_P^2} 
    \end{pmatrix}, \nonumber
\end{align}
where $q$, $r$, $z_{P1}$, and $z_{P2}$ are defined as in Eq.(\ref{UuLTrunc}), and $\mathbf{N}_2$ is defined as in Eq.(\ref{UuRTrunc}). The down-like quark sector, meanwhile, has its truncated coupling matrices given by
\begin{align}\label{ADdCouplingTrunc}
    (\mathbf{C}_{A_D}^d)_L \approx \begin{pmatrix}
    \frac{(\mathbf{W}_d^\dagger \vec{\gamma_P}) \otimes (\vec{\gamma}_P^\dagger \mathbf{W}_d)}{v_P^2} & -\frac{\mathbf{W}_d^\dagger \vec{\gamma}_P}{v_P} & \mathbf{0}_{3\times1}\\
    -\frac{\vec{\gamma}_P^\dagger \mathbf{W}_d}{v_P} & 1 & 0\\
    \mathbf{0}_{1\times3} & 0 & 1
    \end{pmatrix},\\
    (\mathbf{C}_{A_D}^d)_R \approx \begin{pmatrix}
    \frac{(\mathbf{W}_d^\dagger \vec{\gamma_P}) \otimes (\vec{\gamma}_P^\dagger \mathbf{W}_d)}{v_P^2} & \mathbf{0}_{3\times1} & -\frac{\mathbf{W}_d^\dagger \vec{\gamma}_P}{v_P}\\
    \mathbf{0}_{1\times3} & 1 & 0\\
    -\frac{\vec{\gamma}^\dagger_P \mathbf{W}_d}{v_P} & 0 & 1
    \end{pmatrix},\nonumber
\end{align}
where we remind the reader that $\mathbf{W}_d$ is the unitary matrix first described in Eq.(\ref{ULdefs}), and that in the $SU(4)_F$ basis we have chosen, $\mathbf{W}_d$ is equal to the CKM matrix. The corresponding coupling matrix for the charged leptons can be given in complete analogy to Eq.(\ref{ADdCouplingTrunc}), with the sole exception that the matrix $\mathbf{W}_d$ must be replaced by the appropriate rotation matrix given the lepton couplings to the scalars $\Phi_A$ and $\Phi_B$. 

In \textbf{I} and \textbf{II}, it was seen that in the event that the portal matter mixes with electrons, there exists a parity-violating interaction of the right-handed electrons with the dark photon field. Consulting Eq.(\ref{ADdCouplingTrunc}), we observe that the analogous terms emerge at the $O(\vec{\gamma}_P^2/v_P^2)$ level, however, they occur \emph{identically} for the right- and left-handed fields (at least up to corrections due to the super-heavy fermions that we have omitted from our truncated coupling matrices). As a result, no such parity-violating interaction occurs in this model, in contrast to those of \textbf{I} and \textbf{II}. Generically, this can be expected to be the case due to the more left-right symmetric form of the model: The left-handed electrons receive a chiral correction from their coupling from the $SU(2)_L$ doublet portal matter, while the right-handed ones receive the \emph{same} correction from $SU(2)_L$ singlet portal matter. The couplings in the neutrino sector may be qualitatively similar to those that we have already explored, however, we note that the recreation of light neutrino masses and the observed mixing matrix will likely require significant modifications to the neutrino mass matrix, which are beyond the scope of this paper. Notably, we have retained $O(\vec{\gamma}_P^2/v_P^2)$ terms in the components of Eqs.(\ref{ADuCouplingTrunc}) and (\ref{ADdCouplingTrunc}) which correspond to mixing among the SM quarks.\footnote{In spite of the fact that our expressions for $\mathcal{U}^u_{L,R}$ and for $A_D$'s mass eigenvector are only valid up to $O(\vec{\gamma}_P/v_P)$, these expressions (at least for flavor-changing interactions, which are of phenomenological interest here) are numerically valid, because the $O(\vec{\gamma}_P^2/v_P^2)$ corrections to the matrices $\mathcal{U}^u_{L,R}$ don't contribute to the $O(\vec{\gamma}_P^2/v_P^2)$ flavor-changing couplings in the SM, and we shall see that the gauge boson $Z_P$, which may mix with $A_D$ at the $O(\vec{\gamma}_P^2/v_P^2)$ level, has SM flavor-universal couplings up to $O(\vec{\gamma}_P/v_P)$ corrections.} In spite of their minute magnitude, with $\vec{\gamma}_P/v_P \sim 10^{-(3-4)}$, we have kept them here because they facilitate highly-constrained flavor-changing neutral currents mediated by the extremely light $A_D$ boson. In Section \ref{ADFlavorConstraintsSection}, we shall in fact derive significant non-trivial constraints on the model parameter space from these interactions. We do note in passing that similar $O(\vec{\gamma}_P^2/v_P^2)$ terms exist elsewhere in Eqs.(\ref{ADuCouplingTrunc}) and (\ref{ADdCouplingTrunc}), however, we have omitted them here since they have a negligible effect on any observable physics.

Apart from the flavor-changing interactions we observe in the SM, the dark photon also facilitates $O(\vec{\gamma}_P/v_P)$ interactions between the portal matter fermion fields and those uncharged under $U(1)_D$, namely the SM quarks and the heavy top partner. In practice, due to $A_D$'s tiny mass compared to other gauge bosons in the theory, these interactions make meaningful contributions to a number of interesting processes; most notably, they dominate the decay of portal matter to an SM quark, or the decay of a top partner to portal matter (or vice versa, depending on which process is kinematically allowed). The presence of these couplings for the extremely light dark photon field can substantially simplify our later discussions of the couplings of heavier gauge bosons: From the Equivalence Theorem \cite{GoldstoneEquivalence}, and the fact that $|\vec{\gamma}_P|\sim m_{A_D} \sim 0.1-1 \; \textrm{GeV}$, we would anticipate that the overall strength of the $A_D$-facilitated interaction between portal and non-portal matter would undergo a substantial enhancement over what the $\vec{\gamma}_P/v_P$ suppression in its coupling would suggest, since this suppression would be cancelled by $m_{A_D}$ (at least for the longitudinal mode of $A_D$). Since obviously such an enhancement doesn't exist for $\vec{\gamma}_P/v_P$-suppressed couplings for heavier gauge bosons, which in turn mediate the same $U(1)_D$-breaking couplings as we see emerging from $A_D$, we therefore can omit a detailed evaluation of $O(\vec{\gamma}_P/v_P)$-suppressed effects in the couplings of $Z_F$, $Z_P$, the SM electroweak gauge bosons  and the SM-like Higgs; in all cases we consider, these effects are overwhelmed by those arising from $A_D$.

We next turn to the couplings for the ``portal $Z$'' boson, $Z_P$. Again referencing Eq.(\ref{AdZPDefs}), this time to get the approximate mass eigenvector for $Z_P$, we arrive at the coupling matrices
\begin{align}\label{ZPExactCoupling}
    -\frac{g_4}{2\sqrt{6} c_P} (\mathbf{C}^{u,d,e,\nu}_{Z_P})_L = -\frac{g_4}{2\sqrt{6} c_P} \mathcal{U}^{u,d,e,\nu \dagger}_L \begin{pmatrix}
    \mathbf{1}_{6\times6} & \mathbf{0}_{6\times1} & \mathbf{0}_{6\times1} & \mathbf{0}_{6\times1}\\
    \mathbf{0}_{1\times6} & 1-4 c^2_P & 0 & 0\\
    \mathbf{0}_{1\times6} & 0 & 4 s^2_P & 0\\
    \mathbf{0}_{1\times6} & 0 & 0 & 1-4 c^2_P
    \end{pmatrix} \mathcal{U}^{u,d,e,\nu}_L,\\
    -\frac{g_4}{2\sqrt{6} c_P} (\mathbf{C}^{u,d,e,\nu}_{Z_P})_R = -\frac{g_4}{2\sqrt{6} c_P} \mathcal{U}^{u,d,e,\nu \dagger}_R \begin{pmatrix}
    \mathbf{1}_{6\times6} & \mathbf{0}_{6\times1} & \mathbf{0}_{6\times1} & \mathbf{0}_{6\times1}\\
    \mathbf{0}_{1\times6} & 4 s^2_P & 0 & 0\\
    \mathbf{0}_{1\times6} & 0 & 1-4 c^2_P & 0\\
    \mathbf{0}_{1\times6} & 0 & 0 & 1-4 c^2_P
    \end{pmatrix} \mathcal{U}^{u,d,e,\nu}_R. \nonumber
\end{align}
We can then determine what the couplings are for the quarks that are light enough to remain phenomenologically relevant, as we've already done for $A_D$, by inserting our results for $\mathcal{U}^{u,d}_{L,R}$ from Eqs.(\ref{UuLTrunc}), (\ref{UdLTrunc}), (\ref{UuRTrunc}), and (\ref{UdRTrunc}) into Eq.(\ref{ZPExactCoupling}). In fact, up to $O(\vec{\gamma}_P/v_P)$ corrections, which we note are negligible compared to similar interactions arising from $A_D$, we find that we can write the truncated coupling matrices as
\begin{align}\label{ZPTruncCoupling}
    &(\mathbf{C}^{u}_{Z_P})_L \approx diag(1,1,1,1,1-4c^2_P, 4 s^2_P), \nonumber\\
    &(\mathbf{C}^{u}_{Z_P})_R \approx diag(1,1,1,1, 4 s^2_P, 1-4 c^2_P),\\
    &(\mathbf{C}^{d,e}_{Z_P})_L \approx diag(1,1,1,1-4 c^2_P, 4 s^2_P), \nonumber\\
    &(\mathbf{C}^{d,e}_{Z_P})_R \approx diag(1,1,1,4 s^2_P, 1-4 c^2_P), \nonumber
\end{align}
where we have noted that in the absence of corrections due to mixing with heavy partner states and $\vec{\gamma}_P$ terms, the coupling matrix for charged leptons here is the same as that for down-like quarks. We shall not explicitly determine the coupling matrices in the neutrino sector in this work. As noted at the end of Section \ref{FermionSpectrumSection}, a realistic neutrino mass matrix in this model would likely involve vacuum expectation values of still more additional $SU(4)_F \times U(1)_F$ scalars, which would complicate the already highly intricate structure of the model while remaining unlikely to influence most of the new physics at an experimentally observable scale.

We finally address couplings which arise from the gauge boson $Z_F$, the ``flavor $Z$'' that represents the only gauge boson from the $SU(3)_F$ flavor symmetry which possesses a low enough mass to have some phenomenological impact. Consulting Eq.(\ref{ApproxZFGenerator}), we see that the coupling matrix for this gauge boson may be written,
\begin{align}
    &\frac{g_4}{2\sqrt{3}}(\mathbf{C}^{u,d,e,\nu}_{Z_F})_{L,R} = \frac{g_4}{2 \sqrt{3}} \mathcal{U}^{u,d,e,\nu \dagger}_{L,R}\begin{pmatrix}
    \boldsymbol{\Lambda}_{3\times3} & \mathbf{0}_{3\times3} & \mathbf{0}_{3\times3}\\
    \mathbf{0}_{3\times3} & \boldsymbol{\Lambda}_{3\times3} & \mathbf{0}_{3\times3}\\
    \mathbf{0}_{3\times3} & \mathbf{0}_{3\times3} & \mathbf{0}_{3\times3}
    \end{pmatrix} \mathcal{U}^{u,d,e,\nu}_{L,R},\\
    &\boldsymbol{\Lambda} \equiv \begin{pmatrix}
    1 & 0 & 0\\
    0 & 1 & 3 A \lambda^2(\rho-i \eta -1)\\
    0 & 3 A \lambda^2(\rho+i \eta -1) & -2
    \end{pmatrix}, \nonumber
\end{align}
where we remind the reader that $A$, $\lambda$, $\rho$, and $\eta$ are the Wolfenstein parameters. Notably, $Z_F$ couples to SM fields and their heavy partners equivalently, and as such doesn't facilitate any couplings between them. However, we see that this gauge boson can produce flavor-changing neutral currents in the SM quark sector. Focusing on this possibility, we consider what the coupling matrices for the SM quarks look like in our setup, arriving at
\begin{align}\label{ZFSMCoupling}
    (\mathbf{C}^u_{Z_F})_{L,R}^{SM} = \boldsymbol{\Lambda}, \; \; \; (\mathbf{C}^d_{Z_F})_{L,R}^{SM} = \mathbf{W}_d^\dagger \boldsymbol{\Lambda} \mathbf{W}_d \approx \begin{pmatrix}
    1 & 0 & 0\\
    0 & 1 & 3 A \lambda^2 (\rho-i \eta)\\
    0 & 3 A \lambda^2(\rho + i \eta) & -2
    \end{pmatrix}, 
\end{align}
where to derive the expression for $(\mathbf{C}^d_{Z_F})_{L,R}^{SM}$, we have used the fact that $\mathbf{W}_d$ is simply given by the CKM matrix, and used the Wolfenstein paramterization of the CKM given in Eq.(\ref{CKMWolf}), keeping terms up to $O(\lambda^2)$. Note that in spite of its appearance (and explicit dependence on Wolfenstein parameters), the coupling of Eq.(\ref{ZFSMCoupling}) is \emph{not} simply an artifact of our choice of $SU(4)_F$ gauge: Effecting an $SU(4)_F$ transformation here to a frame which, for example, $\mathbf{W}_d$ is equal to the identity matrix and $\mathbf{W}_u$ is non-trivial should correspondingly alter the gauge boson mass matrix so that the resulting coupling is preserved. We do, however, note that the right-handed coupling expressions in Eq.(\ref{ZFSMCoupling}) ignore the sign flips (phase rotations) of various right-handed quark fields observed in Eqs.(\ref{UuRTrunc}) and (\ref{UdRTrunc}). Since they cancel in any phenomenological results we shall discuss, we have omitted them above for the sake of brevity.

\subsection{SM Gauge Bosons and the Higgs}\label{SMCouplingsSection}

Having discussed the couplings of the fermions to new gauge bosons in the theory, we now address the coupling matrices for usual SM fields, namely the $Z$ and $W$ gauge bosons and the light SM-like Higgs doublet embedded in the bidoublet $H$. We begin our discussion with the $Z$ boson. Writing the fermions as 9-component vectors in flavor space as outlined in Eq.(\ref{FermionFlavorBasis}), we can write the coupling matrix of the $Z$ as
\begin{align}\label{ZCouplingMat}
    &\frac{g}{c_w}(\mathbf{C}^{u,d,e,\nu}_{Z})_L = \frac{g}{c_w} \mathcal{U}^{u,d,e,\nu \dagger}_L \begin{pmatrix}
    (T_{3L}-Q s_w^2)\mathbf{1}_{3\times3} & \mathbf{0}_{3\times3} & \mathbf{0}_{3\times1} & \mathbf{0}_{3\times1} & \mathbf{0}_{3\times1} \\
    \mathbf{0}_{3\times3} & -Q s_w^2\mathbf{1}_{3\times3} & \mathbf{0}_{3\times1} & \mathbf{0}_{3\times1} & \mathbf{0}_{3\times1} \\
    \mathbf{0}_{1\times3} & \mathbf{0}_{1\times3} & (T_{3L}-Q s_w^2) & 0 & 0 \\
    \mathbf{0}_{1\times3} & \mathbf{0}_{1\times3} & 0 & -Q s_w^2 & 0 \\
    \mathbf{0}_{1\times3} & \mathbf{0}_{1\times3} & 0 & 0 & -Q s_w^2
    \end{pmatrix} \mathcal{U}^{u,d,e,\nu}_L, \\
    &\frac{g}{c_w}(\mathbf{C}^{u,d,e,\nu}_{Z})_R = \frac{g}{c_w} \mathcal{U}^{u,d \dagger}_R \begin{pmatrix}
    -Q s_w^2\mathbf{1}_{3\times3} & \mathbf{0}_{3\times3} & \mathbf{0}_{3\times1} & \mathbf{0}_{3\times1} & \mathbf{0}_{3\times1} \\
    \mathbf{0}_{3\times3} & -Q s_w^2\mathbf{1}_{3\times3} & \mathbf{0}_{3\times1} & \mathbf{0}_{3\times1} & \mathbf{0}_{3\times1} \\
    \mathbf{0}_{1\times3} & \mathbf{0}_{1\times3} & (T_{3L}-Q s_w^2) & 0 & 0 \\
    \mathbf{0}_{1\times3} & \mathbf{0}_{1\times3} & 0 & -Q s_w^2 & 0 \\
    \mathbf{0}_{1\times3} & \mathbf{0}_{1\times3} & 0 & 0 & -Q s_w^2
    \end{pmatrix} \mathcal{U}^{u,d,e,\nu}_R, \nonumber
\end{align}
up to $O(\epsilon)$ corrections due to kinetic mixing, which won't be phenomenologically significant here, since they'll only represent a uniform small correction to the couplings of the three portal matter states, arising from mixing with the dark photon $A_D$. Here, $T_{3L}$ refers to the left-handed isospin of the SM fermion species and $Q$ refers to its electric charge. We can then find the coupling matrices for the phenomenologically relevant mass eigenstates by simply truncating the above matrices to exclude the extremely heavy states and rotating by the approximate diagonalization matrices given in Eqs.(\ref{UuLTrunc}), (\ref{UdLTrunc}), (\ref{UuRTrunc}), and (\ref{UdRTrunc}). Up to numerically negligible $O(\vec{\gamma}_P/v_P)$ terms, we arrive at
\begin{align}\label{ZuCouplingTrunc}
    &(\mathbf{C}^u_Z)_L \approx \begin{pmatrix}
    \frac{1}{2}-\frac{2}{3}s_w^2 & 0 & 0 & 0 & 0 & 0\\
    0 & \frac{1}{2}-\frac{2}{3}s_w^2 & 0 & 0 & 0 & 0\\
    0 & 0 & \frac{1}{2}-\frac{2}{3}s_w^2 & \frac{r}{2}\frac{m_t}{M_t} & 0 & 0\\
    0 & 0 & \frac{r}{2} \frac{m_t}{M_t} & -\frac{2}{3} s_w^2 & 0 & 0\\
    0 & 0 & 0 & 0 & \frac{1}{2}-\frac{2}{3} s_w^2 & 0\\
    0 & 0 & 0 & 0 & 0 & -\frac{2}{3}s_w^2
    \end{pmatrix},\\
    &(\mathbf{C}^u_Z)_R \approx \begin{pmatrix}
    -\frac{2}{3}s_w^2 & 0 & 0 & 0 & 0 & 0\\
    0 & -\frac{2}{3} & 0 & 0 & 0 & 0\\
    0 & 0 & -\frac{2}{3} s_w^2 & 0 & 0 & 0\\
    0 & 0 & 0 & -\frac{2}{3} s_w^2 & 0 & 0\\
    0 & 0 & 0 & 0 & \frac{1}{2}-\frac{2}{3}s_w^2 & 0\\
    0 & 0 & 0 & 0 & 0 & -\frac{2}{3}s_w^2
    \end{pmatrix}, \nonumber
\end{align}
% \begin{align}
%     (\mathbf{C}^u_Z)_L \approx \begin{pmatrix}
%     \frac{1}{2}-\frac{2}{3}s_w^2 & 0 & 0 & 0 & 0 & 0\\
%     0 & \frac{1}{2}-\frac{2}{3}s_w^2 & 0 & 0 & 0 & 0\\
%     0 & 0 & \frac{1}{2}-\frac{2}{3}s_w^2 & \frac{r}{2}\frac{m_t}{M_t} & 0 & \frac{q r z_{P2}^3}{2(1-z_{P2}^2)}\frac{m_t}{M_t} \frac{(\vec{\gamma}_P)_3}{v_P}\\
%     0 & 0 & \frac{r}{2} \frac{m_t}{M_t} & -\frac{2}{3} s_w^2 & -\frac{r z_{P1}^2}{2(1-z_{P1}^2)}\frac{m_t}{M_t} \frac{(\vec{\gamma}_P)_3}{v_P} & 0\\
%     0 & 0 & 0 & -\frac{r z_{P1}^2}{2(1-z_{P1}^2)}\frac{m_t}{M_t} \frac{(\vec{\gamma}_P)_3^*}{v_P} & \frac{1}{2}-\frac{2}{3} s_w^2 & 0\\
%     0 & 0 & \frac{q r z_{P2}^3}{2(1-z_{P2}^2)}\frac{m_t}{M_t} \frac{(\vec{\gamma}_P)_3^*}{v_P} & 0 & 0 & -\frac{2}{3}s_w^2
%     \end{pmatrix},\\
%     (\mathbf{C}^u_Z)_R \approx \begin{pmatrix}
%     -\frac{2}{3}s_w^2 & 0 & 0 & 0 & 0 & 0\\
%     0 & -\frac{2}{3} & 0 & 0 & 0 & 0\\
%     0 & 0 & -\frac{2}{3} s_w^2 & 0 & x\frac{m_t}{M_t} \frac{(\vec{\gamma}_P)_3}{v_P} & 0\\
%     0 & 0 & 0 & -\frac{2}{3} s_w^2 & w \frac{m_t}{M_t} \frac{(\vec{\gamma}_P)_3}{v_P} & 0\\
%     0 & 0 & x \frac{m_t}{M_t} \frac{(\vec{\gamma}_3^*)}{v_P} & w\frac{m_t}{M_t} \frac{(\vec{\gamma}_P)_3^*}{v_P} & \frac{1}{2}-\frac{2}{3}s_w^2 & 0\\
%     0 & 0 & 0 & 0 & 0 & -\frac{2}{3}s_w^2
%     \end{pmatrix},\\
%     x \equiv sgn(y_{P1})z_{P1}\bigg(-\frac{2}{3}s_w^2 \frac{z_{P1} q^2 r^2}{(1+z_{P1})}+\frac{1}{2}\bigg), \; \; w \equiv sgn(y_{P1})\frac{r z_{P1}}{1-z_{P1}^2}\bigg(-\frac{1}{2}+\frac{2}{3}s_w^2 q^2 z_{P1} (z_{P1}-1) \bigg),
% \end{align}
where we remind the reader that the variable $r$ is defined in Eq.(\ref{UuLTrunc}). We see that at this level of approximation, the sole new coupling for the $Z$ boson (other than its diagonal couplings to the new fermions, which directly follow their $SU(2)_L \times U(1)_Y$ quantum numbers) is between the top quark and its vector-like partner, which can be quite large -- the suppression ratio $m_t/M_t$ for this coupling can be as high as $\sim 0.1$, for top partner masses near 1 TeV. We note that there do exist $O(m_t^2/M_t^2)$ corrections to the $Z \bar{t} t$ coupling, however, since these corrections would be at most on the order of a few percent, they are well within current constraints for modifications of the top-$Z$ coupling \cite{Rontsch:2014cca}. Of course, such small variations in the top quark couplings to the $Z$ may be probed by precision measurements made at future $e^+e^-$ colliders. 
In the down-like and charged lepton sectors, we have an analogous result to Eq.(\ref{ZuCouplingTrunc}), however, assuming that the $b$ and $\tau$ partners are too massive to influence the low-energy phenomenology of the theory, we find no significant departures from the SM behavior of the $Z$ and diagonal coupling matrices.\footnote{As an aside, we note that even in the event of a much lighter $b$ partner, brought about by a percent-level tuning of the Higgs Yukawa coupling to the down quarks, which might then have a mass comparable to the top partner mass $M_t \sim 1 \; \textrm{TeV}$, we would not observe a measurable effect in the tightly-constrained $Z\bar{b}b$ coupling, since this would still be suppressed by $O(m_b^2/M_b^2) \sim 10^{-6}$, assuming $M_b \sim M_t$, which is several orders of magnitude below present constraints \cite{Ciuchini:2014dea}}

Next, we address the $W$ boson couplings, restricting our attentions to the quark sector in order to avoid ambiguities arising in the neutrino sector in this model. We may write our coupling matrices here as
\begin{align}\label{WCouplingMat}
    \frac{g}{\sqrt{2}}(\mathbf{C}_W)_L = \frac{g}{\sqrt{2}} \mathcal{U}^{u \dagger}_L \begin{pmatrix}
    \mathbf{1}_{3\times3} & \mathbf{0}_{3\times3} & \mathbf{0}_{3\times1} & \mathbf{0}_{3\times1} & \mathbf{0}_{3\times1} \\
    \mathbf{0}_{3\times3} & \mathbf{0}_{3\times3} & \mathbf{0}_{3\times1} & \mathbf{0}_{3\times1} & \mathbf{0}_{3\times1}\\
    \mathbf{0}_{1\times3} & \mathbf{0}_{1\times3} & 1 & 0 & 0 \\
    \mathbf{0}_{1\times3} & \mathbf{0}_{1\times3} & 0 & 0 & 0 \\
    \mathbf{0}_{1\times3} & \mathbf{0}_{1\times3} & 0 & 0 & 0
    \end{pmatrix} \mathcal{U}^d_L, \\
    \frac{g}{\sqrt{2}} (\mathbf{C}_W)_R = \frac{g}{\sqrt{2}} \mathcal{U}^{u \dagger}_R \begin{pmatrix}
    \mathbf{0}_{3\times3} & \mathbf{0}_{3\times3} & \mathbf{0}_{3\times1} & \mathbf{0}_{3\times1} & \mathbf{0}_{3\times1} \\
    \mathbf{0}_{3\times3} & \mathbf{0}_{3\times3} & \mathbf{0}_{3\times1} & \mathbf{0}_{3\times1} & \mathbf{0}_{3\times1}\\
    \mathbf{0}_{1\times3} & \mathbf{0}_{1\times3} & 1 & 0 & 0 \\
    \mathbf{0}_{1\times3} & \mathbf{0}_{1\times3} & 0 & 0 & 0 \\
    \mathbf{0}_{1\times3} & \mathbf{0}_{1\times3} & 0 & 0 & 0
    \end{pmatrix} \mathcal{U}^d_R. \nonumber
\end{align}
Up to numerically insignificant $O(\vec{\gamma}_P/v_P)$ corrections, we can derive the coupling matrix for the left-handed phenomenologically relevant fermions using Eqs.(\ref{UuLTrunc}) and (\ref{UdLTrunc}) once again, yielding a coupling matrix (and hence the CKM matrix) of
\begin{align}\label{NewCKM}
    (\mathbf{C}_W)_L \approx \begin{pmatrix}
    (\mathbf{W}_d)_{11} & (\mathbf{W}_d)_{12} & (\mathbf{W}_d)_{13} & 0 & 0\\
    (\mathbf{W}_d)_{21} & (\mathbf{W}_d)_{22} & (\mathbf{W}_d)_{23} & 0 & 0\\
    (1-\frac{r^2}{2}\frac{m_t^2}{M_t^2}) (\mathbf{W}_d)_{31} & (1-\frac{r^2}{2}\frac{m_t^2}{M_t^2}) (\mathbf{W}_d)_{32} & (1-\frac{r^2}{2}\frac{m_t^2}{M_t^2}) (\mathbf{W}_d)_{33} & 0 & 0\\
    r\frac{m_t}{M_t} (\mathbf{W}_d)_{31} & r\frac{m_t}{M_t} (\mathbf{W}_d)_{32} & r\frac{m_t}{M_t} (\mathbf{W}_d)_{33} & 0 & 0\\
    0 & 0 & 0 & 1 & 0\\
    0 & 0 & 0 & 0 & 0
    \end{pmatrix},
\end{align}
where $r$ remains as defined in Eq.(\ref{UuLTrunc}). Note, at least to this order, that the first row of this coupling matrix remains unitary when restricted to the SM quarks. Here, we emphasize that in the truncated coupling matrix, only five down-like quarks remain physically relevant (the three SM quarks and two portal matter quarks), while six up-like quarks do (the three SM quarks, the top partner, and two portal matter quarks), as can be seen from the original definition of the truncated flavor-space vectors in Eq.(\ref{ShortVecs}). We also note that, unlike elsewhere in this work, we have retained the $O(m_t^2/M_t^2)$ terms in the coupling matrix here; we shall see that terms of this order represent the leading contribution of the top partner's loop-induced correction to neutral meson mixing, and as such, we must retain these terms for numerical consistency. The right-handed coupling matrix $(\mathbf{C}_W)_R$ is substantially less phenomenologically interesting-- since the only right-handed fermions which couple to the $W$ boson are portal matter fields, and they do so diagonally (at least in the limit, which holds to $O(10^{-6})$ as discussed in Section \ref{FermionSpectrumSection}, that mixing between the portal matter states is numerically negligible), the $W$ does not exhibit numerically significant couplings to the right-handed SM quarks, nor does it facilitate decays of any of the new fermions in the model.

We conclude our exploration of the couplings in our model by considering the SM Higgs field, or rather the combination of elements of the bidoublet scalar $H$ that corresponds to such a field. As noted in Section \ref{ScalarContentSection}, we can estimate the scalar eigenstate corresponding to the SM Higgs field as in Eq.(\ref{LightHiggsState}), which in turn allows us, with reference to the Yukawa action of Eq.(\ref{GUTYukawa}), to write the Higgs coupling matrices to the quarks as
\begin{align}\label{HCouplingMat}
    y_{u,d} \mathbf{C}^{u,d}_H = y_{u,d} \mathcal{U}^{u,d \dagger}_R \begin{pmatrix}
    \mathbf{0}_{3\times3} & \mathbf{0}_{3\times3} & \mathbf{0}_{3\times1} & \mathbf{0}_{3 \times1} & \mathbf{0}_{3\times1}\\
    \mathbf{1}_{3\times3} & \mathbf{0}_{3\times3} & \mathbf{0}_{3\times1} & \mathbf{0}_{3 \times1} & \mathbf{0}_{3\times1}\\
    \mathbf{0}_{1\times3} & \mathbf{0}_{1\times3} & 0 & 0 & 0 \\
    \mathbf{0}_{1\times3} & \mathbf{0}_{1\times3} & 0 & 0 & 0 \\
    \mathbf{0}_{1\times3} & \mathbf{0}_{1\times3} & 1 & 0 & 0
    \end{pmatrix} \mathcal{U}^{u,d}_L,
\end{align}
where the constants $y_{u,d}$ are given in Eq.(\ref{FermionMassMats}). An analogous matrix for the charged leptons should be identical to that of the down-like quarks, up to radiative corrections to the parameter $y_d$. Rather than relying on our truncated rotation matrices here, which by removing the heavy partners of the up and charm quarks, omits the seesaw mechanism by which these SM quarks acquire mass, it is more instructive here to simply work in the limit where $\vec{\gamma}_{P,d} \rightarrow 0$ using the rotation matrices of Eqs.(\ref{UuL0}) and (\ref{UuR0}).  Doing this, and then truncating the matrix to remove the extremely heavy fermions as earlier yields the coupling matrix
\begin{align}
    y_{u} \mathbf{C}^{u}_H \approx \frac{\sqrt{2}}{v} \begin{pmatrix}
    m_u & 0 & 0 & 0 & 0 & 0\\
    0 & m_c & 0 & 0 & 0 & 0\\
    0 & 0 & m_t & m_t \frac{m_t}{M_t} \bigg( \frac{M_t^2-y_S^2 v_S^2}{y_S^2 v_S^2} \bigg)^{1/2} & 0 & 0\\
    0 & 0 & sgn(y_u u_3) m_t \bigg( \frac{M_t^2-y_S^2 v_S^2}{y_S^2 v_S^2} \bigg)^{1/2} & m_t \frac{m_t}{M_t} \bigg( \frac{M_t^2-y_S^2 v_S^2}{y_S^2 v_S^2} \bigg) & 0 & 0\\
    0 & 0 & 0 & 0 & 0 & 0\\
    0 & 0 & 0 & 0 & 0 & 0
    \end{pmatrix},
\end{align}
where the complete coupling is given by this matrix plus that generated by its Hermitian conjugate. Notably, among the SM quarks the Higgs coupling matrix is simply given by the normal SM Higgs coupling matrix -- this conclusion holds up to $O(m_t^2/M_t^2) \sim O(10^{-2})$ corrections to the $H\bar{t}t$ coupling. There are, however, additional couplings between the top partner and the top quark itself -- the largest of these terms is, in fact $O(1)$. In practice, these couplings will contribute significantly to the decay of the top partner to the top quark, as is often the case in models with additional vector-like quarks mixed with the third generation \cite{DeSimone:2012fs}. In the down-like quark and charged lepton sectors, the results are analogous, however, given the fact that the heavy $b$ and $\tau$ partners are likely too massive to be observed, the approximate Higgs couplings in these sectors precisely matches the SM result, up to insignificant numerical corrections.

\subsection{Kinetic Mixing}\label{KMSection}

Having addressed the fermion and gauge boson spectra here, it is useful at this point to comment on the magnitude and effects of the kinetic mixing between $U(1)_F$ and $U(1)_Y$ that will arise from the one-loop contributions of the model's fermion fields. For simplicity, we shall assume that kinetic mixing vanishes until the scale at which the Pati-Salam group is broken down to the SM and the dark/flavor group remains $SU(4)_F\times U(1)_F$. At this scale, the only Abelian groups which may enjoy kinetic mixing are the $U(1)_F$ and $U(1)_Y$ groups, so we shall compute this mixing here.\footnote{As was noted in Section \ref{MatterContentSection}, in the UV theory as written (with all SM gauge symmetries are contained in the non-Abelian Pati-Salam group factors), kinetic mixing mediated by a vacuum polarization-like diagram is forbidden. However, higher-order operators stemming from insertions of scalar vev's will still generate kinetic mixing here, and a truly concerned reader can assume that the $U(1)_F$ is unified with either the $SU(4)_c$ or $SU(2)_R$, the two Pati-Salam groups which contain the $U(1)_Y$ symmetry, at some higher scale, and some form of symmetry breaking at this scale breaks the resulting theory down to the SM gauge group by $U(1)_F$.} As in \textbf{II}, we note that both SM and portal matter fields will contribute to the kinetic mixing via vacuum polarization-like diagrams at one loop. In the original basis, the SM hypercharge boson $\hat{B}$ will mix with the $U(1)_F$ boson $\hat{B}_F$ via a term of the form,
\begin{align}
    \mathcal{L}_{KM} = \frac{\epsilon}{2 c_w c_P} \hat{B}_{\mu \nu} \hat{B}_F^{\mu \nu},
\end{align}
where $c_w$ is the familiar Weinberg angle and $c_P \equiv \cos \theta_P$ is the cosine of the angle $\theta_P$ described directly above Eq.(\ref{GMassMatZPTheta}), and $\hat{B}_{\mu \nu}$ and $\hat{B}_F^{\mu \nu}$ are the field strength tensors of the $U(1)_Y$ and $U(1)_F$ fields, respectively. Given this normalization convention, the kinetic mixing term $\epsilon$ here becomes
\begin{align}
    \epsilon = \sqrt{\frac{2}{3}}\frac{(g s_w)(g_4 s_P)}{24 \pi^2} \sum_i \bigg( \frac{Y_i}{2} \bigg) Q^{F}_i \log \frac{m_i^2}{\mu^2},
\end{align}
where $s_w$ and $s_P$ are simply the sines of the same angles referenced in $c_w$ and $c_P$, $g$ is the $SU(2)_L$ coupling constant, and $g_4$ is the $SU(4)_F$ coupling constant. The sum over $i$ is performed over all the fermions in the theory, $Y_i/2$ is the SM hypercharge of fermion $i$, $Q^F_i$ is its charge under $U(1)_F$, $m_i$ is its mass, and $\mu$ is an arbitrary mass scale which will cancel out of the final calculation. Ignoring the mixing between various states of different representations under $U(1)_Y$ or $U(1)_F$, the effects of which are numerically negligible anyway, gives
\begin{align}\label{epsilonCalc}
    \epsilon \approx (4.2 \times 10^{-4}) \bigg(\frac{g_4 s_P}{g s_w} \bigg) \bigg[& \log \frac{m_{P1}^e}{m_{P1}^u}+ \log \frac{m_{P1}^\nu}{m_{P1}^d}+2\log \frac{m_{P2}^d}{m_{P2}^u}+2\log \frac{m_{P2}^e}{m_{P2}^u}+\frac{4}{5} \log \frac{m_{P3}^d}{m_{P3}^u}+\frac{4}{5} \log \frac{m_{P3}^e}{m_{P3}^u}\\
    &+\sum_i \bigg(\frac{1}{5}\log \frac{m_{di}}{m_{ui}}+\frac{3}{5} \log \frac{m_{ei}}{m_{ui}}+\frac{1}{5}\log \frac{m_{\nu i}}{m_{ui}}+\frac{4}{5} \log \frac{M_{di}}{M_{ui}}+\frac{4}{5} \log \frac{M_{ei}}{M_{ui}} \bigg)\bigg], \nonumber 
\end{align}
where here we have used the same mass labelling convention as Eq.(\ref{calUDefs}), with the sub/superscripts $u$ referring to the up-like quark sector, $d$ to the down-like quark sector, $e$ to the charged lepton sector, and $\nu$ to the neutrino sector (of course, since the model's neutrino sector remains incomplete, the contributions from it to this mixing are somewhat suspect; we include them here for the sake of completeness). We note that some of logarithms of ratios of the exotic particles' masses may have either sign. In the absence of significant hierarchies, we would anticipate that $\epsilon$ would be of $O(10^{-4})$ if $g_4 s_P$ were approximately equal to $g s_W$, however, we note that there exist several mass ratios in Eq.(\ref{epsilonCalc}) that are necessarily quite hierarchical. For example, if we assume that each SM neutrino possesses a mass of approximately $0.1 \; \textrm{eV}$ and that the masses of the the portal matter fields $m_{P1,P2,P3}^{u,d,e,\nu}$ are close to degenerate (that is, $m_{Pi}^u \sim m_{Pi}^d \sim m_{Pi}^e \sim m_{Pi}^\nu$ for $i=1,2,3$, at least for the purposes of computing the logs of their ratios), then we can estimate the magnitude of Eq.(\ref{epsilonCalc}) by noting that $\log (m_{di,ei}/m_{ui}) \sim -\log (M_{di,ei}/M_{ui})$, which holds as long as there are no significant hierarchies between the couplings to the scalar $H$ among the charged leptons, up-like quarks, and down-like quarks. We then arrive at a rough numerical estimate of
\begin{align}
    \epsilon \sim (3 \times 10^{-3}) \bigg(\frac{g_4 s_P}{g s_w} \bigg),
\end{align}
which is in fact an order of magnitude larger than the numerical coefficient in front of Eq.(\ref{epsilonCalc}) might suggest. While this level of kinetic mixing is still not unfeasible if $g_4 s_P \simeq g s_w = e$, it does suggest that this coupling is unlikely to be much greater than this, and that to produce smaller $O(10^{-4})$ values for the kinetic mixing parameter, the coupling combination $g_4 s_P$ should likely be somewhat smaller, perhaps closer to $\sim 0.1 e$.

With $\epsilon$ computed, the treatment of kinetic mixing is ultimately entirely analogous to that of \textbf{I} and \textbf{II} (albeit with no additional mixing due to scalars charged under both the dark and SM gauge groups, which occur in \textbf{II} but not here), with dark photon couplings to SM fields of $\epsilon e Q$, where $e$ is the proton charge and $Q$ is the electromagnetic charge of a given fermion. As these results are well-known, we do not reproduce them here.

\subsection{Low-Energy Parameters From the High-Energy Model}\label{LowEnergyParamsSection}

With the field content, mass spectra, and coupling terms for the model now determined, we have only one remaining task before being able to explore this setup's phenomenological implications: Properly identifying the parameters with which we might conduct a probe of the model space. Over the course of our development of the model in Section \ref{ModelSetupSection}, we have noted that in spite of the large number of new particles present in the model at high energy, only a handful of these can be expected to have any significant effect at scales that can be experimentally probed now or in the near future. It stands to reason, then, that the large number of parameters in our model at high energy can in fact be reduced to a more manageable quantity at low energy. For the sake of clarity, we shall distinguish now between these two pictures: The ``high-energy'' model shall refer to the complete model with the field content outlined in Section \ref{ModelSetupSection}. The ``low-energy'' model shall refer to the model in which only the fields of mass $\lsim O(10 \; \textrm{TeV})$, namely those which might have an observable effect on current and upcoming experiments, are retained.

In Section \ref{FermionSpectrumSection}, we found that the quark sector of the model is uniquely specified by the SM Higgs Yukawa-coupling parameters $y_H \sim O(1)$ and $\alpha \in [-\pi, \pi]$, the two portal matter Yukawa couplings $y_{P1} \sim O(1)$ and $y_{P2} \sim O(1)$, the vector-like mass term $M$, the scalar vev parameters $\vec{\gamma}_P$, $\vec{\gamma}_d$, $v_P$, and $y_S v_S$\footnote{For the purposes of this analysis, we have combined the coupling $y_S$ with the vev parameter $v_S$ of the singlet scalar $\Phi_S$. Because the only instances of these parameters occurring separately happen when considering interactions of the physical scalar arising from $\Phi_S$, a detailed analysis of which we have omitted here, this simplification is sufficient for our purposes.}, and finally the signs of the eigenvalues of the matrices $\mathbf{u}$ (defined in Eq.(\ref{ULdefs})) and $\mathbf{d}$ (its analogous quantity in the down-like sector). In the limit where particles which we estimate to be too heavy to be experimentally observable decouple from the theory, our work simplifies substantially. Referencing our expressions for the masses and eigenvectors of the fermion fields in Section \ref{FermionSpectrumSection}, we see that up to signs of various quantities (which we shall see do not affect any physical results up to numerically small corrections), we note that we can completely specify the physics of the accessible quark sector, namely the masses in Eqs.(\ref{uLSMMat}) and (\ref{uLPortalMat}), their analogous values in the down-like sector, and the mass eigenvectors given in Eqs.(\ref{UuLTrunc}), (\ref{UdLTrunc}), (\ref{UuRTrunc}), and (\ref{UdRTrunc}), simply with the vev parameters $v_P$, $y_S v_S$, and $\vec{\gamma}_P$, the mass of the top quark partner $M_t$, and the masses of the accessible portal matter fields, $m^u_{P1} = m^d_{P1}$ and $m^u_{P2} = m^d_{P2}$. 

Moving on to the gauge boson sector, we see that can enjoy a similar drastic reduction in independent parameters. Consulting Section \ref{FlavorGaugeCouplingsSection}, we see that the coupling matrices of $Z_F$, $Z_P$, and $A_D$, the three new gauge bosons that are light enough to potentially have experimentally observable effects, depend only on the CKM matrix, the mass eigenvectors of the fermions, the $SU(4)_F$ coupling constant $g_4$, and the angle $\theta_P$ which functions as a Weinberg-like angle for the group $SU(4)_F \times U(1)_F$. Turning to Section \ref{GaugeSpectrumSection}, we note that while the mass of $Z_P$, given in Eq.(\ref{PortalZMass}), is entirely specified by the parameters $g_4$, $v_P$, and $\theta_P$, the masses of the gauge bosons $Z_F$ and $A_D$ arise as complicated functions of parameters which aren't otherwise relevant in the low-energy theory, namely the parameters $u_{1,2}$ (defined in Eq.(\ref{ULdefs})) and $d_{1,2,3}$ (their counterparts in the down-like quark sector), the Yukawa couplings $y_A$ and $y_B$, and, in the case of $A_D$, the $U(1)_D$-breaking vev components $\vec{\gamma}_d$. Because the mass of $A_D$ depends on one set of parameters that the mass of $Z_F$ doesn't, meanwhile, there is no obvious rigid relationship between these masses. It is therefore easier to simply specify the masses of $Z_F$ and $A_D$ as independent low-energy parameters in their own right, for the purposes of probing the phenomenology of the model.

Finally, we note that the above treatment has neglected to include any discussion of the leptons in the model. While we have noted earlier that the neutrino sector will lie largely unaddressed in this work, we cannot afford the same luxury with the charged leptons, so the fact that we are only introducing phenomenological parameters which cover the emergence of new physics in the quark sector bears some discussion. In the charged lepton sector, we note that we would require specifying some additional parameters, since we've posited that the charged lepton spectrum will be generated by the same mechanism as that of the down-like quarks, up to radiative corrections in the couplings to the scalars $\Phi_A$, $\Phi_B$, the parameters $y_H$ and $\alpha$, and the mass term $M$. Among the accessible elements of the theory, these radiative corrections would result in a modification of the rotation matrix $\mathbf{W}_l$ (defined analogously to $\mathbf{W}_u$ is defined in the up-like sector in Eq.(\ref{ULdefs})), rather than it being precisely equal to $\mathbf{W}_d$, its counterpart in the down-like quark sector which must be approximately given by the CKM matrix. Referencing the couplings of various interactions in Section \ref{FlavorGaugeCouplingsSection}, we see that the matrix $\mathbf{W}_l$ plays a role in the lepton-flavor-changing currents facilitated by the dark photon $A_D$, and the couplings of the accessible leptonic portal matter states to SM leptons (which in turn will govern the lifetimes and branching fractions of these new particles). However, we note that constraints on lepton-flavor-violating two-body decays of the $\tau$ lepton are much less restrictive than those in the quark sector \cite{Zyla:2020zbs}, while the most restricted leptonic flavor-changing processes mediated by $A_D$, $\mu^- \rightarrow e^- A_D$, is essentially always kinematically disallowed in the range of dark photon masses that we consider ($m_{A_D} >100 \; \textrm{MeV}$).\footnote{It should be noted that for lighter dark photon masses, the two-body decay of a muon to a dark photon and an electron would have a distinctive experimental signal, with a sharp peak in the electron energy spectrum. Limits on flavor violating decays in the quark sector remain more stringent than these constraints for now, but null results in searches for $\mu^+ \rightarrow e^+ X$ decays, where $X$ is some undetected boson, from the upcoming Mu3e experiment can be expected to constrain this ratio to $\sim 10^{-8}$, which may allow it to begin to compete with current limits on the model from flavor-violating $K$ meson decays \cite{Perrevoort:2018ttp}.} Meanwhile, discovery limits on the color-singlet leptonic portal matter will likely be far less constraining than those of the colored quark portal matter particles. As a result, we note that the new physics arising from additional parameters we must include to describe the charged lepton sector is likely beneath any notice, and we therefore have no need to expand our parameter space beyond what is necessary to specify the kinematically accessible fermions in the quark sector.

Our full parameterization of the lower energy observable sector of the model then simply consists of two vev parameter $v_P$ and $y_S v_S$, a three-component complex vector of vev's $\vec{\gamma}_P$, two gauge coupling parameters $g_4$ and $\theta_P$, and five particle masses: the mass $M_t$ of the heavy top partner, the masses $m^u_{P1} = m^d_{P1}$ and $m^u_{P2} = m^d_{P2}$ of the accessible portal matter fields, the mass $m_{Z_F}$ of the ``flavor $Z$'' boson $Z_F$, and the mass $m_{A_D}$ of the dark photon. For the reader's convenience, we have listed the parameters that are relevant in the high- and low-energy theories, as well as their approximate ranges, in several tables. Table \ref{table:LowEnergyParams} contains the parameters which are given in the underlying high-energy theory, and retained unchanged as independent parameters in our probe of the sector of the theory which is experimentally accessible, as well as the ranges that we have assumed here.

Table \ref{table:HighEnergyParams} contains the parameters which must be specified to generate the complete high-energy theory, but can be substituted for other parameters in the lower-energy theory. Finally, in Table \ref{table:LowEnergyMasses}, we present the parameters which may be used in lieu of those of Table \ref{table:HighEnergyParams} when probing the model parameter space where only phenomenologically accessible new particles are included. Notably, while there are simple expressions for $m_{P1}^{u,d}$ and $m_{P2}^{u,d}$ in terms of the parameters of Table \ref{table:HighEnergyParams} in Eq.(\ref{uLPortalMat}), allowing for an easy estimate of the range these parameters might take on in our model, the natural ranges for the other three parameters in Table \ref{table:LowEnergyMasses}, namely the masses $M_t$, $m_{A_D}$, and $m_{Z_F}$, are not immediately obvious: There are no such compact expressions for these masses, at least in terms readily corresponding to those in Table \ref{table:HighEnergyParams}. Since $m_{A_D}$ is the only parameter which directly depends on the magnitude of $\vec{\gamma}_P$, rather than the ratio $\vec{\gamma}_P/v_P$, we find that by specifying $v_P$, $\vec{\gamma}_P$, and $\vec{\gamma}_d$ properly we can reproduce virtually any $m_{A_D}$ between 0.1 and 1 GeV, so it is not unreasonable to treat this as a free parameter in this range, but the masses $M_t$ and $m_{Z_F}$ are more restricted. We can, however, produce estimates for the ranges of these parameters via a simple numerical probe of the high-energy parameter space. 

Before beginning this exercise, however, we may simplify our task with a handful of observations. For $m_{Z_F}$, we see from Eqs.(\ref{GMassMat}), (\ref{PhiMats}), and (\ref{PhiIdentities}) that the mass matrix which produces $Z_F$ (namely, the first 8 rows and columns of the matrix given in Eq.(\ref{GMassMat}), up to tiny $O(\vec{\gamma}_{P,d})$ corrections) should only have terms which depend on the matrices $\mathbf{u}$ (as defined in Eq.(\ref{ULdefs})) and $\mathbf{d}$, its down-like quark counterpart -- even dependence on the parameter $M$ cancels out in this portion of the gauge boson mass squared matrix. Furthermore, inspection of Eq.(\ref{uLSMMat}), in particular the SM fermion mass expression which may be rewritten (in the up-like sector, with a corresponding expression applying in the down-like sector) as $u_i^2 \approx y_S^2 v_S^2 (y_u^2 v^2/(2 m_{ui}^2)-1)$, that the eigenvalues $u_{1,2,3}$ and $d_{1,2,3}$ of $\mathbf{u}$ and $\mathbf{d}$ are all directly proportional to $y_S v_S$. That is, for all other model parameters held constant, the matrices $\mathbf{u}$ and $\mathbf{d}$ scale directly with $y_S v_S$. This in turn allows us to say that the mass matrix which governs $m_{Z_F}$, which in turn consists solely of combinations of $\mathbf{u}$ and $\mathbf{d}$, must be directly proportional to $y_S v_S$, and therefore the mass $m_{Z_F}$ is directly proportional to $y_S v_S$ as well. Meanwhile, a similar argument can be made regarding the mass of the top partner $M_t$, noting in Eq.(\ref{uLSMMat}) that we may rewrite the equation for the mass of this fermion as $M_t^2 = y_S^2 v_S^2 y_u^2 v^2/(2 m_t^2)$, and hence this mass is also directly proportional to $y_S v_S$. Therefore, when probing the masses $m_{Z_F}$ and $m_{A_D}$, we only need to consider the range of these parameters at one specified value for $y_S v_S$ and extrapolate from there, and not probe the entire range we list in Table \ref{table:LowEnergyParams}. Similarly, we need not probe the entire range of the $SU(4)_F \times U(1)_F$ gauge coupling parameters $g_4$ and $\theta_P$, rather noting that $m_{Z_F}$, consisting up to tiny corrections entirely of $SU(4)_F$ bosons, is proportional to $g_4$.
% , while Eq.(\ref{DarkPhotonMass}) tells us that $m_{A_D}$ is proportional to $g_4 s_P$. Meanwhile, we can see from the expression for the dark photon mass $m_{A_D}$ of Eq.(\ref{DarkPhotonMass}) that this mass has a single term which depends on a parameter which is still included in our low-energy probe, $\vec{\gamma}_P$, and it is provable (and, indeed, physically necessary) that the remaining terms in the expression must still be positive. As a result, we can claim a minimum value for $m_{A_D}$ derived from the value of $\vec{\gamma}_P^* \cdot \vec{\gamma}_P$, specifically,
% \begin{align}
%     m_{A_D}^2 \geq (m_{A_D})^2_{\textrm{min}} = \frac{7 g_4^2 s^2_P}{12}(\vec{\gamma}_P^* \cdot \vec{\gamma}_P),
% \end{align}
% and perform a numerical probe of the parameter space to determine the natural range of $m_{A_D}-(m_{A_D})_\textrm{min}$.

\begin{figure}[htbp]
\centerline{\includegraphics[width=5.0in,angle=0]{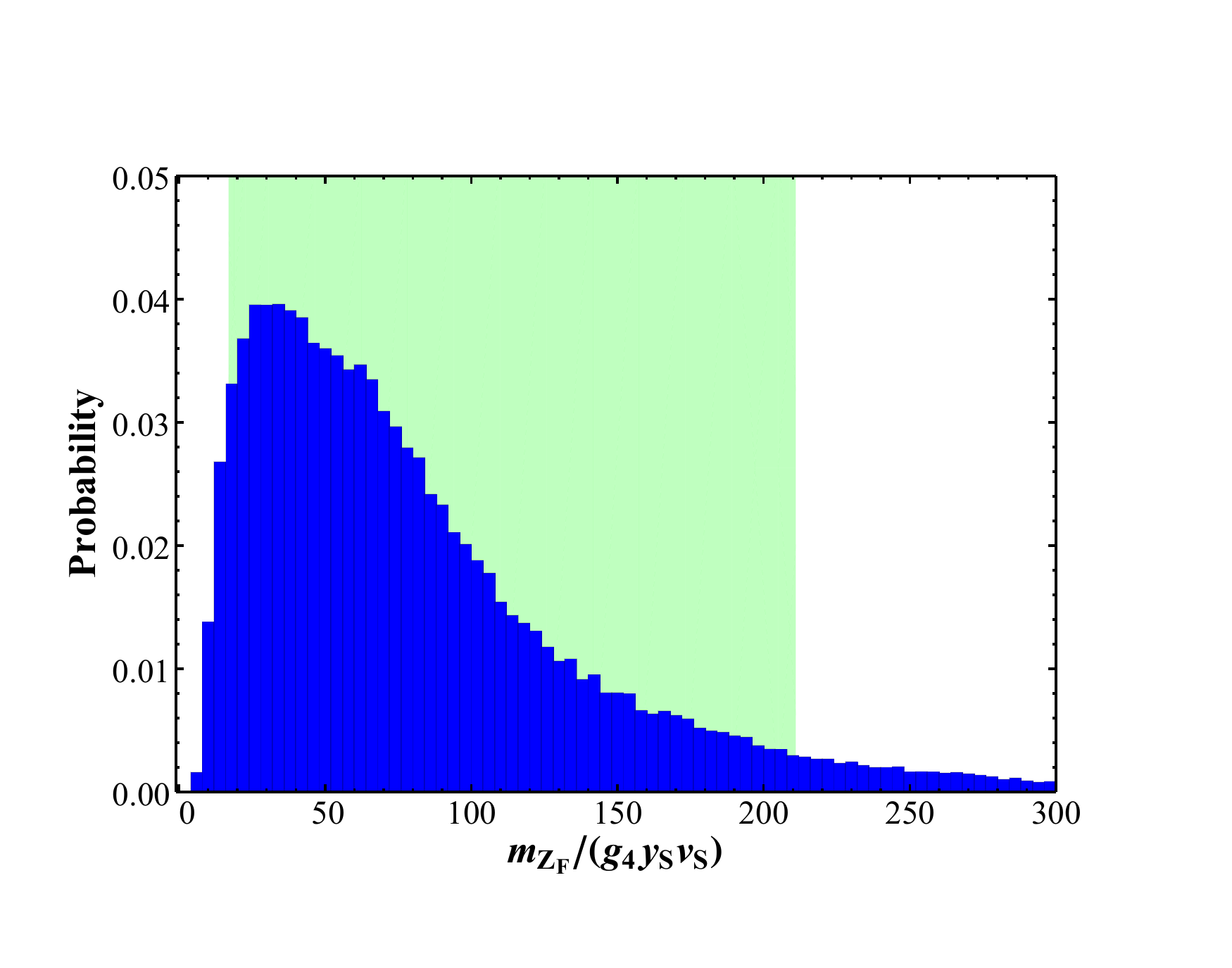}}
\vspace*{-2.0cm}
\centerline{\includegraphics[width=5.0in,angle=0]{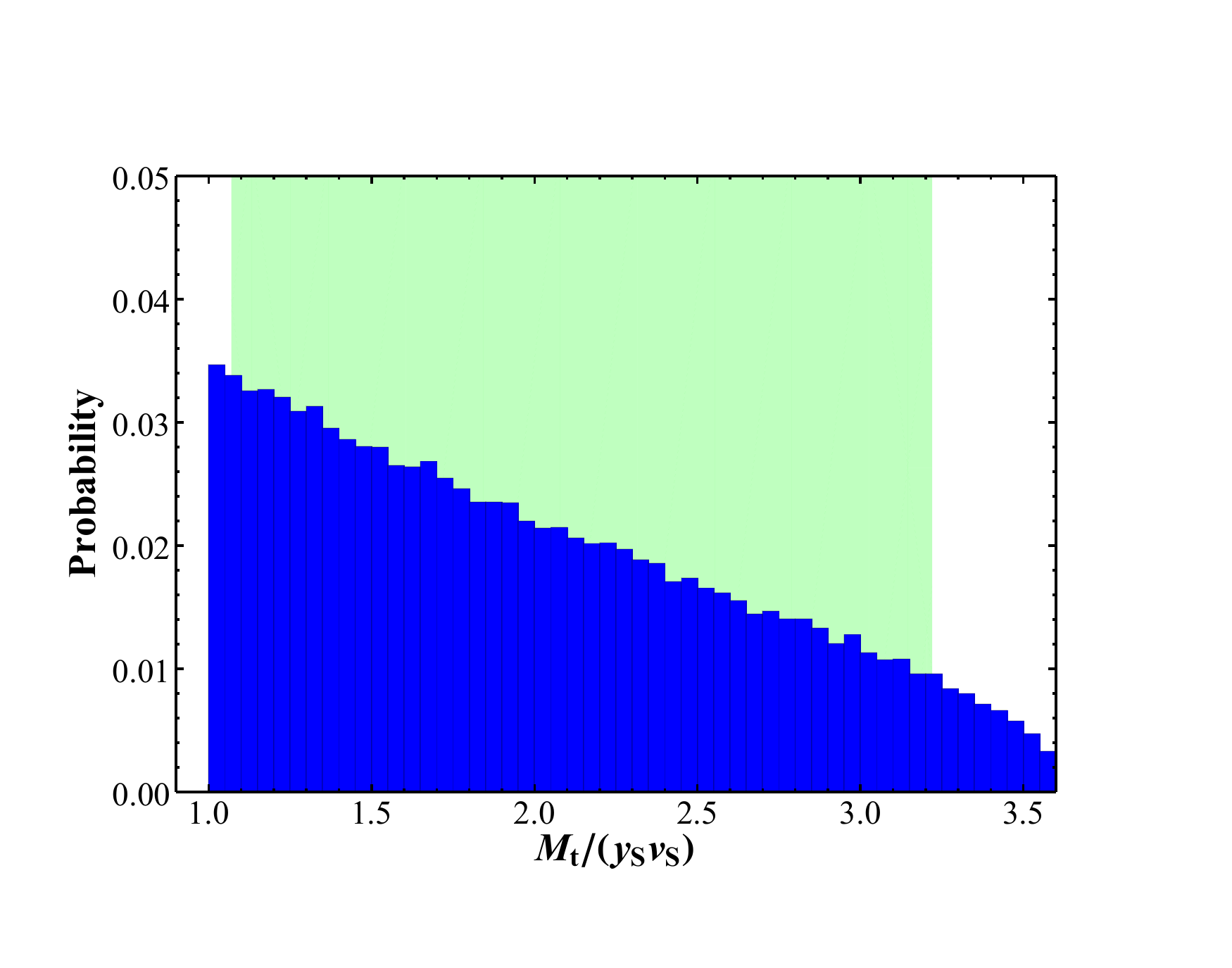}}
\vspace*{-1.0cm}
\caption{(Top) A probability histogram of the values of $m_{Z_F}$ in units of $g_4 y_S v_S$ obtained in our numerical sample, described in the text. The green region denotes the values of $m_{Z_F}$ between the 5\% and 95\% quantiles.
(Bottom) Same as above, but the probability histogram now depicts values of $M_t/(y_S v_S)$.}
\label{figMtmZF}
\end{figure}

To perform our numerical examination, we next generate a sample of $10^5$ points  in parameter space (we note that larger samples grant us the same results for our parameter ranges, indicating that this sample size is likely large enough to fully probe this setup), specifying $y_S v_S = 1 \; \textrm{TeV}$, and selecting $y_H$, $\alpha$, $\beta$, $M$, $y_A$, and $y_B$ randomly (\ie, with flat priors) in the ranges described in Table \ref{table:HighEnergyParams}. We also impose several additional conditions on $y_H$, $\alpha$, and $\beta$, the three parameters which govern the SM-like Higgs bidoublet sector. Specifically, we require that $y_u$ and $y_d$, as defined in Eq.(\ref{FermionMassMats}), are such that $y_u^2>(2 m_t^2)/v^2$ and $y_d^2 > 1/100$. The condition on $y_u$ ensures that the SM quark spectrum can be reproduced using the relations of Eq.(\ref{uLSMMat}). The condition on $y_d$ ensures that this Yukawa coupling is not hierarchically small, which would in turn yield a heavy $b$ partner with a mass roughly comparable to that of the heavy top partner, a region of parameter space that we have already established lies beyond the scope of this paper. To achieve a full sample of $10^5$ points which meet the the aforementioned constraints on $y_u$ and $y_d$, we generate $y_H$, $\alpha$, and $\beta$, then re-generate these points (again with flat priors) if the original set fails them, until a point is found that meets the constraints; this in turn ensures that we uniformly sample the parameter space of $y_H$, $\alpha$, and $\beta$ that meet the $y_u$ and $y_d$ constraints without bias.
%The parameters $\vec{\gamma}_P$ and $\vec{\gamma}_d$ are then generated as vectors with a random magnitude between 0.1 and 1 GeV, with random orientation in their three-dimensional space and random phases between $-\pi$ and $\pi$ assigned to their components.
Finally, we note that outside of numerically negligible corrections, the particular values of the parameters $\vec{\gamma}_P$, $\vec{\gamma}_d$, $v_P$, $y_{P1}$, and $y_{P2}$ have no effect on the masses we wish to probe here, so there is no need to specify them. Given these conditions, we estimate the expected ``natural'' ranges of $M_t$ and $m_{Z_F}$, as between the $5\%$ and $95 \%$ quantile of their values in this scan -- because we engage in a sampling of all of the parameters in our high energy theory with flat priors, we expect the results of our numerical probe to at least qualitatively reflect the necessary fine-tuning to effect certain values of these parameters. The results are then given in the ranges quoted in Table \ref{table:LowEnergyMasses}. We should note that although our sampling of the high-energy parameters is uniform over their ranges, the same cannot be said for the ranges of the parameters $M_t$ and $m_{Z_F}$. To get a sense of the shapes of the $M_t$ and $m_{Z_F}$ distributions, we depict probability histograms of them in Figure \ref{figMtmZF}. Consulting this Figure, we see that while $M_t$ has a straightforward (if nonuniform) distribution in which smaller values of $M_t/(y_S v_S)$ are slightly favored over larger ones, the distribution of $m_{Z_F}$ is somewhat more complex, with a large peak appearing near $m_{Z_F} \sim 30 g_4 y_S v_S$ followed by a steady drop-off. As a result of this long tail in the distribution, the range between the 5\% and 95\% quantiles of our sample given in Table \ref{table:LowEnergyMasses} is unusually large. As we shall see in 
Section \ref{ZFMesonSection}, however, the phenomenologically interesting region of parameter space lies in the region where $m_{Z_F} \lsim 60 g_4 y_S v_S$, so we shall restrict our attentions there, merely noting that ample parameter space exists for significantly larger $m_{Z_F}$ without significant fine-tuning.

\vspace{2.0cm}

\begin{table}[h!]
\centering
\begin{tabular}{|c | c | c | c | c|} 
 \hline
 $y_S v_S$ & $v_P$ & $\vec{\gamma}_P$ & $g_4$ & $\theta_P$\\
 \hline
 1-10 TeV & 1-10 TeV & 0.1-1 GeV & (0.1-0.6) & 0-$\frac{\pi}{2}$\\
 \hline
\end{tabular}
\caption{The parameters which must be set to specify a unique point in parameter space in both the complete high-energy model and the low-energy model, as described in the text. Here, $g$ denotes the electroweak coupling constant. The philosophy behind the chosen ranges of $y_S v_S$, $v_P$, and $\vec{\gamma}_P$ are discussed in Section \ref{FermionSpectrumSection}, while we assume that $g_4 \lesssim g$ to keep the magnitude of the kinetic mixing $\lesssim O(10^{-3})$, as discussed in Section \ref{KMSection}}
\label{table:LowEnergyParams}
\end{table}

\vspace{1.0cm}

\begin{table}[h!]
\centering
\begin{tabular}{|c | c | c | c | c | c | c | c | c|} 
 \hline
 $y_H$ & $\alpha$ & $\beta$ & $y_A$ & $y_B$ & $y_{P1}$ & $y_{P2}$ & $M/(y_S v_S)$ & $\vec{\gamma}_d$\\
 \hline
 $\pm [1/3,3]$ & $[-\pi,\pi]$ & $[0, \arctan(0.8)]$ & $\pm [1/3,3]$ & $\pm [1/3,3]$ & $\pm [1/3,3]$ & $\pm [1/3,3]$ & $[-10^6, 10^6]$ & 0.1-1 GeV \\
 \hline
\end{tabular}
\caption{The parameters which must be set to specify a unique point in parameter space in the complete high-energy model, but which can be eliminated in favor of a simpler set of parameters in the low-energy model. The ranges quoted here assume that the Yukawa couplings are all of $O(1)$ and $\beta$ is restricted based on the requirement for perturbativity in the left-right-symmetric model Higgs sector. The magnitudes of $M$ and $\vec{\gamma}_d$ are discussed in Section \ref{FermionSpectrumSection}.}
\label{table:HighEnergyParams}
\end{table}

\vspace{1.0cm}

\begin{table}[h!]
\centering
\begin{tabular}{|c | c | c | c | c |} 
 \hline
 $M_t$ & $m_{P1}^u=m_{P1}^d$ & $m_{P2}^u=m_{P2}^d$ & $m_{A_D}$ & $m_{Z_F}$\\
 \hline
 $(1.1-3.2) y_S v_S$ & $(1/3-3) v_P$ & $(1/3-3) v_P$ & 0.1-1 GeV & $g_4 y_S v_S (17-210)$\\
 \hline
\end{tabular}
\caption{The parameters which which may be used to specify a unique point in the parameter space of the low-energy model in lieu of the high-energy model parameters in Table \ref{table:HighEnergyParams}. The large range of $m_{Z_F}$, and the shape of the $m_{Z_F}$ distribution in the numerical probe, is discussed in the text and Figure \ref{figMtmZF}.}
\label{table:LowEnergyMasses}
\end{table}

To better illustrate the transition from the enormously complicated high-energy model to the comparatively compact low-energy model, and to get an idea of the sensitivity (or lack thereof) of low-energy parameters to the specific realization of the complicated pattern of $SU(3)_F$ breaking in the high-energy model, it is useful to explicitly summarize the scales of symmetry breaking effected in the high-energy model and compare it to the scale of the new physics fields which survive in the low-energy model. Specifically, we note that in the high-energy setup, the gauge symmetry $SU(4)_F \times U(1)_F$ undergoes the symmetry breaking pattern
\begin{align}
    \mathbf{4}_F \mathbf{1}_F \xrightarrow{\gg \textrm{TeV}} \mathbf{1}_F' \mathbf{1}_F \xrightarrow{ O(\textrm{TeV})} \mathbf{1}_D \xrightarrow{O(\textrm{GeV})} \textrm{Nothing}.
\end{align}

In particular, we see that the $SU(3)_F$ symmetry breaking, which required prodigious fine-tuning in Section \ref{ModelSetupSection}, generally takes place at the multi-TeV scale. In fact, the \emph{only} fields in the low-energy model which remain sensitive to the $SU(3)_F$ breaking are the ``flavor $Z$'' field $Z_F$ and the vector-like top quark partner. Given that the emergence of a vector-like top partner of mass $M_t \sim O(\textrm{TeV})$ is well-motivated as long as a seesaw mechanism is used to generate the quark mass hierarchy, we can comfortably assert that, with the exception of any observable effects of the $Z_F$ boson, \emph{the phenomenology of the low-energy model is applicable to a wide range of high-energy constructions and agnostic to the complicated specifics of $SU(3)_F$ breaking}.

\section{Phenomenology: Flavor-Changing Neutral Currents}\label{FCNCSection}

Having set up the model and determined the relevant parameters we can begin to analyze the sector of the model within experimental reach, the sole remaining task left to us is to actually do the necessary exploration. To begin, we note that our significant expansion of the model's flavor sector has resulted in substantial additional sources of flavor-changing couplings, in particular flavor-changing neutral currents (FCNC's), which have the potential to contribute to highly constrained processes. To examine these flavor-changing effects, then, we explore the impact of the model's FCNC's in two sectors: rare meson decays and neutral meson oscillation. 
%we can  now move to a probe of the model's parameter space to determine what manner of constraints and signals it may possess. While we have established in Section \ref{ModelSetupSection} that our model has a significant number of finely-tuned parameters in order to recreate the SM flavor structure, we also have noted that a large quantity of the new physics the model introduces (and which depend on these parameters) are entirely too massive to give rise to experimentally observable signals in the foreseeable future. In particular, we have found that in the fermion sector, only the SM fermions, the heavy top partner, and the portal matter fields should be expected to have significant observable effects, while in the gauge boson and scalar sectors, we have limited our attentions to the SM gauge bosons, the SM Higgs, and three new gauge bosons arising from the $SU(4)_F \times U(1)_F$ symmetry: An $SU(3)_F$ boson $Z_F$ of mass $O(10^5 \; \textrm{GeV})$, a heavy ``portal $Z$'' $Z_P$ with a mass of $O(1 \; \textrm{TeV})$, and a dark photon $A_D$ of mass $O(0.1-1 \; \textrm{GeV})$. Our task is then simple: First we must determine what parameters of our original theory affect the physics at a low enough energy to be observed, and then we must determine what constraints on these parameters exist at present, and may appear in the foreseeable future.

\subsection{Flavor-Changing $A_{D}$ Interactions}\label{ADFlavorConstraintsSection}

We begin our discussion of potential phenomenological effects with a brief foray into the most constrained couplings which emerge here, namely the light flavor-changing neutral currents mediated by the dark photon. In particular, the tiny flavor-changing couplings among the SM quarks in Eq.(\ref{ADdCouplingTrunc}) allow for $b \rightarrow s + A_D$, $b \rightarrow d + A_D$, and $s \rightarrow d + A_D$ transitions. As in \textbf{II}, we shall assume that the on-shell long-lived light dark photon will escape our detector and/or decay to dark matter,\footnote{the dark photon decay to dark matter will be the dominant decay if it is kinematically accessible, since its coupling isn't suppressed by a small kinetic mixing factor, so this assumption is both completely reasonable and trivially realizable with the addition of a dark matter particle of mass less than $m_{A_D}/2$, for example in the form of a scalar singlet under the SM gauge group.} and therefore these interactions can facilitate meson decays which mimic to some extent the highly-constrained rare decay channels $B \rightarrow K \nu \overline{\nu}$, $B \rightarrow \pi \nu \overline{\nu}$, and $K \rightarrow \pi \nu \overline{\nu}$.

It should be noted that the naive bounds from the three-body $B \rightarrow K \nu \overline{\nu}$, $B \rightarrow \pi \nu \overline{\nu}$, and $K \rightarrow \pi \nu \overline{\nu}$ are not precisely applicable to the two-body decay of a $B$ and $K$ mesons to a lighter meson and a long-lived dark photon: The latter will result in a sharply peaked momentum distribution for the visible light meson in the final state. This difference has a significant effect on our analysis of $K \rightarrow \pi \, A_D$ constraints in particular: Searches in the ``golden channel'' $K^+ \rightarrow \pi^+ \nu \overline{\nu}$, which are normally highly constraining, are substantially weakened in cases where $A_D \sim m_\pi$ due to high backgrounds from the $K^+ \rightarrow \pi^+ \pi_0$ channel \cite{Artamonov:2009sz,Fuyuto:2014cya,Ahn:2018mvc}. Instead, it is more instructive to consider the constraints from $K_L \rightarrow \pi_0 X$ searches such as \cite{Ahn:2018mvc}, where $X$ is simply some light long-lived invisible particle. Because the background $K_L \rightarrow 2\pi_0$ is CP-violating (and therefore suppressed), unlike the decay $K^+ \rightarrow \pi^+ \pi_0$, searches for $K_L \rightarrow \pi_0 X$ are not subject to the same ruinous kinematic cuts when the mass of the $X$ particle is close to the mass of the pion. So, to derive for our model's constraints from $K_L \rightarrow \pi_0 X$ branching fractions, we compute the branching fraction for this decay arising from $A_D$ emission and compare it to the 90\% CL upper limits from \cite{Ahn:2018mvc} obtained for various mass values. In the case of the other flavor-changing decays we consider, a more careful analysis of $B^+ \rightarrow \pi^+,K^+ \nu \overline{\nu}$ searches, which lack the same troubling kinematic cuts as appear in $K^+ \rightarrow \pi^+ \nu \overline{\nu}$ searches, doesn't result in significant differences from the constraints derived by simply quoting the 90\% CL upper bounds of $\mathcal{B}(B^+ \rightarrow \pi^+ \nu \overline{\nu}) \leq 0.8 \times 10^{-5}$ \cite{Grygier:2017tzo} and $\mathcal{B}(B^+ \rightarrow K^+ \nu \overline{\nu}) \leq 1.6 \times 10^{-5}$ \cite{Lees:2013kla} as direct limits on the corresponding flavor-changing decays mediated by $A_D$ (although it should be noted that we should treat these constraints as order-of-magnitude, not precise, limits barring a more detailed analysis of the final-state light meson spectrum). Since we shall see that these constraints are much less rigorous than those which are derived from the $K_L \rightarrow \pi_0 \, A_D$ system, we also note that the changes in restrictions on the parameter space from $O(1)$ changes in these constraints are not qualitatively significant.

A straightforward calculation employing the coupling constants in Eq.(\ref{ADdCouplingTrunc}) gives us the decay width of the process
$K_L \rightarrow \pi_0 A_D$
\begin{align}\label{KDecay}
    &\Gamma_{K_L \rightarrow \pi_0 \, A_D} = g_4^2 s^2_P \frac{|\xi_1|^2|\xi_2|^2}{24 \pi}\frac{m_K^3}{m_{A_D}^2} \bigg(1+\frac{(m_{A_D}^2-m_\pi^2)^2}{m_K^4} -\frac{2(m_{A_D}^2+m_\pi^2)}{m_K^2}\bigg)^{3/2}|f_+^{K^0 \pi^0}|^2,\\
    &\xi_i \equiv \frac{(\mathbf{W}_d \vec{\gamma}_P)_i}{v_P} \nonumber
\end{align}
where $m_K$ and $m_\pi$ are the masses of the neutral $K$ and $\pi$ mesons, respectively, $f_+^{K^0 \pi^0}$ is a hadronic form factor we can extract from \cite{Leutwyler:1984} (up to percent level corrections, this factor is equal to 1 for the process $K_L \rightarrow \pi_0 A_D$), and we remind the reader that, also up to percent level corrections, $\mathbf{W}_d$ is equal to the CKM matrix. Expressions for $B^+ \rightarrow K^+ A_D$ and $B^+ \rightarrow \pi^+ A_D$ decays can be easily determined by replacing the appropriate indices of $\xi$ and meson masses in Eq.(\ref{KDecay}), and extracting the appropriate hadronic form factors from \cite{Ball:2004ye}. Consulting the definition of $\xi_i$ in Eq.(\ref{KDecay}), we see that these decay processes should naturally undergo substantial suppression, since $\vec{\gamma}_P \sim 0.1-1 \; \textrm{GeV}$ and $v_P \gsim1 \; \textrm{TeV}$. We might expect, therefore, that a natural value of $\xi_i$ would be in the realm or $10^{-4}$, leading to a $\sim 10^{-16}$ suppression of these decay processes, as was noted in \textbf{II}. In spite of this suppression, we shall find, as in \textbf{II}, that constraints on these decays are severe enough to provide meaningful limits on the values of the $\xi_i$'s beyond even their naturally small magnitudes.

\begin{figure}
    \centerline{\includegraphics[width=3.5in]{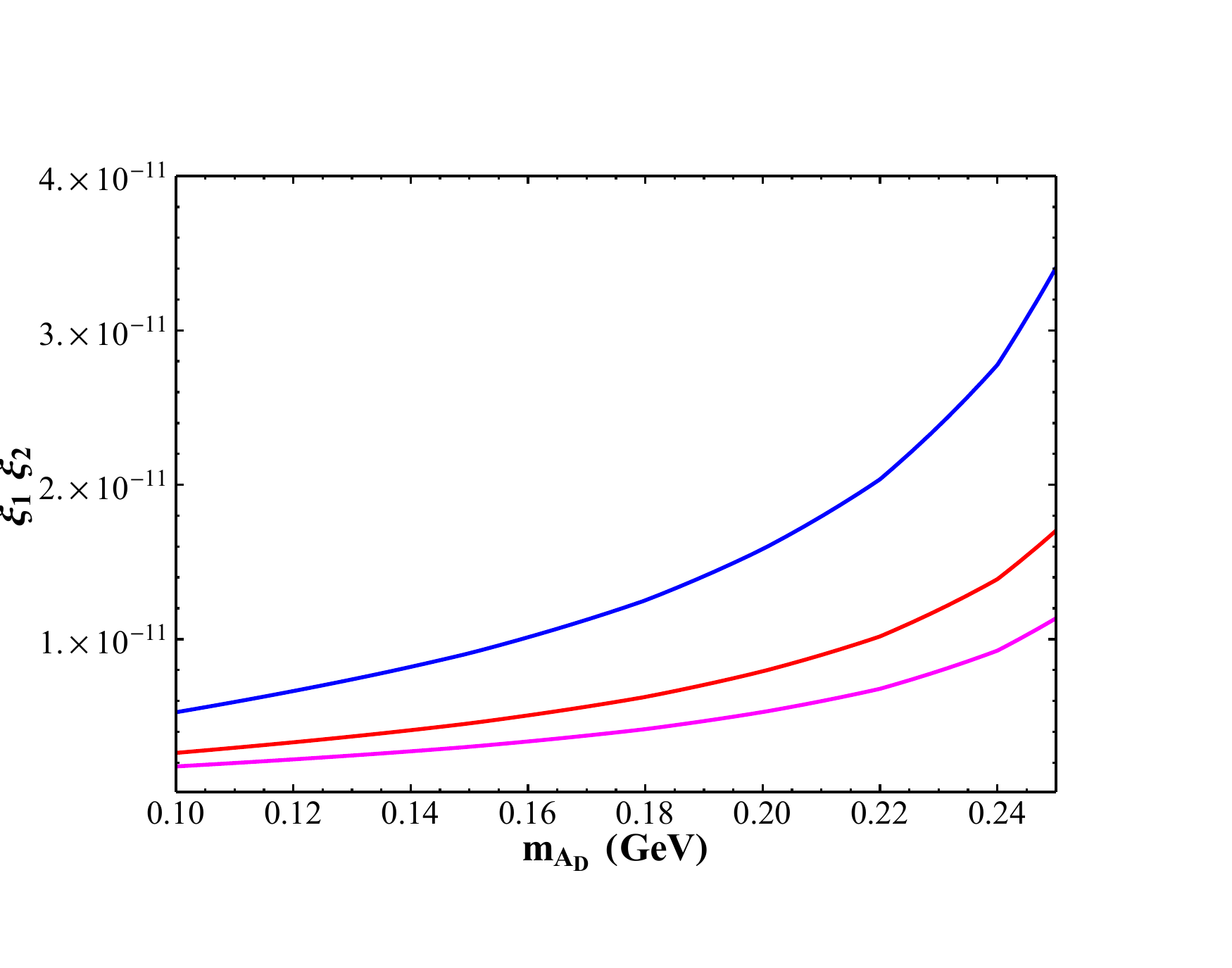}
    \hspace{-0.75cm}
    \includegraphics[width=3.5in]{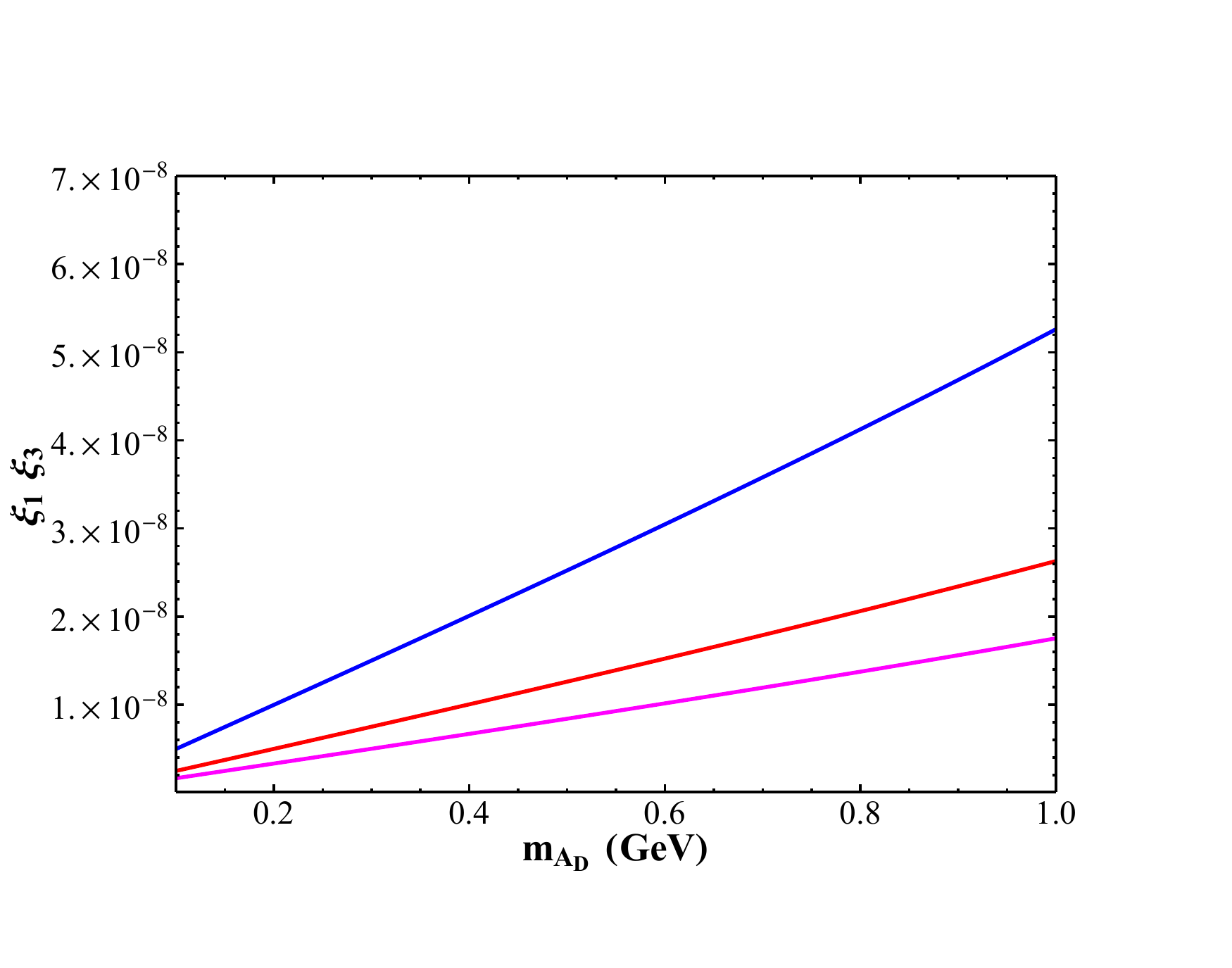}}
    \vspace*{-0.25cm}
    \centering
    \includegraphics[width=3.5in]{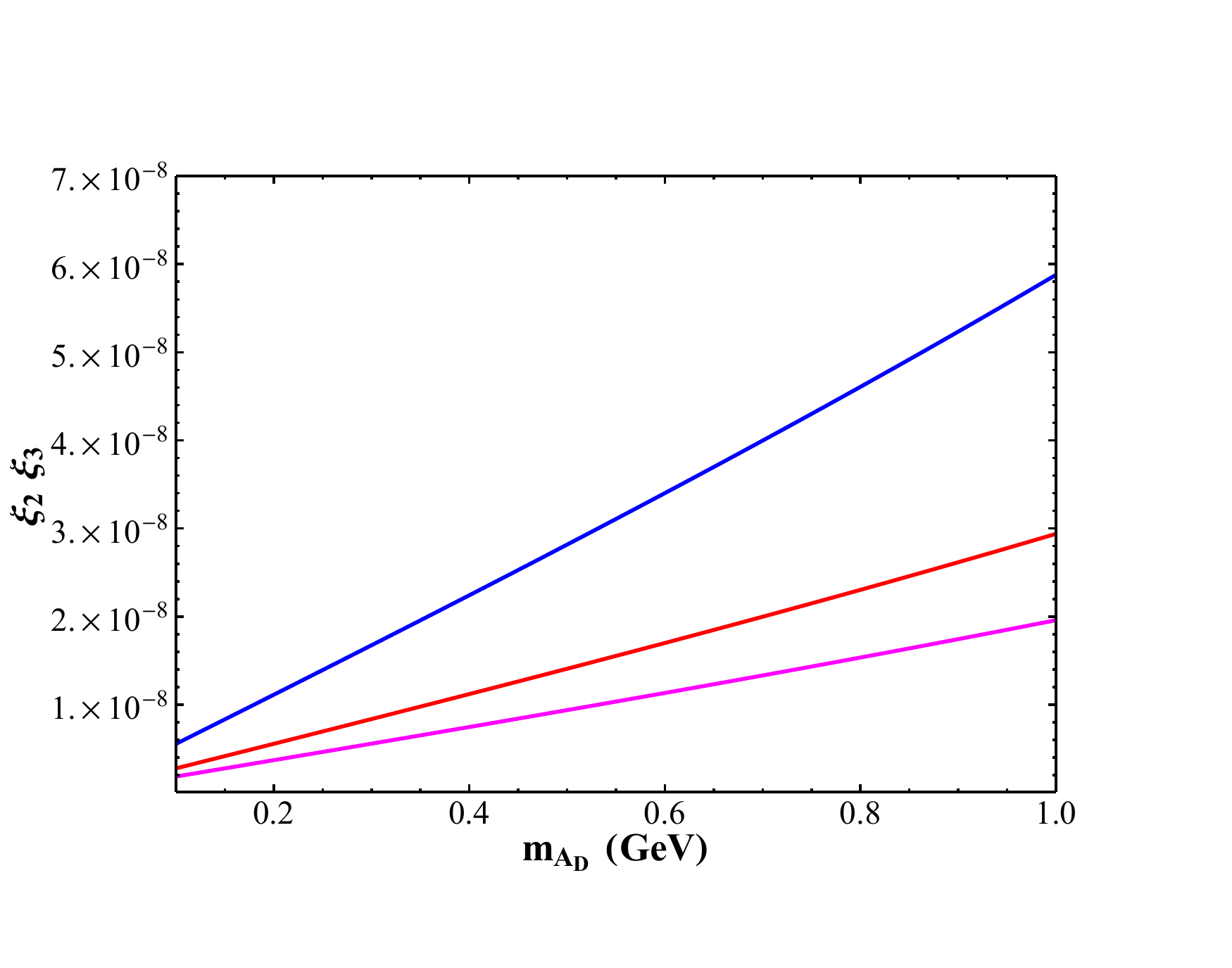}
    \caption{Maximum values of the products $\xi_1 \xi_2$ (top left), $\xi_1 \xi_2$ (top right), and $\xi_2 \xi_3$ (bottom) based on the 90\% CL measured limits on $\mathcal{B}(K_L \rightarrow \pi_0 X)$ (where $X$ is some invisible long-lived particle) \cite{Ahn:2018mvc}, $\mathcal{B}(B^+ \rightarrow \pi^+ \nu \overline{\nu})$ \cite{Grygier:2017tzo}, and $\mathcal{B}(B^+ \rightarrow K^+ \nu \overline{\nu})$ \cite{Lees:2013kla}, respectively. Benchmark values of $g_4 s_P$ are taken as 0.1 (blue), 0.2 (red), and 0.3 (magenta). The $\xi$ parameters are described in the text.}
    \label{fig1}
\end{figure}

We can straightforwardly derive expressions for the maximum allowed magnitudes for products of $\xi_i$ values, upon which the constrained meson decays depend, for various representative choices of the parameters $m_{A_D}$ and $g_4 s_P$ by comparing our results to upper experimental limits on the branching ratios of these mesons to the corresponding two-neutrino final states (or in the case of $K_L\rightarrow \pi_0 \, A_D$, the mass-dependent limits on the branching ratios $\mathcal{B}(K_L \rightarrow \pi_0 \, X)$). The maximum values of the relevant components of $(\mathbf{W}_d^\dagger \vec{\gamma}_P)/v_P$ as functions of $m_A$ at various values of $g_4 s_P$ are depicted in Figure \ref{fig1}.

From Figure \ref{fig1}, we can derive some approximate constraints on the parameters $\xi_i$, which suggest that some of their magnitudes must be well below the natural value of $\sim O(10^{-4})$ that we might expect. Up to $O(1)$ variations as depicted in this Figure, we arrive at the rough constraints
\begin{align}\label{XiConstraints}
    \xi_1 \xi_2 \lsim 10^{-11}, \;\;\; \xi_1 \xi_3 \lsim 10^{-8}, \;\;\; \xi_2 \xi_3 \lsim 10^{-8},
\end{align}
where the first (and most stringent) constraint, coming from $K_L \rightarrow \pi_0 A_D$, only applies when $m_{A_D} \lsim 260 \; \textrm{MeV}$, past which the study of \cite{Ahn:2018mvc} doesn't offer constraints due to overly large $K_L \rightarrow \pi^+ \pi^- \pi_0$ backgrounds. The much less severe constraints from the other decay processes have no such similar restriction on the dark photon mass for which they apply; these processes remain kinematically accessible over the entire range of $m_{A_D}$ we consider. Because the constraints in Eq.(\ref{XiConstraints}) only apply to the products of $\xi_i$ terms, there is a fair amount of flexibility in how these constraints might be satisfied. For example, we might meet the harsh constraint on the product of $\xi_1 \xi_2$ by allowing one of either $\xi_1$ or $\xi_2$ to be $\sim 10^{-7}$, several orders of magnitude below its expected magnitude of $\sim 10^{-4}$, while the other stays at the natural higher magnitude, or we can require that $\xi_1 \sim \xi_2 \sim 3 \times 10^{-6}$.
%We also note that regardless of our suppressions of various individual $\xi_i$'s, we do anticipate that at least one value of $\xi_i$ must be approximately at the natural value of $\sim 10^{-4}$, since otherwise, couplings that mediate the decay of the portal matter, such as those in Eq.(\ref{ADuCouplingTrunc}), would be highly suppressed, and likely depend on subleading effects such as those stemming from $\vec{\gamma}_d$ terms.
Given that the other constraints here are or comparatively little significance, we can outline several phenomenologically allowed benchmark values of the parameters $\xi_i$, listed in Table \ref{table:XiBenchmarks}. An astute reader may be concerned that, in suppressing various $\xi_i$ (and hence $\vec{\gamma}_P$) values to such low values, we may anticipate that contributions to fermion mixing from $\vec{\gamma}_d$, which we have neglected in our treatment of the low-energy phenomenology, may be significant. However, in practice we find that even assuming all components of $\xi_i$ are as small as $O(10^{-10})$, the $\vec{\gamma}_P$ effects still dominate all phenomenologically significant couplings in the model.

\begin{table}[h!]
\centering
\begin{tabular}{|c | c | c | c | c|} 
 \hline
 Benchmark & $\xi_1$ & $\xi_2$ & $\xi_3$\\
 \hline
 A1 & $\lsim 10^{-4}$ & 0 & 0\\
 \hline
 A2 & 0 & $\lsim 10^{-4}$ & 0\\
 \hline
 A3 & 0 & 0 & $\lsim 10^{-4}$\\
 \hline
 B1 & $\lsim 10^{-4}$ & 0 & $\lsim 10^{-4}$\\
 \hline
 B2 & 0 & $\lsim 10^{-4}$ & $\lsim 10^{-4}$\\
 \hline
 C1 & $\lsim 10^{-4}$ & $\lsim 10^{-4}$ & $\lsim 10^{-4}$\\
 \hline
 C2 & $\lsim 10^{-6}$ & $\lsim 10^{-6}$ & $\lsim 10^{-6}$\\
 \hline
\end{tabular}
\caption{The different benchmarks for $\xi_i$ magnitudes which can satisfy the rough constraints of Eq.(\ref{XiConstraints}). Scenarios A1, A2, and A3 correspond to cases where one $\xi_i$ is hierarchically larger than the others, with the number denoting which generation, first, second, or third, has the dominant $\xi_i$. Scenario B1(B2) assumes that either $\xi_1(\xi_2)$ is roughly equal in magnitude to $\xi_3$, while $\xi_2(\xi_1)$ is hierarchically smaller. Scenario C1, in which all $\xi_i$'s are approximately equal, is only viable if $m_{A_D} \gtrsim 250 \; \textrm{MeV}$, and hence the decay $K_L \rightarrow \pi_0 A_D$ is either hidden by $K_L \rightarrow \pi^+ \pi^- \pi_0$ backgrounds or is kinematically inaccessible; scenario C2 is the equivalent of C1 in the event that $m_{A_D}\lesssim 250 \; \textrm{MeV}$.}
\label{table:XiBenchmarks}
\end{table}

Before moving on, we note that, as mentioned in Section \ref{ScalarContentSection}, we expect the model should contain a physical scalar of mass $\sim m_{A_D}$, \ie, the `dark Higgs', which should generically have similar flavor-changing couplings as the dark photon. If we likewise assume that the light scalar is long-lived or can primarily only decay to dark photons as we would expect, we might anticipate that it will make additional contributions to the branching fractions we have discussed in this section, of comparable magnitude to those facilitated by the dark photon. However, given the complexity of the scalar sector within this model, computing the exact magnitude of these contributions is highly non-trivial, and will likely depend on a number of additional parameters in the scalar potential. For the sake of simplicity, then, we shall not compute the light scalar contributions to these flavor-changing decays. We can assume that at worst, these likely manifest $O(1)$ corrections to the constraints appearing in Figure \ref{fig1}, and given the highly non-trivial nature of the scalar sector, it is just as plausible that large regions of parameter space in the scalar potential exist such that these decays are either kinematically forbidden or highly suppressed by a small phase space factor.

Finally, it should be noted that similar transitions exist in the up-like quark and charged lepton sectors (\eg, $t \rightarrow c A_D$, $c \rightarrow u A_D$, $\mu \rightarrow e A_D$). However, in the case of up-like quarks the experimental constraints on decays mediated by these processes are much weaker, while the case against including restrictions in the leptonic sector was discussed in Section \ref{LowEnergyParamsSection}. In either case, these processes provide no significant limit on our model parameters at present. 

\subsection{Neutral Meson Oscillation}

While the dark photon's flavor-changing couplings are capable of facilitating distinctive and highly-constrained meson decays, they are hardly the only potentially dangerous flavor-changing couplings in the theory. In particular, we find that $\overline{B}_d-B_d$ and $\overline{B}_s-B_s$ oscillations can suffer significant contributions to their dispersive amplitudes, given by $(M_{12})_q = \langle \overline{B}^0_q | \mathcal{H}^{\Delta B = 2}_{eff} | B^0_q \rangle/(2 m_{B_q})$ (where $\mathcal{H}^{\Delta B = 2}_{eff}$ is simply the effective Hamiltonian for the relevant flavor-changing interactions) from two additional sources: The flavor-changing $Z_F$ couplings, identified in Eq.(\ref{ZFSMCoupling}), and the one-loop contributions of the heavy top partner to the box diagram which generates meson mixing in the SM. We note that effects in $\overline{K}^0-K^0$ mixing, by contrast, are generally quite muted: The contribution to this process due to the top partner loops is suppressed by CKM factors, while the corresponding FCNC's mediated by $Z_F$ vanish at $O(\lambda^4)$. To obtain our results, we shall compute the new physics contributions to $(M_{12})_q$ and parameterize them as \cite{Charles:2013aka,Charles:2020dfl,Hocker:2006xb}
\begin{align}
    (M_{12})_q = (M_{12})_q^{SM}(1+h_q e^{2 i \sigma_q}),
\end{align}
where $(M_{12})_q^{SM}$ is the SM contribution to $B_q$ meson mixing, while $h_q$ and $\sigma_q$ are real parameters for the contribution of new physics to these processes. $(M_{12})_q^{SM}$ is in turn given by \cite{Buras:1984pq}
\begin{align}\label{M12SM}
    &(M_{12})_q^{SM} \approx \frac{G_F^2 m_W^2}{12 \pi^2}f_{B_q}^2 m_{B_q} B_{B_q} (\lambda^{(q)}_t)^2 S_0(x_t),\nonumber\\
    &S_0(x_t) \equiv \bigg( \frac{x_t (4-11 x_t+x_t^2)}{4(1-x_t)^2}-\frac{3 x_t^3 \log(x_t)}{2(1-x_t)^3} \bigg),\\
    &\lambda^{(q)}_i \equiv V_{i b}^* V_{i q}, \; \; \; x_i \equiv \frac{(m_i^\textrm{pole})^2}{m_W^2}, \nonumber
\end{align}
We can then compare the new physics contributions to the limits in a fit of CKM observables to new physics in the meson mixing sector in \cite{Charles:2020dfl} in order to extract approximate constraints on our model parameters. In particular, this fit gives
\begin{align}\label{CharlesConstraints}
    h_d \leq 0.26, \; \; h_s \leq 0.12
\end{align}
at the 95\% CL level as of Summer 2019. We note that simply referring to these ranges represents only a very approximate assessment of the constraints afforded by flavor-changing neutral currents in this model; for example, the above ranges assume uncorrelated $h_d$ and $h_s$ values, which we shall see is not the case here, and ignore non-trivial constraints on the phase $\sigma_d$, which we shall see may substantially tighten constraints on $h_d$. In spite of these limitations, however, we find that the limits in Eq.(\ref{CharlesConstraints}) will afford us an approximate picture of the effect our model has on meson oscillation parameters.
%For the sake of simplicity, we shall consider the effects of the $Z_F$ boson and the heavy top partner on these processes separately, without performing RG running together. In practice, we find no significant inaccuracy stemming from this approach: The $Z_F$ effects are fairly minimal without some modest degree of fine-tuning of high-energy model parameters, and furthermore, we note that the tree-level $Z_F$ interaction might be highly sensitive to physics in the neutrino sector, such as high-vev flavorful scalars that might provide sterile neutrinos with large Majorana masses, which would likely render this boson entirely too massive for its effects to be observed.

\subsubsection{Neutral Meson Oscillation: Top Partner Loop Contribution}\label{TopLoopSection}

\begin{figure}[htbp]
\centerline{\includegraphics[width=5.0in,angle=0]{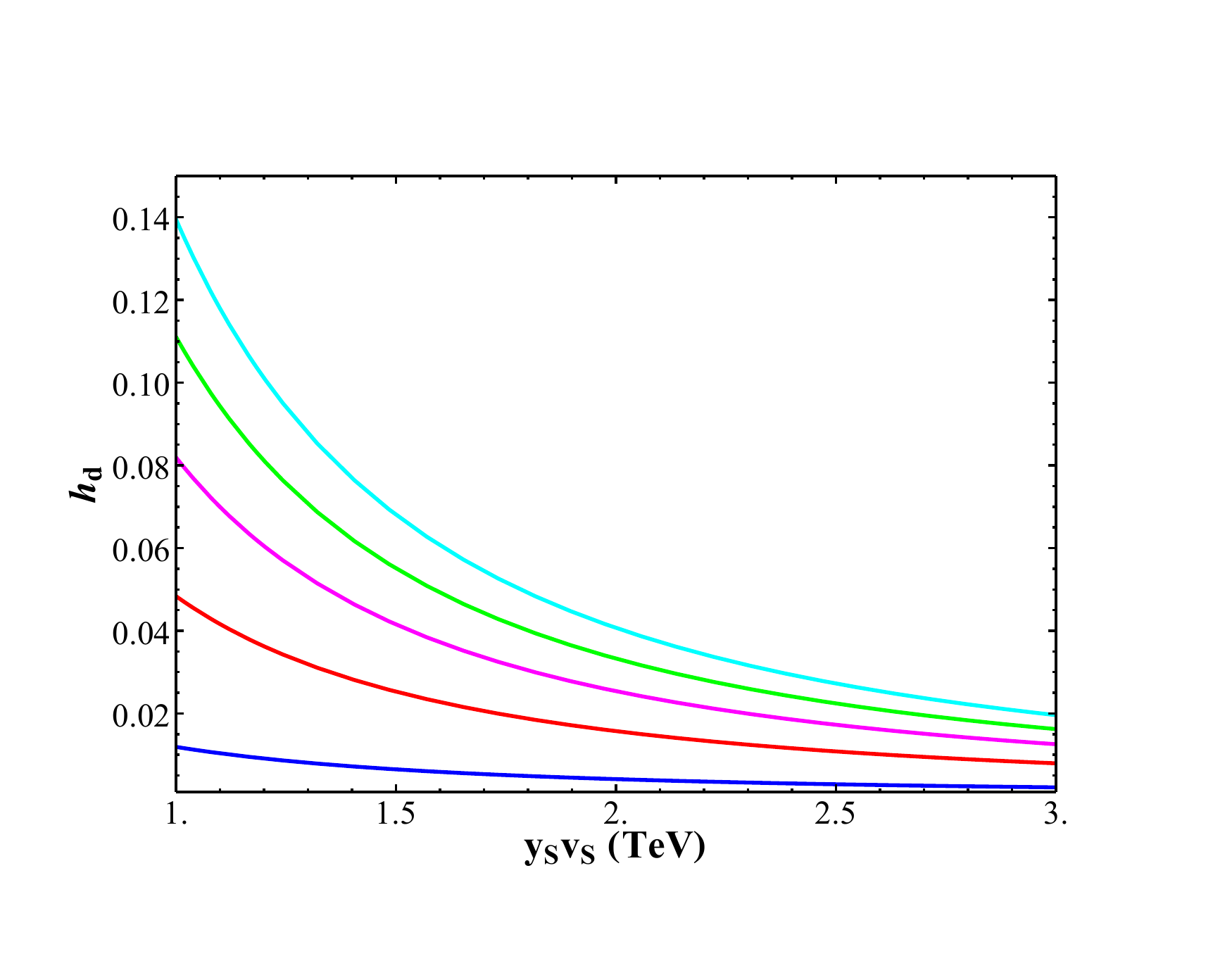}}
\vspace*{-2.0cm}
\centerline{\includegraphics[width=5.0in,angle=0]{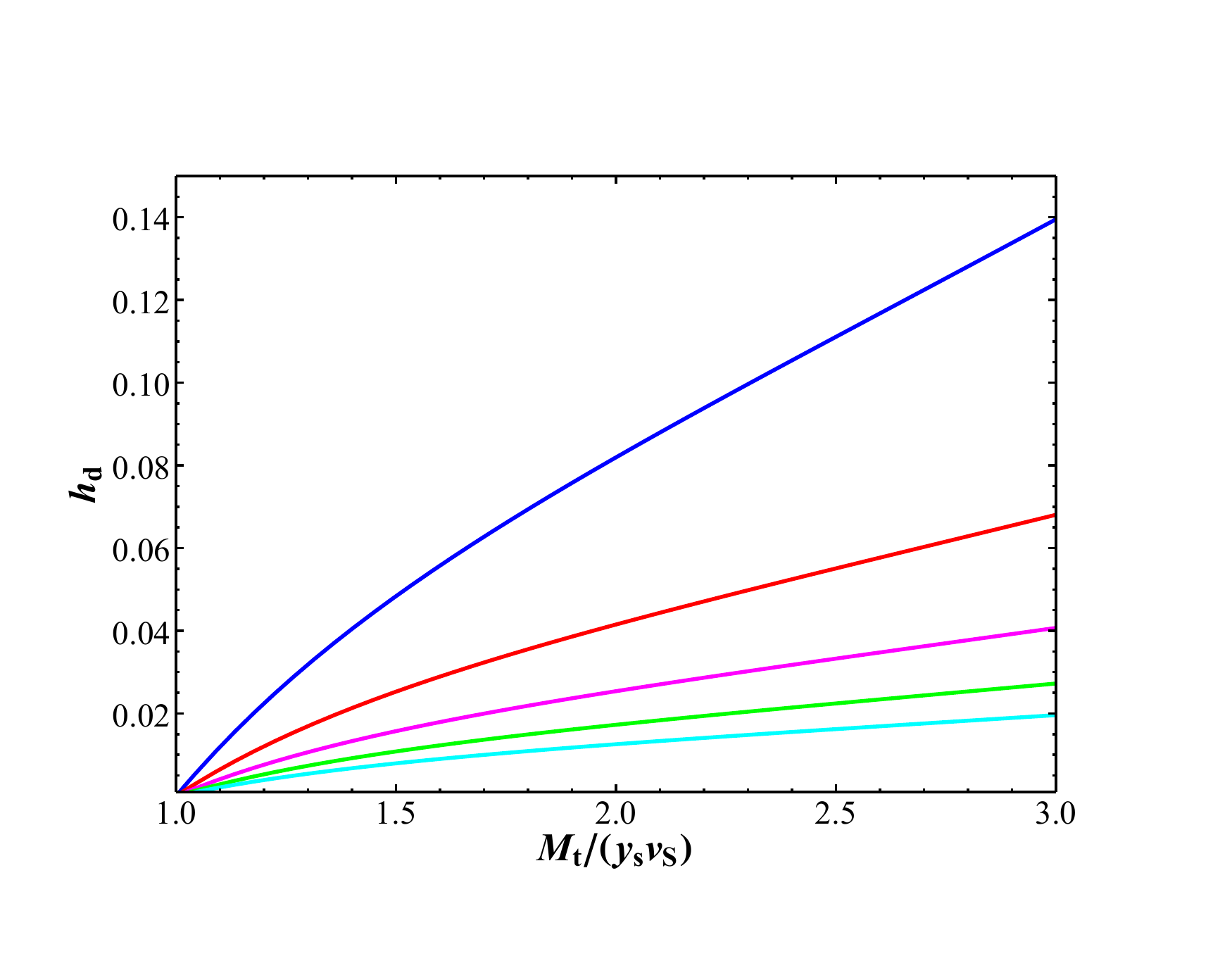}}
\vspace*{-1.0cm}
\caption{(Top) Value of $h_{d}$ as a function of $y_S v_S$ for, from top to bottom, $M_t/(y_S v_S) =$ 3, 2.5, 2, 1.5, and 1.1, respectively.
(Bottom) Value of $h_{d}$ as a function of $M_t/(y_S v_S)$ for, from top to bottom, $y_S v_S =$ 1, 1.5, 2, 2.5, and 3 TeV, respectively. }
\label{fig2}
\end{figure}

We now move on from the flavor effects of the dark photon $A_D$ to the effects of heavier new physics on flavor observables, in particular the highly constrained measurements related to neutral meson oscillation. To start, we consider the effect of the heavy top partner fermion. In spite of being a loop-level interaction, the contribution of the heavy top quark partner to neutral meson mixing processes, in particular those of the $B_s$ and $B_d$ mesons, is significant enough to warrant discussion. We note that the new physics contributions to meson mixing from the top partner take two forms: Additional loop diagrams featuring the heavy top quark, and, as can be seen in Eq.(\ref{NewCKM}), tree-level modifications of the SM CKM matrix elements $V_{td}$, $V_{ts}$, and $V_{tb}$. As discussed in \cite{Charles:2020dfl}, the fit performed in that work extrapolates these CKM matrix elements from unitarity, rather than directly, and as such the tree-level modifications to these parameters must be included as ``new physics'' that contributes to $h_{d,s}$ and $\sigma_{d,s}$. 
We can now determine the contribution of the top partner here in direct analogy to the SM calculation done in, for example, \cite{Buras:1984pq}. Inserting Eq.(\ref{NewCKM}) for the CKM couplings of the top partner, we find that up to $O(10^{-2})$ corrections due to subdominant loops featuring up and charm quarks, the top partner's contribution to $(M_{12})_q$ is
\begin{align}\label{TopPartnerLoopM}
    &\frac{(M_{12})_q^{\textrm{T}}}{(M_{12})_q^{SM}} = h_{q}^T e^{2 i \sigma^T_{q}} \approx \frac{r^2}{S_0(x_t)}\frac{m_t^2}{M_t^2} (-2 \Tilde{S}(x_t,x_t)+2 \Tilde{S}(0,x_t)+2 \Tilde{S}(x_t, x_T)+r^2 \frac{m_t^2}{M_t^2} \tilde{S}(x_T, x_T)),\\
    &\Tilde{S}(x_i,x_j) = \frac{4-7 x_i x_j}{4(1-x_i-x_j+x_i x_j)}+\frac{(4-(8-x_i)x_j)x_i^2 \log(x_i)}{4(x_i-1)^2(x_i-x_j)}+\frac{(4-(8-x_j)x_i)x_j^2 \log(x_j)}{4(x_j-1)^2(x_j-x_i)}, \nonumber \\
    &x_T \equiv \frac{M_t^2}{m_W^2}, \;\;\; x_t \equiv \frac{(m_t^{\textrm{pole}})^2}{m_W^2}, \nonumber
\end{align}
where we have assumed that the QCD corrections for the heavy top partner loops are approximately equal to those of the top quark itself (as done in, for example, \cite{Vatsyayan:2020jan}), $S_0(x_t)$ is the Inami-Lim function given here in Eq.(\ref{M12SM}), and we remind the reader that $r$ is the variable first defined in Eq.(\ref{UuLTrunc}). It is useful to point out that here, as elsewhere in this work, when we refer to an SM quark mass such as $m_t$, we are explicitly referring to its value at a scale of $\sim 1 \; \rm TeV$ (as we noted in Section \ref{FermionSpectrumSection}), the scale at which we assumed the SM fermion masses were generated via a seesaw mechanism. Hence, $m_t$ in Eq.(\ref{TopPartnerLoopM}) refers to the top quark mass \emph{at this scale}, or in this case $m_t (\sim 1 ~\rm TeV) \approx 144 \; \textrm{GeV}$, and not its pole mass, which is explicitly denoted by $m_t^{\textrm{pole}} \approx 173 \; \textrm{GeV}$. As both the pole mass $m_t^{\textrm{pole}}$ and the RG-evolved mass $m_t$ appear in Eq.(\ref{TopPartnerLoopM}), some care must be taken in order to use the formula correctly. Since the scale $1 \; \textrm{TeV}$ is dramatically closer to the top partner pole mass than it is to that of the top quark, we neglect any effects of RG running in $M_t$. In the limit where $x_T \gg x_t$, which applies here, we can combine Eqs.(\ref{M12SM}) and (\ref{TopPartnerLoopM}) as
\begin{align}\label{TopPartnerLoopMApprox}
    &h_{d,s}^T \approx \frac{r^2 x_t}{2 S_0(x_t)}\frac{m_t^2}{M_t^2} \bigg[ \frac{6 x_t}{(-1+x_t)^2}-\frac{3(-1+3 x_t)}{(-1+x_t)^3}\log(x_t)-1+\frac{r^2}{2}\bigg( \frac{m_t}{m_t^{\textrm{pole}}}\bigg)^2 -2\log \bigg( \frac{m_t^{\textrm{pole}}}{M_t}\bigg)\bigg],\\
    &\sigma_{d,s}^T = 0. \nonumber
\end{align}
Inserting $m_t^{\textrm{pole}} \approx 173 \; \textrm{GeV}$, $m_t \approx 144 \; \textrm{GeV}$, and $x_t \approx 4.63$, we arrive at
\begin{align}\label{TopPartnerLooph}
    h_{d,s}^T \approx 0.064 \bigg( \frac{1 \; \textrm{TeV}}{y_S v_S}\bigg)^2 \bigg( 1-\frac{y_S^2 v_S^2}{M_t^2}\bigg) \bigg[ 1+0.102 \bigg( \frac{M_t^2}{y_S^2 v_S^2}-1\bigg)-0.592 \log \bigg(\frac{1 \; \textrm{TeV}}{M_t} \bigg)\bigg],
\end{align}
where for clarity we have explicitly inserted the equation $r^2=M_t^2/(y_S^2 v_S^2)-1$ in the above expression. Notably, we see that the heavy top partner contributes the \emph{same} correction to both $B_d$ and $B_s$ oscillation, and also that this correction can be quite significant, even of $O(10 \%)$. Obtaining a realistic range for $M_t/(y_S v_S)$ from Table \ref{table:LowEnergyMasses}, we can then consider the correction $h_{d,s}^T$ as a function of the parameters $y_S v_S$ and $M_t/(y_S V_S)$. In the case of $h_d$, the top partner loops represent the dominant new physics contribution to this parameter, so our computation of this variable can be completed here. We depict the results for $h_d$ for various values of $y_S v_S$ and $M_t/(y_S v_S)$ in Figure \ref{fig2} (we remind the reader that $M_t> y_S v_S$, which can be straightforwardly seen from the equation for $M_t^2=M_{u3}^2$ in the last line of Eq.(\ref{uLSMMat})).

From Figure \ref{fig2}, we can see several interesting characteristics of this new physics contribution. First, comparing the magnitude of the effects in this Figure to the constraints in Eq.(\ref{CharlesConstraints}), we see that for the region of parameter space we are considering, the effects of the top partner loops on meson mixing parameters should lie within current experimental constraints. We do, however, note that for regions with the largest effect, which can reach $h_{d} \sim 0.14$,  the results exceed the expected sensitivity of LHCb and Belle-II data in the coming years, which may be sensitive to $h_d \sim 0.07$ within the next decade \cite{Charles:2020dfl}. Furthermore, we note that a quick reference to Eq.(\ref{TopPartnerLoopM}) allows us to see that $\sigma_d = 0$ from this contribution (indicating a \emph{positive} proportional contribution to $(M_{12})_d$), while the best-fit value of $\sigma_d$ in the fit of \cite{Charles:2020dfl} is $\sigma_d = -1.40^{+0.97}_{-0.23}$ (or roughly corresponding to a \emph{negative} proportional contribution to the mixing matrix-- see \cite{Charles:2013aka} for a full chart of allowed values in the $h_d-\sigma_d$ plane, albeit with slightly outdated data but exhibiting similar favored regions of $h_d-\sigma_d$ as the present constraints). We might therefore expect that constraining our model more rigorously, with a CKM fit of our own, might result in non-trivial constraints on the scenarios with low $y_S v_S$ values from measurements of $B_d$ oscillation. However, the sharp dropoff of these contributions as $y_S v_S$ increases likely removes this sensitivity as $y_S v_S \gsim 1.75 \; \textrm{TeV}$, unless these constraints are an order of magnitude more sensitive than the generic boundaries found in \cite{Charles:2020dfl}. In practice, therefore, we find that the loop-level contributions of the top partner to neutral meson mixing are presently not subject to any substantial constraints; however additional data from LHCb and Belle-II in the coming years will likely constrain the parameter space near the 
lower end of the range $y_S v_S \sim 1 \; \textrm{TeV}$ from effects in this sector.

\subsubsection{Neutral Meson Oscillation: $Z_F$ Contribution}\label{ZFMesonSection}

\begin{figure}
    \centerline{\includegraphics[width=3.5in]{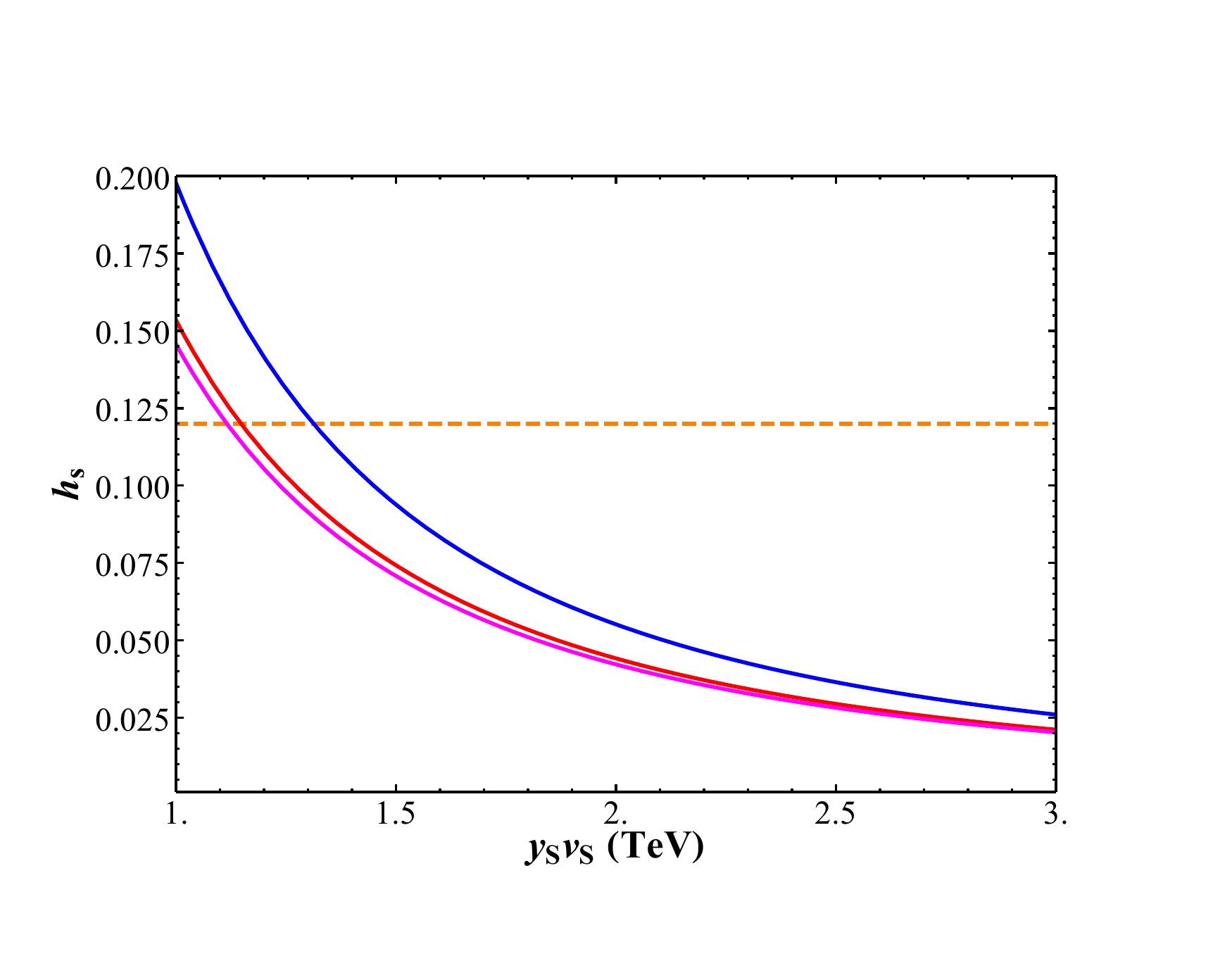}
    \hspace{-0.75cm}
    \includegraphics[width=3.5in]{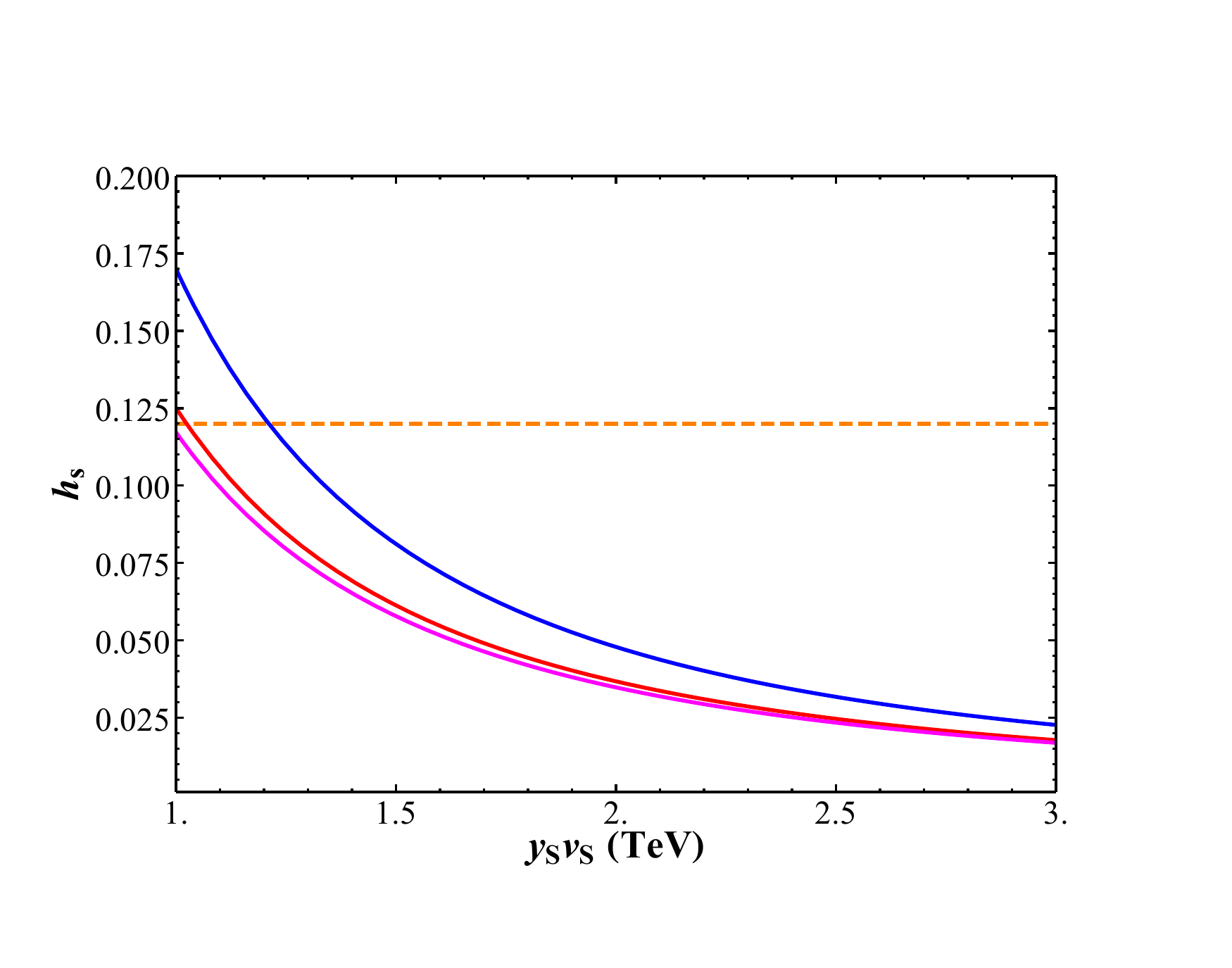}}
    \vspace*{0.0cm}
    \centerline{\includegraphics[width=3.5in]{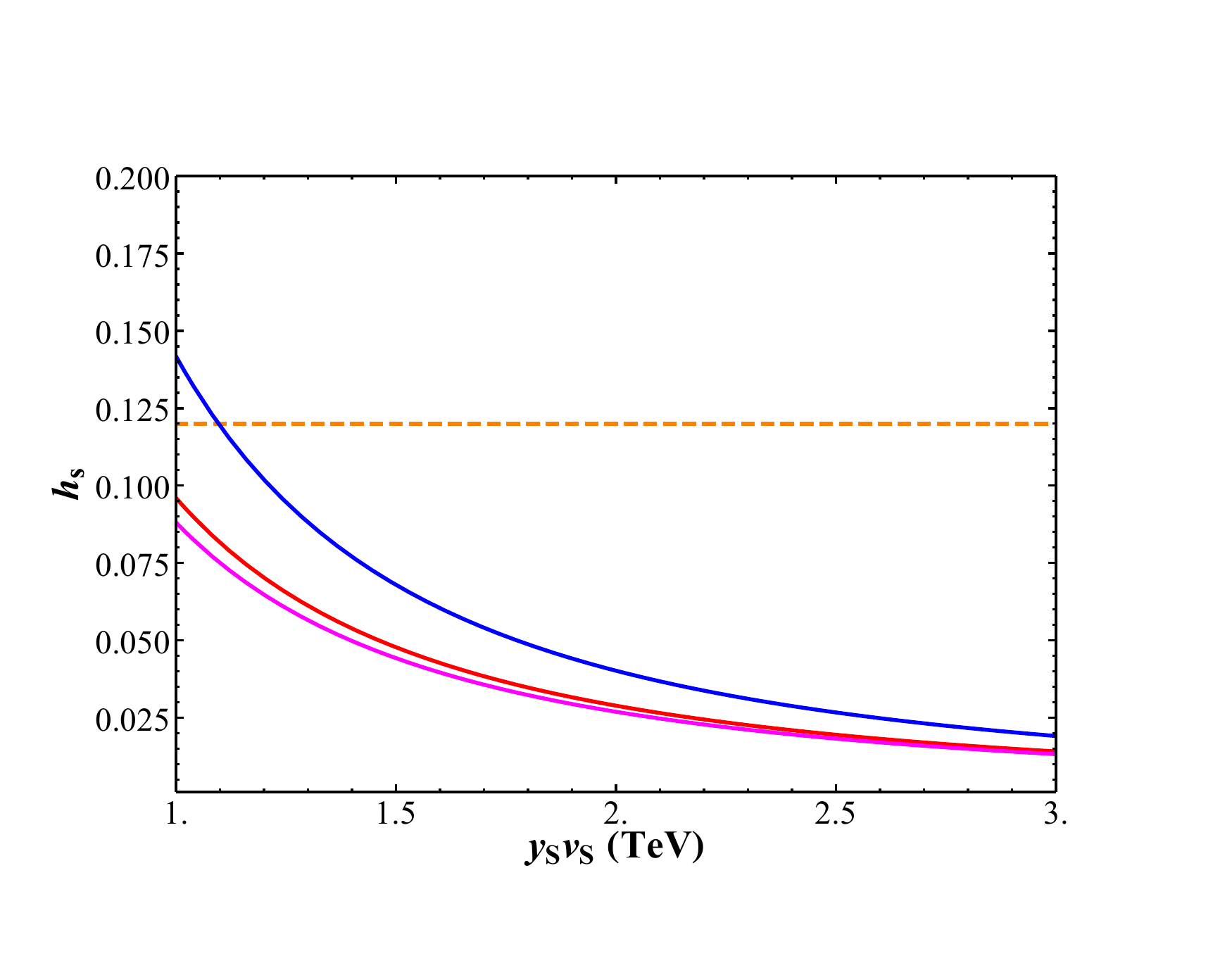}
    \hspace{-0.75cm}
    \includegraphics[width=3.5in]{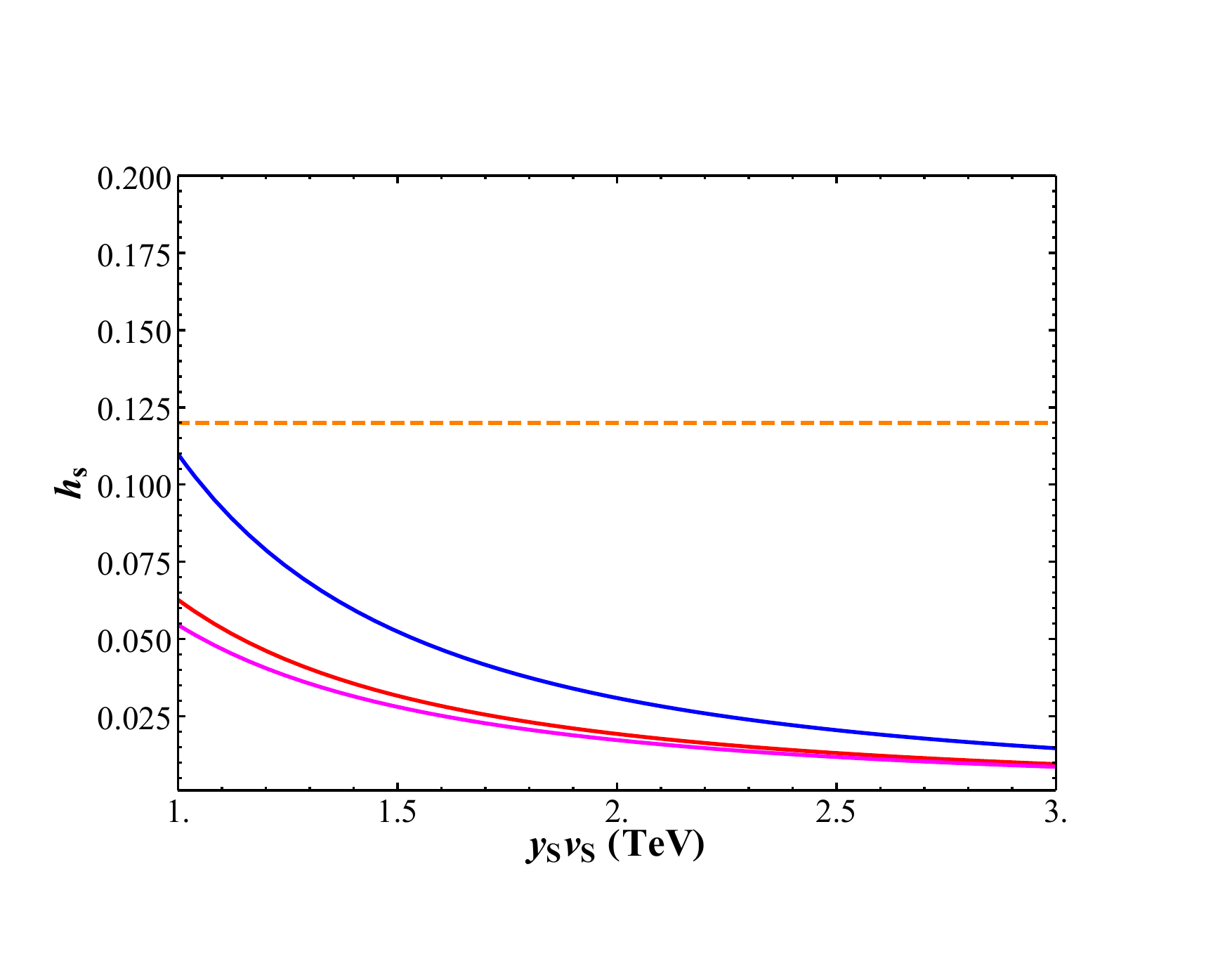}}
    \caption{The full value of $h_s$, including tree-level $Z_F$ and top partner loop contributions, for $m_{Z_F}/(g_4 y_S v_S)=20$ (blue), $40$ (red), and $60$ (magenta). These plots  assume that $M_t/(y_S v_S) = 3$ (top left), $2.5$ (top right), $2$ (bottom left), and $1.5$ (bottom right). For comparison, the 95\% CL limit on $h_s$ is depicted in each plot as a dashed orange line.}
    \label{fig3}
\end{figure}

While the top partner loops represent by far the dominant effect in the $B_d$-meson mixing system, the $B_s$ mixing system suffers additional contributions from another source, \ie, the ``flavor $Z$'' boson, $Z_F$. In particular, we can see from the coupling in Eq.(\ref{ZFSMCoupling}) that the dominant contribution of the $Z_F$ boson will be to $B_s$ mixing alone. We can compute this contribution straightforwardly by following the treatment of, \eg, \cite{Bazavov:2016nty}. In this case, we can write the effective flavor-changing Hamiltonian as
\begin{align}
    \mathcal{H}^{\Delta B = 2}_{Z_F} = \mathcal{C}_1 (m_{Z_F}) \mathcal{O}_1 (m_{Z_F})+\Tilde{\mathcal{C}}_1 (m_{Z_F}) \Tilde{\mathcal{O}}_1 (m_{Z_F})+\mathcal{C}_4 (m_{Z_F}) \mathcal{O}_4(m_{Z_F})+\mathcal{C}_5 (m_{Z_F}) \mathcal{O}_5(m_{Z_F}),
\end{align}
with
\begin{align}\label{BBBarOperators}
    \mathcal{O}_1 = (\overline{b}_L^\alpha \gamma_\mu q_L^\alpha)(\overline{b}_L^\beta \gamma_\mu q_L^\beta), \;\; \Tilde{\mathcal{O}}_1 = (\overline{b}_R^\alpha \gamma_\mu q_R^\alpha)(\overline{b}_R^\beta \gamma_\mu q_R^\beta), \;\; \mathcal{O}_4 = (\overline{b}_R^\alpha s_L^\alpha)(\overline{b}_L^\beta s_R^\beta), \;\; \mathcal{O}_5 = (\overline{b}_R^\alpha s_L^\beta)(\overline{b}_L^\beta s_R^\alpha),
\end{align}
using operators defined identically to those of \cite{Bazavov:2016nty}. Our computation of $(M_{12})_s$ stemming from these effects then simply requires identifying the Wilson coefficients $\mathcal{C}_{i}$ and extracting the expectation values of the operators $\mathcal{O}_i$ from \cite{Bazavov:2016nty}. Using Eq.(\ref{ZFSMCoupling}), we can straightforwardly see that at some high scale $\mu_\textrm{high}$, we have
\begin{align}\label{ZFWilsonCoeffs}
    \mathcal{C}_1 (\mu_\textrm{high}) = \Tilde{\mathcal{C}}_1 (\mu_\textrm{high}) = \frac{3 g_4^2}{4 m_{Z_F}^2} A^2 \lambda^4 (\rho+i \eta)^2, & ~~\mathcal{C}_4(\mu_\textrm{high}) = 0, & \mathcal{C}_5 (\mu_\textrm{high}) = -\frac{3 g_4^2}{m_{Z_F}^2}A^2 \lambda^4(\rho+i \eta)^2.
\end{align}
We can then run these coefficients down to hadronic scales using anomalous dimension matrices which can be extracted from \cite{Bagger:1997gg}. For the sake of simplicity, we shall assume that $\mu_\textrm{high} = 10 \; \textrm{TeV}$, which should be within a factor of $\sim O(\textrm{a few})$ from $m_{Z_F}$, consulting the range for this parameter given in Table \ref{table:LowEnergyMasses} and assuming $g_4 \sim 0.3$, $y_S v_S \sim 1 \; \textrm{TeV}$. We also assume, consulting the likely ranges for these masses in Table \ref{table:LowEnergyMasses}, that at the scale $\sim 10 \; \textrm{TeV}$, only five non-SM quark flavors are dynamical: The top partner, two up-like portal matter quarks, and two down-like portal matter quarks, and that for the purposes of our RGE calculation all five of these extra flavors can be integrated out at the same scale, which we take to be 1 TeV. Running the coupling constants of Eq.(\ref{ZFWilsonCoeffs}) down to the $b$ quark mass, extracting the expectation value of the operators $\mathcal{O}_{1,4,5}$ (and the bag parameter as well as the  decay constant from the SM contribution to $B_s$ oscillation given in Eq.(\ref{M12SM})) from \cite{Bazavov:2016nty}, we arrive at 
\begin{align}\label{ZFh}
    \frac{(M_{12})_s^{Z_F}}{(M_{12})_s^{SM}} = h_s^{Z_F} e^{2 i \sigma_s^{Z_F}} \approx (0.024)e^{2 i (-0.099)\pi}\bigg( \frac{10 \; \textrm{TeV}}{m_{Z_F}}\bigg)^2 \bigg( \frac{g_4}{0.3}\bigg)^2.
\end{align}
We note that while this correction has a magnitude which varies depending on the $SU(4)_F$ coupling $g_4$ and the mass $m_{Z_F}$, its phase remains fixed. This is the result of the coupling constants in Eq.(\ref{ZFWilsonCoeffs}) all having the same complex phase, that of $\rho + i \eta$, which is of course fixed by the Wolfenstein parameters. The result of Eq.(\ref{ZFh}) suggests, as in the case of the loop-level corrections due to the top partner, that we may anticipate roughly $O(\textrm{a few} \%)$ corrections from $Z_F$ exchange: That is, these two processes have roughly comparable effects. We should, then, estimate total contribution to $h_s$ as
\begin{align}\label{hsFull}
    h_s = |h_s^{Z_F}e^{2 i \sigma_s^{Z_F}}+h_s^T e^{2 i \sigma_s^T}|,
\end{align}
with $h_s^{Z_F}$ and $\sigma_s^{Z_F}$ given by Eq.(\ref{ZFh}), $h_s^T$ given by Eq.(\ref{TopPartnerLooph}), and $\sigma_s^T=0$, as noted in Section \ref{TopLoopSection}. Taking numerical results of Eq.(\ref{hsFull}) and comparing the results to the constraints on $h_s$ in Eq.(\ref{CharlesConstraints}), we can also extract some rough constraints on the parameters $y_S v_S$ for various choices of the proportional mass parameters $M_t/(y_S v_S)$ and $m_{Z_F}/(g_4 y_S v_S)$. In Figure \ref{fig3}, we depict the combined effect of the $Z_F$ and top partner contributions to the variable $h_s$.

Comparing the results of Figure \ref{fig3} to those of Figure \ref{fig1}, which should be identical to $h_s^T$, or the new physics parameter in the absence of any $Z_F$ contributions, we see that for lower masses of $Z_F$, in particular $m_{Z_F} = 20 g_4 y_S v_S$ (the lower end of the range we consider), the effect of the $Z_F$ boson on $h_s$ is significant, on the order of $0.06$ for $y_S v_S = 1 \; \textrm{TeV}$. For lower values of $M_t/(y_S v_S)$, the $Z_F$ contribution at these low masses is even the same magnitude as the contribution due to vector-like quarks. However, some caveats are required here. First, a $Z_F$ mass as low as $20 g_4 y_S v_S \sim 7 y_S v_S$ is the result of somewhat finely-tuned parameters in the high-energy theory: Our numerical probe of the high-energy parameter space in Section \ref{LowEnergyParamsSection} places such a mass only just above the 5\% quantile. Larger values of $m_{Z_F}$, such as $60 g_4 y_S v_S \sim 20 y_S v_S$, which is close to the median $m_{Z_F}$ of our sample, generally offer only an $O(10\%)$ correction to the mixing result purely from considering top partner loops. For larger values of $m_{Z_F}$, which can reach as high as $200 g_4 y_S v_S \sim 70 y_S v_S$ in our sample, the effects of the $Z_F$ boson quickly become negligible. We also note that, as discussed in Section \ref{GaugeSpectrumSection}, the $Z_F$ boson's mass may be sensitive to modifications in the model to accomodate the SM neutrino sector, or generally to a broad array of changes to the specifics of the high energy model. In particular, the mass of any gauge bosons corresponding to the $SU(3)_F$ flavor group may increase dramatically. As a result, we might see any observable new physics contribution from $Z_F$ vanish.

These concerns aside, however, we do see in Figure \ref{fig3} that any constraint from $B_s$ meson mixing is quite mild
-- while there do exist points in parameter space which can exceed the 95\% CL upper bound for $h_s \simeq 0.12$, we see that they tend to only occur in the ``optimal'' scenario for these corrections: The only scenarios we depict in which the present $h_s$ cannot be evaded by simply requiring $y_S v_S \gsim 1.1 \; \textrm{TeV}$, a mere 10\% increase from the lowest value of $y_S v_S$ we consider, occur when we have assumed $m_{Z_F} = 20 g_4 y_S v_S$, near its 5\% quantile value in our numerical sample, and $M_t \geq 2.5 y_S v_S$. As in the case of $B_d$ mixing, however, we can note that future improvements to $h_s$ observations from Belle-II and LHCb measurements (and improvements in lattice computations of hadronic matrix elements), which might be expected to reduce the 95\% CL bound on $h_s$ to $\sim 0.06$ \cite{Charles:2020dfl}, may begin to place more meaningful constraints on our model parameter space for the $Z_F$ and top partner masses.

\section{Phenomenology: Portal Bosons and Exotic Matter}\label{ExoticMatterSection}

Having discussed the low energy flavor-violating implications of the model, we can turn to some other simple processes in the model, in particular focusing on where our results differ from those of \textbf{I} and \textbf{II}. We shall organize this discussion by considering each of the likely kinematically accessible heavier exotic particles, the $Z_P$, top partner, and the portal matter separately.

\subsection{Phenomenology: Portal Matter Fermions}\label{PortalMatterSection}

We shall begin our discussion of the phenomenology of our new exotic particles with the portal matter: The fermion fields which possess non-trivial $U(1)_D$ charge. From Section \ref{FermionSpectrumSection}, we note that our theory anticipates two such fields for each sector of the theory (up-like quarks, down-like quarks, charged leptons, and neutrinos) that we can expect to have masses potentially accessible to experiment in the foreseeable future. For convenience, we shall label the two portal matter states based on their masses: $P^u_1$, for example, will be the up-like portal matter quark with mass $m_{P1}^u$ (as given in Eq.(\ref{uLPortalMat})), while $P^u_2$ will be the up-like portal matter quark with mass $m_{P2}^u$ (again given in Eq.(\ref{uLPortalMat})). Analogous definitions exist for the portal matter in the down-like quark and leptonic sectors. For the sake of simplicity, and because the constraints on color triplet new particles are dramatically stronger than those for color singlets, we shall limit our discussion here to the portal matter in the quark sector, $P_{1,2}^{u,d}$.

Much of the behavior of the portal matter phenomenology in the present setup is analogous to that of \textbf{I} and \textbf{II}, and as such does not bear particular repeating. However, the additional structure present in this model does introduce some non-trivial complications to the existing frameworks discussed there. In particular, the couplings of the portal matter states to various generations of SM matter are entirely controlled by the orientation of the vector $\vec{\gamma}_P$ in three-dimensional flavor space: Couplings of $i^{th}$-generation up-like quarks are proportional to $(\vec{\gamma}_P)_i/v_P$, while couplings of the $i^{th}$-generation down-like quarks are proportional to $\xi_i = (\mathbf{W}_d^\dagger \vec{\gamma}_P)_i/v_P$, where we remind the reader that $\mathbf{W}_d$ is well-approximated by the CKM matrix. Since the CKM matrix is nearly diagonal, this in turn limits the freedom with which we may couple different quark portal matter states to various generations: If the benchmark A1 is taken from the $\xi_i$ benchmarks in Table \ref{table:XiBenchmarks}, for example, the branching fractions of the up-like portal matter to the second- and third-generation up-like quarks will be suppressed by CKM factors of $\lsim O(10^{-2})$. Finally, we note that the universal dependence on $\vec{\gamma}_P$ to generate couplings between portal matter and SM states limits the freedom with which discrepancies may appear between the mixing of $P_1$ states with the SM and that of $P_2$ states: With the exception of the third generation of the up-like sector, where corrections occur due to mixing between the SM top quark and its non-portal partner, the magnitude of $P_1$ and $P_2$ couplings to SM states are the same, with $P_1$ states coupled to left-handed ($SU(2)_L$ doublet) SM states and $P_2$ states coupled to right-handed ($SU(2)_L$ singlet) ones. So, for example, the branching fractions of $P_1^d$ and $P_2^d$ portal matter particles to each of $b$, $s$, and $d$ quarks are always \emph{equal}. Meanwhile, the fact that the $P_1$ states overwhelmingly couple to $SU(2)_L$ doublet SM fields and the $P_2$ overwhelmingly couple to $SU(2)_L$ singlet fields is unsurprising: Given that the $P_1$ states are $SU(2)_L$ doublets, mixing between $P_1$ states and right-handed ($SU(2)_L$ singlet) fields violates $SU(2)_L$, and so only occurs due to mass terms in Eqs.(\ref{MMatuL}) and (\ref{MMatuR}) proportional to the $SU(2)_L$-breaking vev, $v/\sqrt{2} \sim 170 \; \textrm{GeV}$. Mixing between $P_1$ states and left-handed fermions, however, has no such restriction, and can be proportional to the significantly larger vev terms $v_S$ and $v_P$. An analogous argument applies for $P_2$ states preferring to couple to $SU(2)_L$ singlet fields over doublets.

Regarding the production of portal matter states, the QCD pair production cross section of these fermions will be the same as the results already given \textbf{I} and \textbf{II} for 13 TeV and 14 TeV center-of-mass energy at the LHC, respectively. Additionally, as in \textbf{II}, there exist potentially significant non-QCD contributions to the $q\overline{q} \rightarrow \bar P^u_i P^u_i$ process from $t-$channel dark photon exchange. We note that the dark photon exchange contribution has one piece of additional structure compared to that of \textbf{II}: Because all three generations may mix with the portal matter states, depending on the orientation of the vector $\vec{\gamma}_P$, the contribution of the $t$-channel $A_D$ exchange to portal matter pair production may stem from portal matter couplings to multiple different SM quarks. This is in contrast to \textbf{II}, in which each portal matter field was only coupled to a single SM quark, and therefore portal matter pair production would only suffer a $t$-channel dark photon contribution from that specific quark flavor. However, we also note that the effect of this exchange varies radically depending on the generation in question, with portal matter coupling to lighter quarks having a dramatically stronger effect than coupling with heavier quarks, due to the greater content of the lighter quarks in proton parton distribution functions. We can then see, consulting the results of \textbf{II} for the down-like sector of our model, that restricting considerations of mixing in this contribution to just that of the lightest generation for which the mixing is non-trivial (in the language of Section \ref{ADFlavorConstraintsSection}, the $\xi_i$ for the lowest index $i$ that isn't hierarchically smaller than any other $\xi_i$) will likely be accurate to within $O(10\%)$ corrections, at worst. Given the even larger hierarchies among quark masses in the up-like sector, we anticipate this approximation to be even more accurate for up-like portal matter.

Finally, we can discuss the decay branching fractions of portal matter fields to SM states. As in \textbf{I} and \textbf{II}, the dominant decay process is the emission of a dark photon to decay into an SM state with the same SM quantum numbers as the portal matter field. Consulting the couplings in Eqs.(\ref{DarkPhotonCoupling}), (\ref{ADuCouplingTrunc}), and (\ref{ADdCouplingTrunc}), we find that the decay widths for these processes are given by 
\begin{align}
    &\Gamma_{P_{1,2}^d \rightarrow d_i A_D} = \frac{g_4^2 s_P^2 (m_{P1,P2}^d)^3}{48 \pi m_{A_D}^2}\frac{|(\mathbf{W}_d^\dagger \vec{\gamma}_P)_i|^2}{v_P^2}, \; i=1,2,3,\nonumber\\
    &\Gamma_{P_{1}^u \rightarrow u_i A_D} = \frac{g_4^2 s_P^2 (m_{P1}^u)^3}{48 \pi m_{A_D}^2}\frac{|(\vec{\gamma}_P)_i|^2}{v_P^2}, \; i=1,2,3,\\
    &\Gamma_{P_{2}^u \rightarrow u_i A_D} = \frac{g_4^2 s_P^2 (m_{P2}^u)^3}{48 \pi m_{A_D}^2} \frac{|(\vec{\gamma}_P)_i|^2}{v_P^2}, \; i=1,2, \nonumber\\
    &\Gamma_{P_{2}^u \rightarrow t A_D} = \frac{g_4^2 s_P^2 (m_{P2}^u)^3}{48 \pi m_{A_D}^2}\frac{|(\vec{\gamma}_P)_3|^2}{v_P^2} \bigg( \frac{M_t^2-y_S^2 v_S^2}{M_t^2} \bigg), \nonumber
\end{align}
where $d_1 \equiv d$, $d_2 \equiv s$, $d_3 \equiv b$, analogous definitions hold for the up-like quarks, and we have neglected the effects of the top quark mass (which should only introduce $O(m_t^2/(m^u_{P1,2})^2) \sim O(10^{-2})$ corrections). For clarity, we note that the enormous ratios $(m^{u,d}_{P1,P2})^2/(m_{A_D})^2$ in these expressions are cancelled by the comparably tiny ratios $|(\vec{\gamma}_P)_i|^2/v_P^2$, leading to results with reasonable magnitudes -- this is in fact analogous to the discussion in Section 3.3 of \textbf{II} in which the dark photon coupling between SM and portal matter states is strongly suppressed, but the suppression is counteracted by the dark photon's small mass to give an $O(1)$ overall strength for the interaction. In addition to these decay widths, we would expect equal (at least in the limit where $m_{A_D} \rightarrow 0$) contributions from a light scalar, corresponding to a physical scalar mode associated with $U(1)_D$ breaking, from Goldstone boson equivalence. As the scalar sector is not well-explored here, and such changes would not affect the branching fractions of the portal matter anyway, we shall ignore this scalar from here on out, with one exception discussed in Section \ref{TopPartnerSection}. As discussed in \textbf{I}, assuming that the dark photon is long-lived or only decays to DM (which we will assume here),  the signal for a portal matter pair production event here should consist of two (possibly fat) jets, which may be from $t$, $b$, or lighter quarks depending on $\vec{\gamma}_P$'s orientation in flavor space, plus a significant amount of missing energy stemming from the dark photon decays. Perhaps more interesting in this model are the possibilities afforded us by the existence of the heavy top partner: In addition to decays to SM particles, decays with comparable rates should exist for the portal matter fields into the top partner, provided the latter is lighter enough that these channels are kinematically accessible. We should note that, because the masses $M_t$, $m_{P1}^u$, and $m_{P2}^u$ are simply free parameters in our theory with masses of at least $O(\textrm{TeV})$, we have little a priori guidance about whether the top partner is heavier than the portal fields or vice versa. It therefore behooves us to consider both the possibility that a portal matter field can decay into a top partner, and, in Section \ref{TopPartnerSection}, whether the reverse process might occur in a different region of parameter space. Turning our attention back to the decay of portal matter to a top partner, up to $O(m_t^2/M_t^2)$  corrections and assuming that $m_{A_D}$ and $\vec{\gamma}_P$ are much smaller than any other mass scale in the expression, we find that these decay widths are given by
\begin{align}
    \Gamma_{P^u_1 \rightarrow T A_D} \approx \frac{g_4^2 s_P^2 (m^u_{P1})^3}{48 \pi m_A^2}\bigg(1-\frac{M_t^4}{(m_{P1}^u)^4} \bigg) \frac{m_t^2}{y_S^2 v_S^2} \frac{|(\vec{\gamma}_P)_3|^2}{v_P^2},\\
    \Gamma_{P^u_2 \rightarrow T A_D} \approx \frac{g_4^2 s_P^2 (m^u_{P2})^3}{48 \pi m_A^2}\bigg(1-\frac{M_t^4}{(m^u_{P2})^4} \bigg)\frac{y_S^2 v_S^2}{M_t^2} \frac{|(\vec{\gamma}_P)_3|^2}{v_P^2}. \nonumber
\end{align}

\begin{figure}[htbp]
\centerline{\includegraphics[width=5.0in,angle=0]{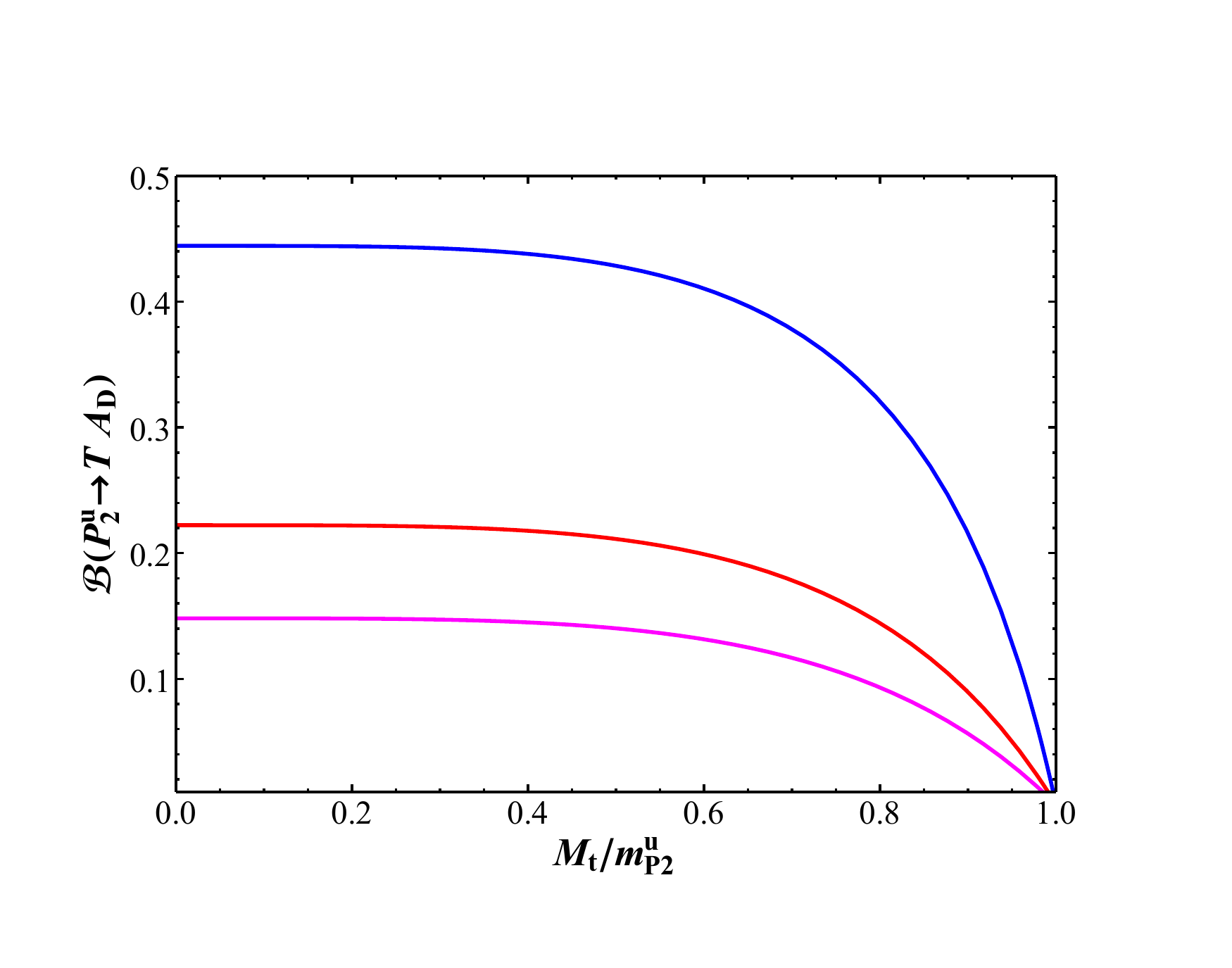}}
\vspace*{-2.0cm}
\centerline{\includegraphics[width=5.0in,angle=0]{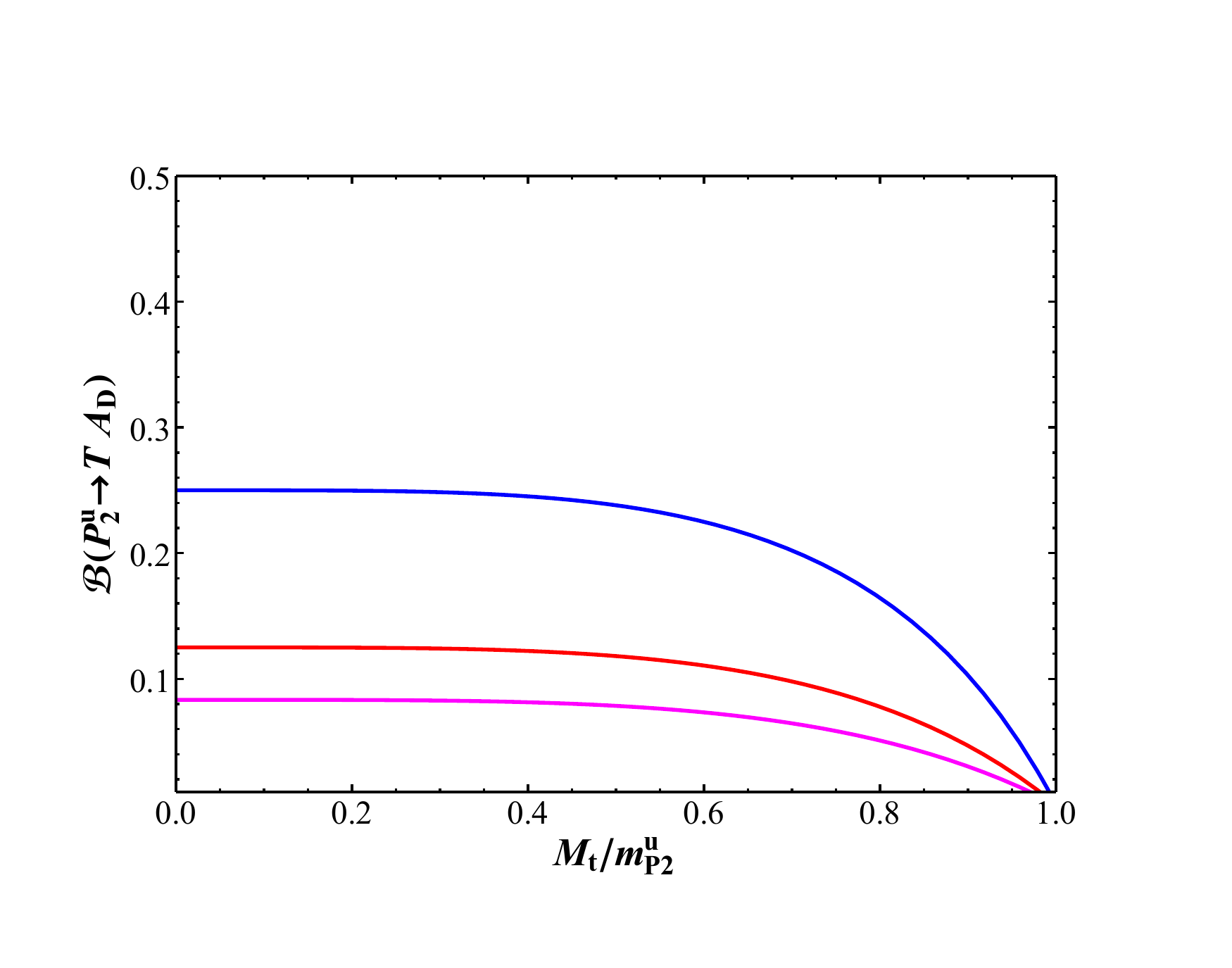}}
\vspace*{-1.0cm}
\caption{(Top) The branching fraction of the $P^u_2$ quark to the top partner, assuming that $M_t/(y_S v_S)=1.5$ and that $\vec{\gamma}_P$ is oriented according to the benchmarks $A_3$ (blue), $B_1$ (red), and $C_1$ (magenta), described in Table \ref{table:XiBenchmarks}.
(Bottom) Same as above, but assuming that $M_t/(y_S v_S)=2$. }
\label{fig6}
\end{figure}

The somewhat strange dependence of these decay widths on the fourth power of the ratio $M_t/m_{P1,2}^u$ bears a brief discussion here: It occurs because the phase space factor in the decay (which approaches $1-M_t^2/(m^u_{P1,P2})^2$ in the limit where $m_{A_D}\rightarrow 0$) multiplies the combination of coupling constants appearing in the squared amplitude, which are proportional to $1+M_t^2/(m^u_{P1,P2})^2$. Meanwhile, we also note that the decay of $P_1^u$ to a top partner quark is suppressed by a factor of $m_t^2/(y_S^2 v_S^2) \lsim 10^{-2}$, and so is unlikely to have much phenomenological impact. However, the decay of $P_2^u$ to the heavy top partner can be quite significant. The branching fraction for this decay should be given by
\begin{align}\label{PortalToPartnerBR}
    \mathcal{B}(P_2^u \rightarrow T \, A_D) = \frac{\bigg(1 - \frac{M_t^4}{(m_{P2}^u)^4}\bigg)\frac{y_S^2 v_S^2}{M_t^2}|(\vec{\gamma}_P)_3|^2}{(\vec{\gamma}_P^* \cdot \vec{\gamma}_P)-\frac{M_t^4}{(m_{P2}^u)^4}\frac{y_S^2 v_S^2}{M_t^2}|(\vec{\gamma}_P)_3|^2}.
\end{align}
Note that because $M_t< (m^u_{P2})$,  $y_S v_S < M_t$, and $|(\vec{\gamma}_P)_3|^2$ will always be less than $\vec{\gamma}_P^* \cdot \vec{\gamma}_P$ here, the above branching fraction will always be positive. To get a feel for the magnitude of this effect, we show this branching fraction as a function of the ratio $M_t/m_{P1}^u$ in Figure \ref{fig6} for some of the benchmarks of $\vec{\gamma}_P$ orientation outlined in Table \ref{table:XiBenchmarks}. We note that the benchmarks in Table \ref{table:XiBenchmarks} list values of $\xi_i \equiv (\mathbf{W}_d^\dagger \vec{\gamma}_P)_i/v_P$, not $\vec{\gamma}_P$, however, the corrections due to letting $\xi_i \approx (\vec{\gamma}_P)_i/v_P$ to Eq.(\ref{PortalToPartnerBR}) will be suppressed by $O(\lambda^2) \sim 10^{-2}$ CKM factors and so are negligible for demonstrative purposes.

In Figure \ref{fig6}, we can see that when kinematically allowed, the branching fraction of $P_2^u$ to the top partner can be quite significant. In cases where the portal matter mixes predominantly with the top quark, it can even account for nearly half of possible decays. As we shall see in Section \ref{TopPartnerSection}, in this scenario the top partner will then decay via the emission of an SM $Z$, $W$, or Higgs into an SM top quark, reflecting the more conventional decay channels for vector-like quarks. This behavior in turn can yield interesting collider signals: Rather than simply producing a pair of jets plus missing energy, as will overwhelmingly occur for the other portal fields in this model, one of the $P^u_2$ fields decaying via the top partner might result in, for example, missing energy, a pair of fat top quark jets, and a pair of leptons from a $Z$ boson. This is an atypical signature.

\subsection{Phenomenology: Top Quark Partner}\label{TopPartnerSection}

Other than the portal fields, the top partner $T$ is the sole other exotic fermion field that may be accessible at current or planned experiments in this model. Anticipating that the top partner here will behave similarly to conventional up-like vector-like quarks, we have quoted the constraint of \cite{Aaboud:2018pii}, which places a 95\% CL limit on the mass of a generic electroweak singlet vector-like quark with this electric charge as $> 1.31 \; \textrm{TeV}$. However, we note that this constraint assumes that the top partner will always decay via the standard processes $T \rightarrow h \, t$, $T \rightarrow Z \, t$, or $T \rightarrow W^+ \, b$. In our present construction, additional decay channels may also be present, so it is useful to determine the extent to which this assumption is valid.

The interactions of this top partner with the SM closely follow the familiar behavior of vector-like quarks in other models; most notably, as we can see from the couplings in Section \ref{SMCouplingsSection}, the SM $Z$, $W$, and Higgs all mediate interactions between the top partner and the SM top, facilitating the  $T \rightarrow Z \, t$, $T \rightarrow W^+ \, b$, and $T \rightarrow h \, t$ decay processes which are assumed in \cite{Aaboud:2018pii}, and which are by contrast highly suppressed for the portal matter fields. Specifically, for these processes we have (up to $O(m_t^2/M_t^2) \sim 10^{-2}$ corrections)
\begin{align}\label{TopPartnerSMWidths}
    \Gamma_{T \rightarrow W b}=2 \Gamma_{T \rightarrow Z t} = 2 \Gamma_{T \rightarrow h t} = \frac{G_F m_t^2}{\sqrt{2} ~8 \pi} \bigg( \frac{M_t^2-y_S^2 v_S^2}{y_S^2 v_S^2}\bigg)M_t,
\end{align}
where $G_F= 1.1663787\times 10^{-5} \; \textrm{GeV}^{-2}$ is the Fermi constant. Notably, while there are \emph{no} decays of this quark into other SM up-like quarks (at least up to highly-suppressed flavor-changing neutral currents from, for example, the dark photon), there are CKM-suppressed decays of the form $T \rightarrow W^+ \, d$ and $T \rightarrow W^+ \, s$; these are, however, suppressed by at least $\lambda^4 \sim O(10^{-3})$ compared to the $T \rightarrow W^+ \, b$ decay, and therefore have no observable experimental effect at present.

\begin{figure}[htbp]
\centerline{\includegraphics[width=5.0in,angle=0]{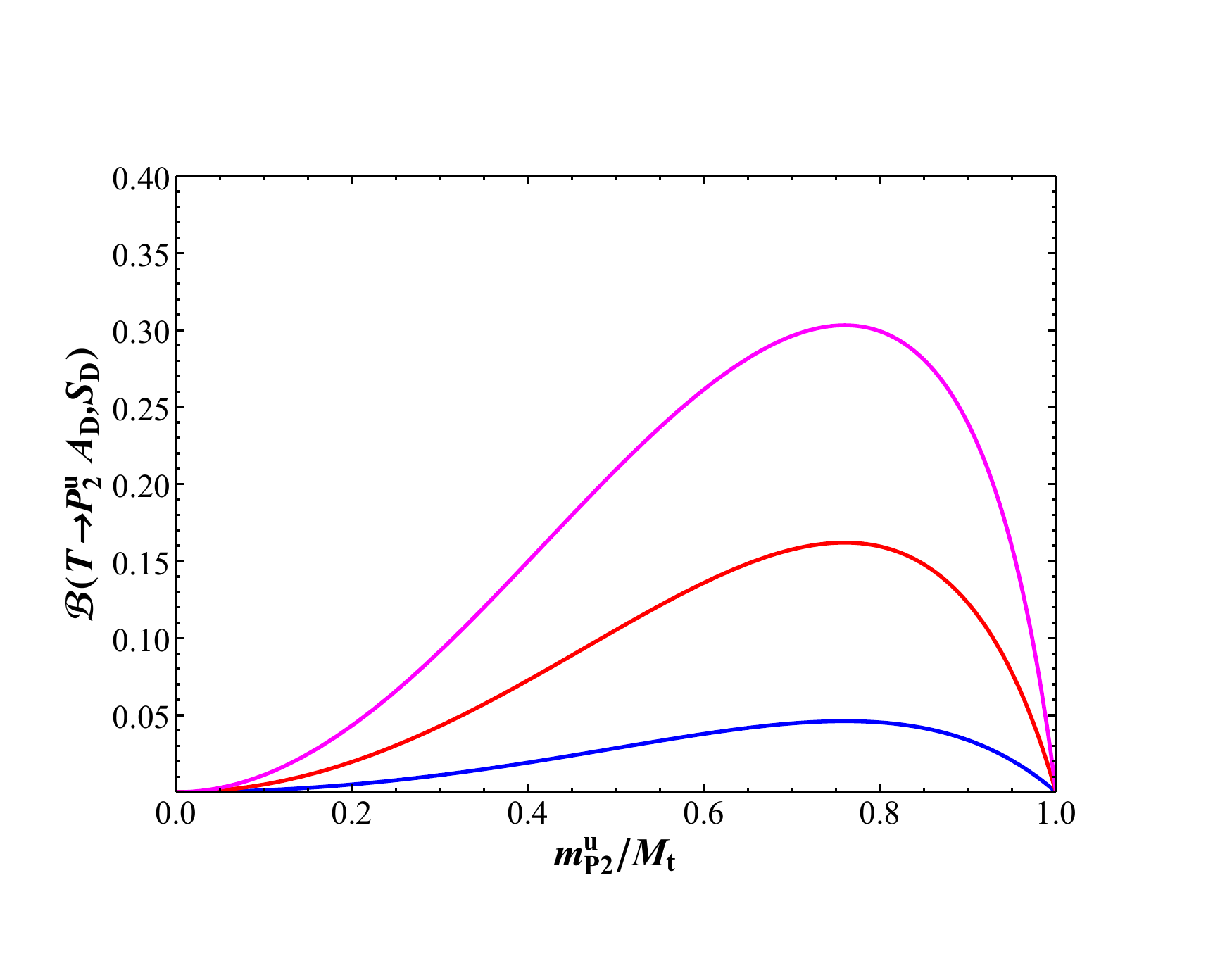}}
\vspace*{-2.0cm}
\centerline{\includegraphics[width=5.0in,angle=0]{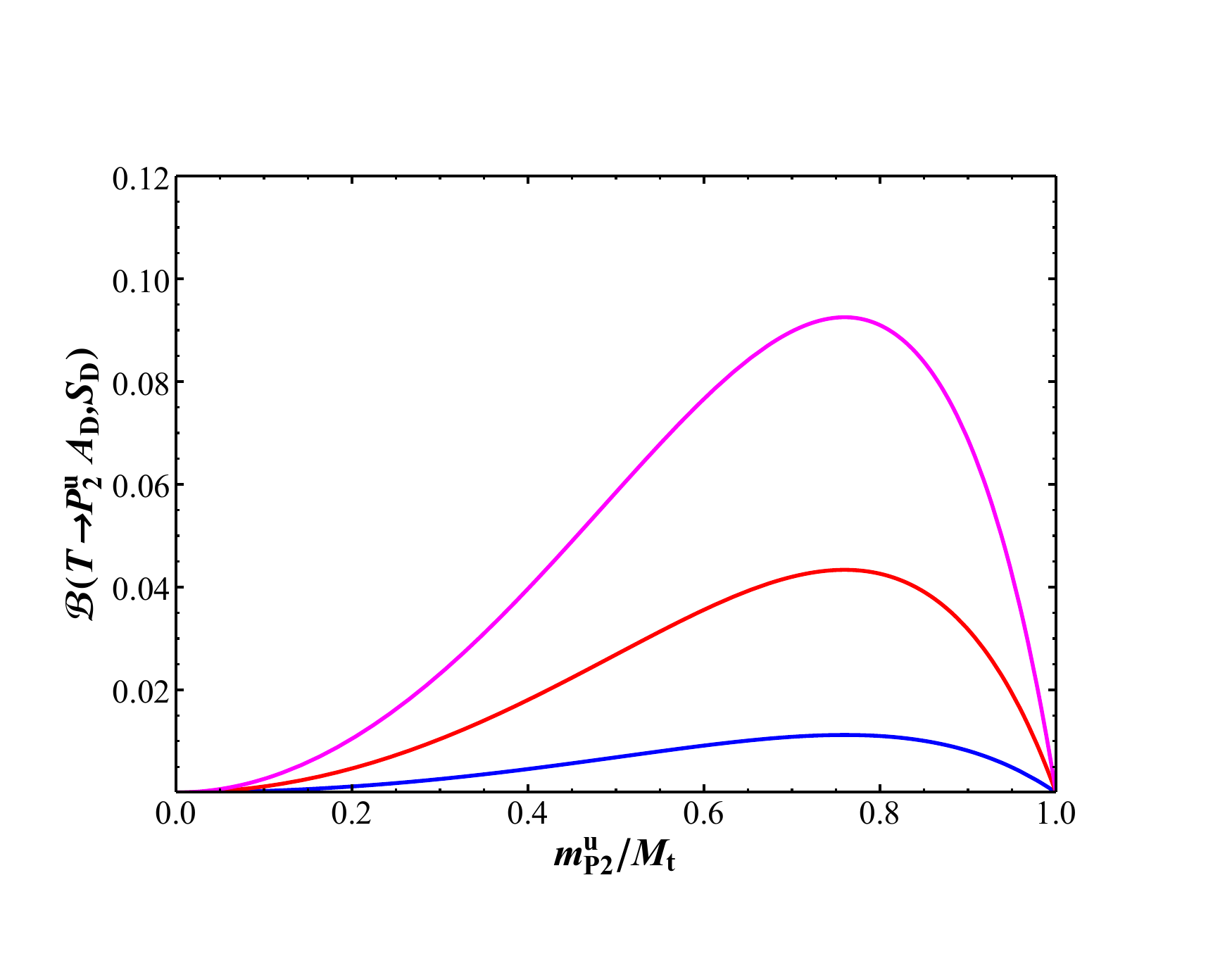}}
\vspace*{-1.0cm}
\caption{(Top) The branching fraction of the top partner quark to $P^u_2$ as a function of the ratio of their masses, assuming that $g_4 s_P =0.3$, $m_{A_D}= 0.1 \; \textrm{GeV}$, $|(\vec{\gamma}_P)_3|/v_P = 10^{-4}$, and $M_t/(y_S v_S)=1.5$, for $M_t = 2$ TeV (blue), 4 TeV (red), and 6 TeV (magenta).
(Bottom) Same as above, but assuming that $M_t/(y_S v_S)=2$. }
\label{fig7}
\end{figure}

In the event that all portal matter fields in the model are more massive than $T$, the above exhausts the possible decay channels for the top partner, indicating that it strongly resembles a conventional vector-like quark. The scenario can become more complicated, however, when the up-like portal matter is light enough to allow for the top partner to decay into it. In this case, the widths for decays of $T$ to $P^u_1$ and $P^u_2$ are
\begin{align}\label{TopPartnerPortalWidths}
    &\Gamma_{T \rightarrow P^u_1 \, A_D} = \frac{g_4^2 s_P^2 (m_{P1}^u)^2}{48 \pi m_{A_D}^2} \bigg( 1- \frac{(m_{P1}^u)^4}{M_t^4} \bigg) \bigg( 1- \frac{y_S^2 v_S^2}{M_t^2}\bigg) \frac{m_t^2}{y_S^2 v_S^2} M_t \frac{|(\vec{\gamma}_P)_3|^2}{v_P^2},\\
    &\Gamma_{T \rightarrow P^u_2 \, A_D} = \frac{g_4^2 s_P^2 (m_{P2}^u)^2}{48 \pi m_{A_D}^2} \bigg( 1- \frac{(m_{P2}^u)^4}{M_t^4} \bigg) \frac{y_S^2 v_S^2}{M_t^2} M_t \frac{|(\vec{\gamma}_P)_3|^2}{v_P^2}. \nonumber
\end{align}
We see that, analogous to the case for the reversed decay pattern discussed in Section \ref{PortalMatterSection}, the decay width from $T$ to $P_1^u$ is suppressed by a factor of $m_t^2/(y_S^2 v_S^2) \lsim 10^{-2}$, and so is likely not numerically significant, while the decay width from $T$ to $P_2^u$ has no such suppression and may be large. Furthermore, we note (as discussed in Section \ref{PortalMatterSection}), we would anticipate from Goldstone boson equivalence that a physical scalar,  the `dark' Higgs, which we shall call $S_D$, associated with the breaking of $U(1)_D$ will likely also exist, with a decay width such that $\Gamma_{T\rightarrow P^u_2 \, S_D}=\Gamma_{T\rightarrow P^u_2 \, A_D}$, at least in the limit where $m_{A_D}\rightarrow 0$. Unlike in the case of portal matter decays, for which each numerically significant decay channel has such an identical channel with $S_D$, and hence branching fractions are unaffected by ignoring the scalar, the top partner's dominant SM decays $T \rightarrow Z \, t$, $T \rightarrow h \, t$, and $T \rightarrow W^+ \, b$ do not have an equivalent $S_D$ channel, while the decay to portal matter does. Therefore, when discussing the decay of the top partner to portal matter, we must include the decay width of the process $T \rightarrow P^u_2 \, S_D$ explicitly, arriving at the combined branching fraction from Eqs.(\ref{TopPartnerSMWidths}) and (\ref{TopPartnerPortalWidths}) of
\begin{align}\label{TopPartnerPortalBR}
    \mathcal{B}(T \rightarrow P^u_2 \, A_D, S_D) \approx \frac{2 \Gamma_{T \rightarrow P^u_2 \, A_D}}{4\Gamma_{T \rightarrow h t}+2\Gamma_{T \rightarrow P^u_2 \, A_D}}.
\end{align}
To get a sense of the possible magnitude of the branching fraction given in Eq.(\ref{TopPartnerPortalBR}), we depict it for various choices of $M_t$ and $M_t/(y_S v_S)$ in Figure \ref{fig7}, assuming that $|(\vec{\gamma}_P)_3|/v_P = 10^{-4}$, the ``natural'' magnitude we discussed in Section \ref{ADFlavorConstraintsSection}.

In Figure \ref{fig7}, we have selected values of $g_4 s_P$ and $m_{A_D}$ so as to maximize the branching fraction to the portal matter fields, selecting $m_{A_D}$ at the bottom of the range we consider, while selecting $g_4 s_P$ close to the upper end of the range that still permits $O(10^{-3})$ kinetic mixing (as discussed in Section \ref{KMSection}). Given that the decay width of the top partner to $m_{P1}$ is proportional to the square of $g_4 s_P$, and the inverse square of $m_{A_D}$, we can anticipate that larger values of $m_{A_D}$ and smaller couplings $g_4 s_P$ will result in dramatically smaller branching fractions -- letting $m_{A_D}=0.3 \; \textrm{GeV}$, for example, will reduce the branching fraction by a factor of 9. Even in the maximal case we depict in Figure \ref{fig7}, we generally only attain limited branching fractions of $T$ to $P^2_u$ here, however -- we achieve a $\sim 30 \%$ branching fraction at the largest when $M_t = 6 \; \textrm{TeV}$, half that when $M_t = 4 \; \textrm{TeV}$, and in all other cases we depict, the branching fraction is less than $10\%$. This branching fraction can be enhanced if we assume a larger value of $|(\vec{\gamma}_P)_3|/v_P$ (for example, enhancing this parameter by a factor of 3 would increase the decay width of $T$ to portal matter by a factor of 9), however, considering a value of $|(\vec{\gamma}_P)_3|/v_P$ much larger than $10^{-4}$, as we have taken it to be in Figure \ref{fig7}, begins to be in tension with constraints on flavor-changing $B$ meson decays discussed in Section \ref{ADFlavorConstraintsSection}. In short, in contrast to the scenario with the possible decay of portal matter to the top partner the reverse scenario is likely to have a minimal effect on the signal of top partner production, except in certain special corners of parameter space. In these corners, however, we might anticipate some interesting potential top partner events -- for example, top partner pair production might give us one `conventional' top partner which decays via the channels $T \rightarrow Z \, t$, $T \rightarrow h \, t$, or $T \rightarrow W \, b$, and the other that decays to a top (or, depending on the orientation of $\vec{\gamma}_P$ in flavor space, a lighter up-like quark) through sequentially emitting two dark photons, which would appear as missing energy. Determining the degree to which the latter decay path might be kinematically distinguishable from, for example, $T \rightarrow Z \, t$ with an invisibly decaying $Z\to \bar \nu \nu$ may be of interest, but lies beyond the scope of this work.

\subsection{Phenomenology: $Z_P$ Production}

\begin{figure}[htbp]
\centerline{\includegraphics[width=3.5in]{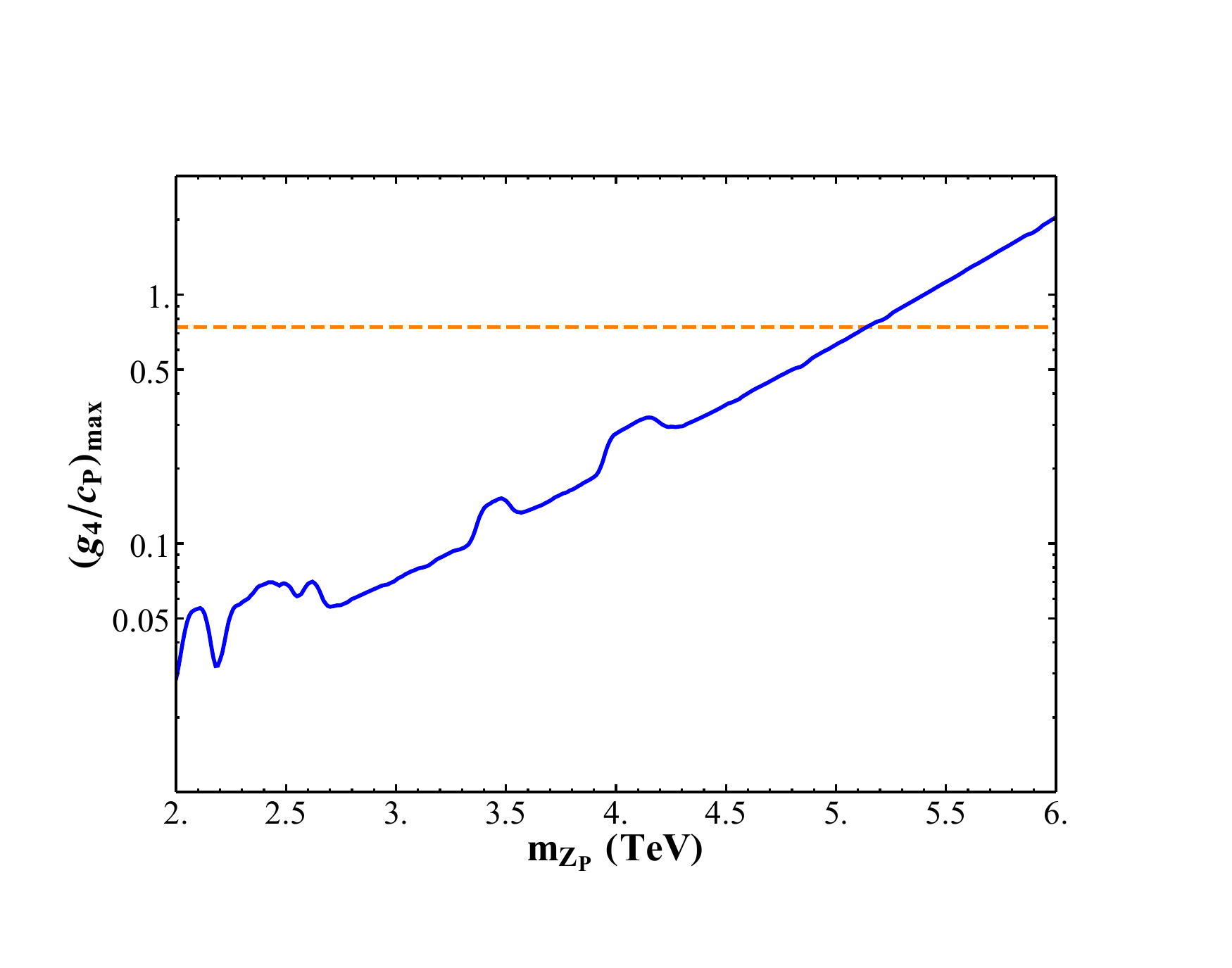}
\hspace{-0.75cm}
\includegraphics[width=3.5in]{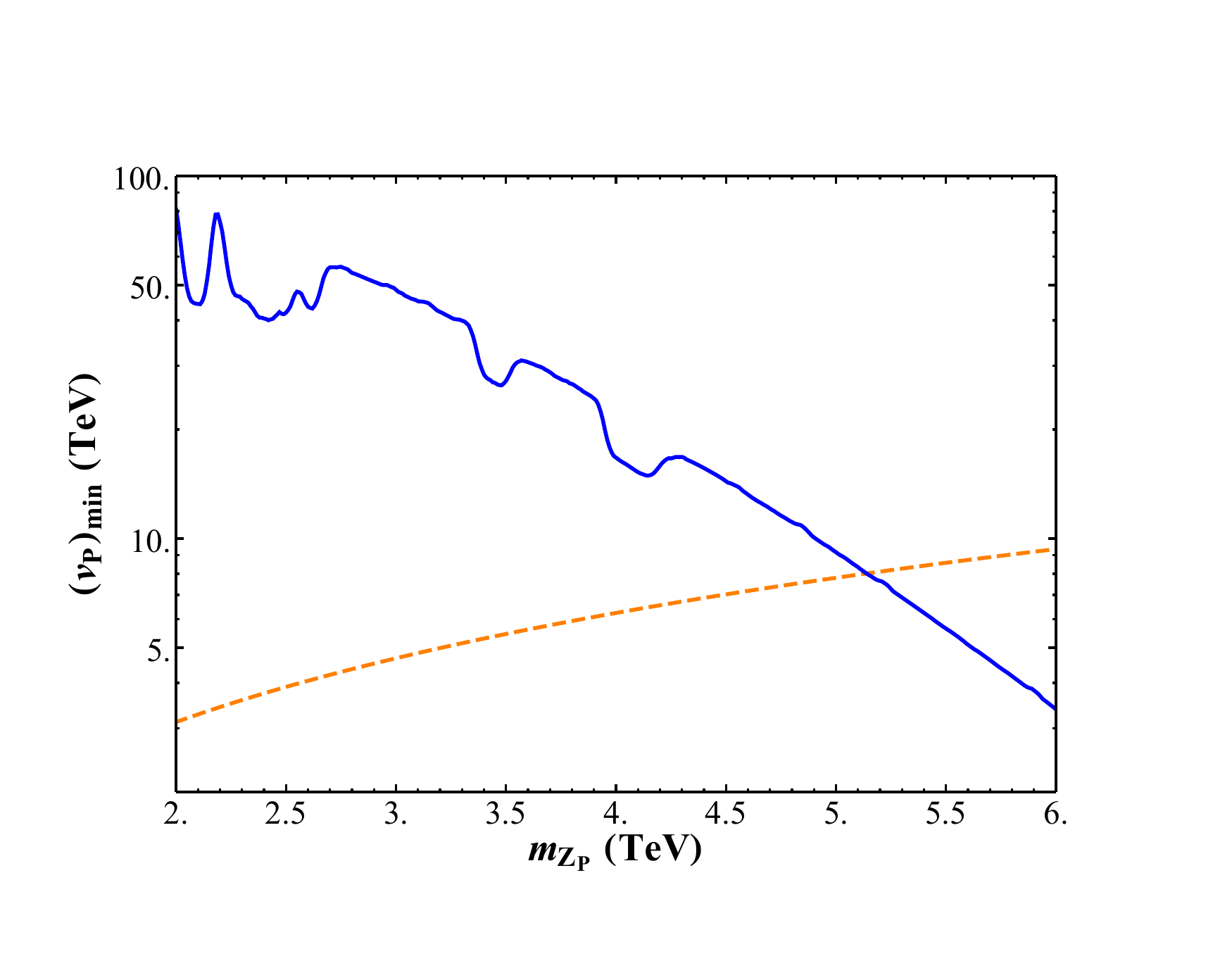}}
\vspace*{0.0cm}
\centerline{\includegraphics[width=3.5in]{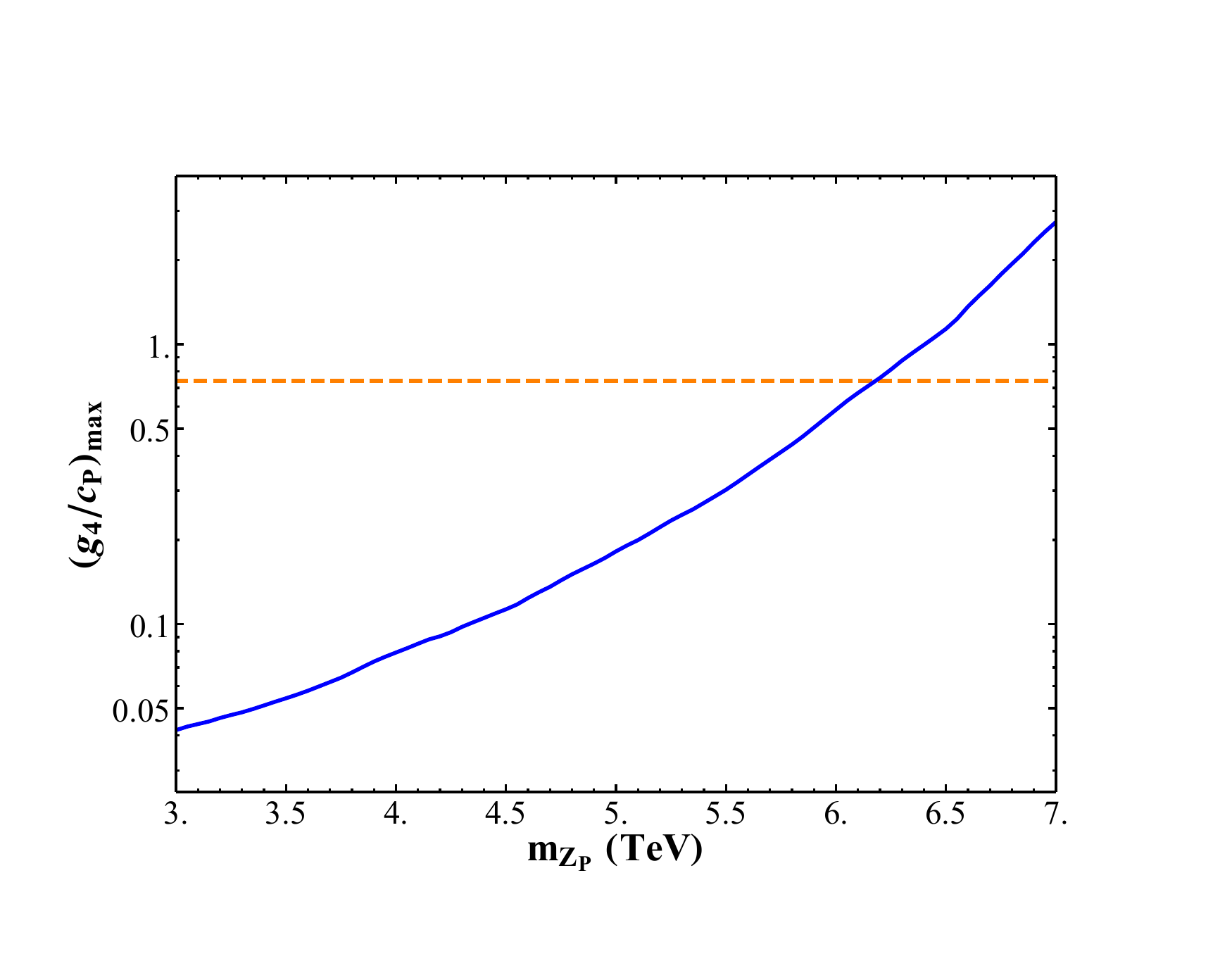}
\hspace{-0.75cm}
\includegraphics[width=3.5in]{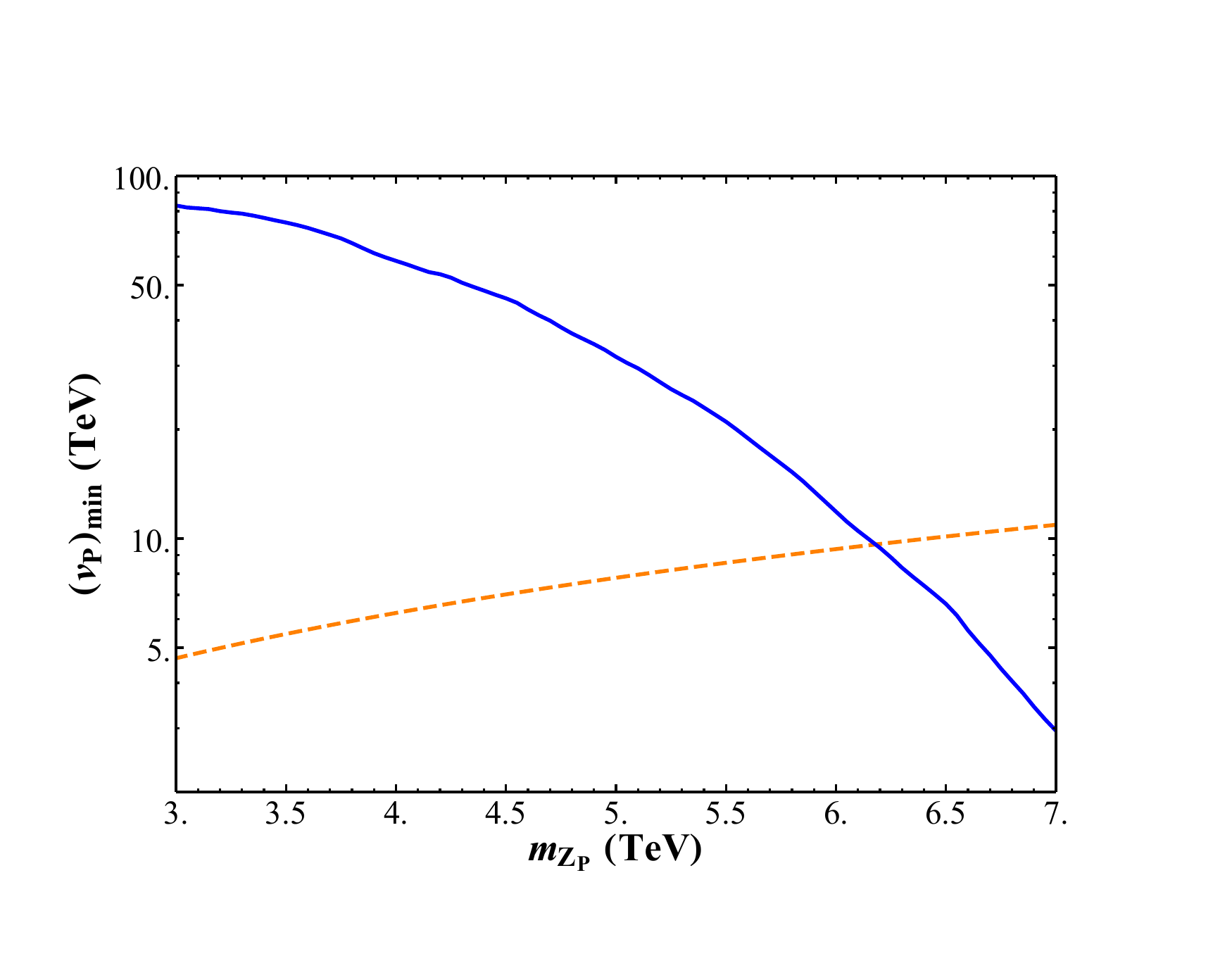}}
\caption{(Top) The maximum allowed value of $g_4/c_P$ (left) and the minimum allowed $v_P$ (right) for different values of $m_{Z_P}$, based on limits from dilepton-channel searches with 139 $\textrm{fb}^{-1}$ of 13 TeV LHC data \cite{Aad:2019fac}, assuming $Z_P$ only decays into SM final states. For comparison, the electroweak coupling $g/c_w$ (left) and the value of $v_P$ assuming $g_4/c_P=g/c_w$ for a given $m_{Z_P}$ value (right) have been depicted as dashed orange lines. (Bottom) Same as above, but using the expected constraints from a null result with 3 $\textrm{ab}^{-1}$ of 14 TeV LHC data \cite{ATLAS:2018tvr}}
\label{fig4}
\end{figure}

To start, we shall consider the production of the ``portal $Z$'' boson $Z_P$, representing the heavy counterpart to the lighter dark photon, with a mass of $O(1 \; \textrm{TeV})$. Notably, since the other gauge bosons with which $Z_P$ might couple (specifically, those associated with $SU(4)_F$ generators other than $t^{15}$, so in principle any gauge boson except $A_D$) are all much heavier than $Z_P$ itself, there is no region of parameter space in which $Z_P$ might decay into any other gauge bosons at tree-level. Our sole concern at this point, then, would be the decay of the $Z_P$ into fermion pairs. Consulting the coupling matrix for $Z_P$ given in Eqs.(\ref{ZPExactCoupling}) and (\ref{ZPTruncCoupling}), we can then straightforwardly determine the decay widths of the $Z_P$ boson. In particular, for decay into any particular SM fermion, we have
\begin{align}
    \Gamma_{Z_P \rightarrow \overline{f}f} \approx C_f \frac{g_4^2/c^2_P}{288 \pi}m_{Z_P}\bigg[ 1 + O \bigg( \frac{m_f^2}{m_{Z_P}^2} \bigg) \bigg],
\end{align}
where $C_f$ is simply a color factor (3 for quarks, 1 for charged leptons and neutrinos). Depending on the relative masses of the top partner and the portal matter fields to $m_{Z_P}$, however, $Z_P$ may also decay into these states. For the top partner, the decay width is
\begin{align}
    \Gamma_{Z_P \rightarrow \overline{T} T} = \frac{g_4^2/c^2_P}{96 \pi}m_{Z_P} \bigg(1+ \frac{2 M_t^2}{m_{Z_P}^2} \bigg)\sqrt{1-\frac{4 M_t^2}{m_{Z_P}^2}}.
\end{align}
The phenomenologically relevant portal matter fields, meanwhile, yield decay processes with widths
\begin{align}\label{ZPPortalWidth}
    &\Gamma_{Z_P \rightarrow \overline{P}_i^{u}P_i^{u}} = 3~\frac{g_4^2/c^2_P}{64 \pi} m_{Z_P} \bigg[\bigg( 1+\frac{8}{9} x_P \bigg) \bigg(1-\frac{(m_{Pi}^u)^2}{m_{Z_P}^2}\bigg)+\frac{8}{3} x_P \frac{(m_{Pi}^u)^2}{m_{Z_P}^2} \bigg]\sqrt{1-\frac{4 (m_{Pi}^u)^2}{m_{Z_P}^2}},\\
    &x_P \equiv s^2_P(1-4 c^2_P), \nonumber
\end{align}
where $P_i^u$ refers to the portal matter field of mass $m_{Pi}^u$, and analogous expressions exist for the portal matter in the down-like quark, charged lepton, and neutrino sectors (note that the factor of 3 in front of Eq.(\ref{ZPPortalWidth}) is a color factor, and will hence be equal to one for the corresponding case of neutrino and charged lepton portal matter).
Because the couplings of the $Z_P$ to SM states are flavor-universal, this model makes easy contact with existing searches for high-mass spin-1 resonances. 
To estimate constraints on our model, we employ the limits on leptonic cross sections from \cite{Aad:2019fac}, a dilepton (dielectron and dimuon)-channel null search using 139 $\textrm{fb}^{-1}$ of 13 TeV data, and the analogous expected limits from a null result with 3000 $\textrm{fb}^{-1}$ of LHC data at 14 TeV \cite{ATLAS:2018tvr}. In Figure \ref{fig4}, we depict the maximum value of $g_4/c_P$ allowed as a function of $m_{Z_P}$ and the corresponding minimum allowed value of $v_P$, both for current experimental limits and those expected from a null result at the High-Luminosity LHC, assuming that the $Z_P$ boson only decays into SM final states (in which case the branching ratios for the dielectron or dimuon channels are simply 1/24). 

\begin{figure}[htbp]
\centerline{\includegraphics[width=5.0in,angle=0]{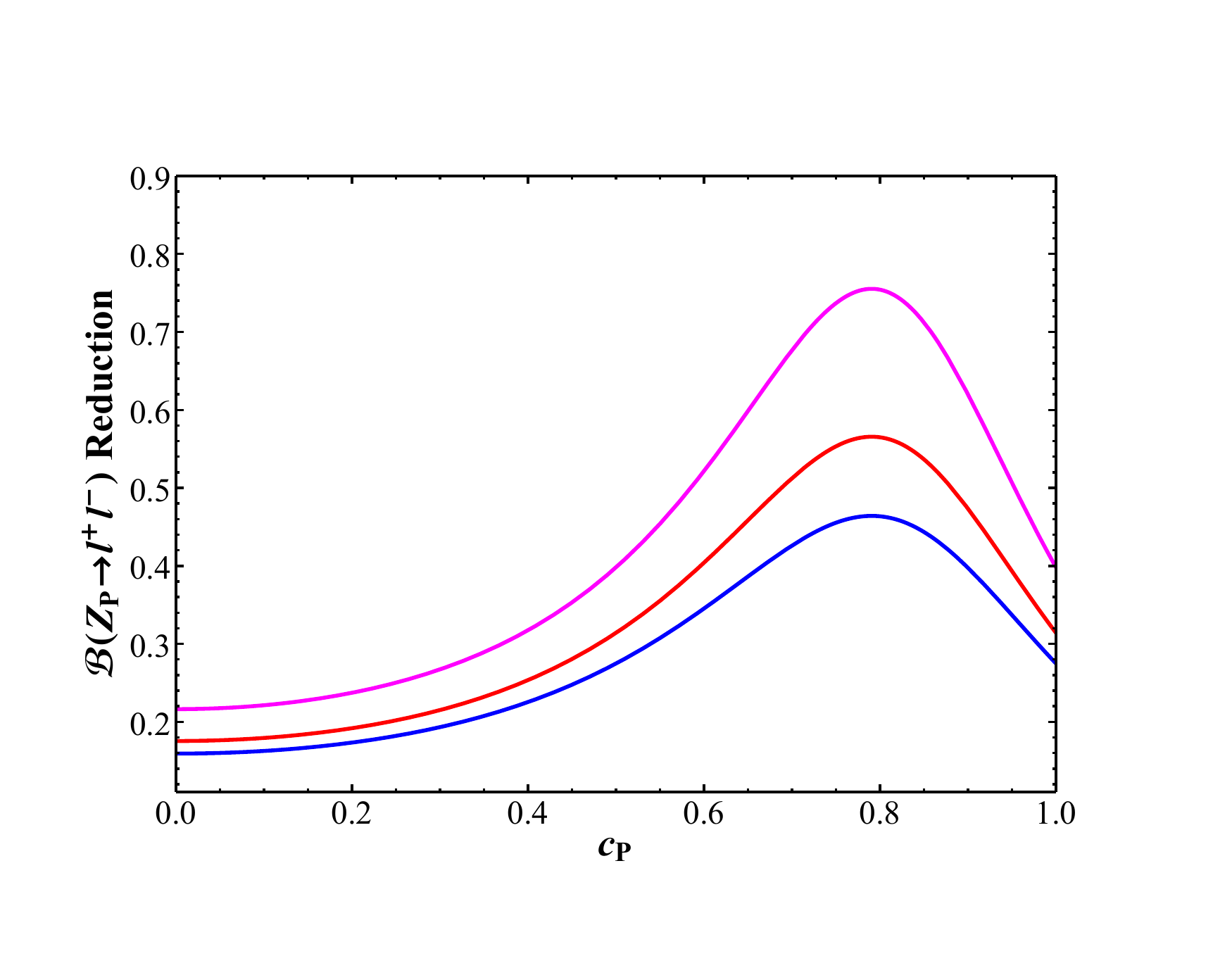}}
\vspace*{-2.0cm}
\centerline{\includegraphics[width=5.0in,angle=0]{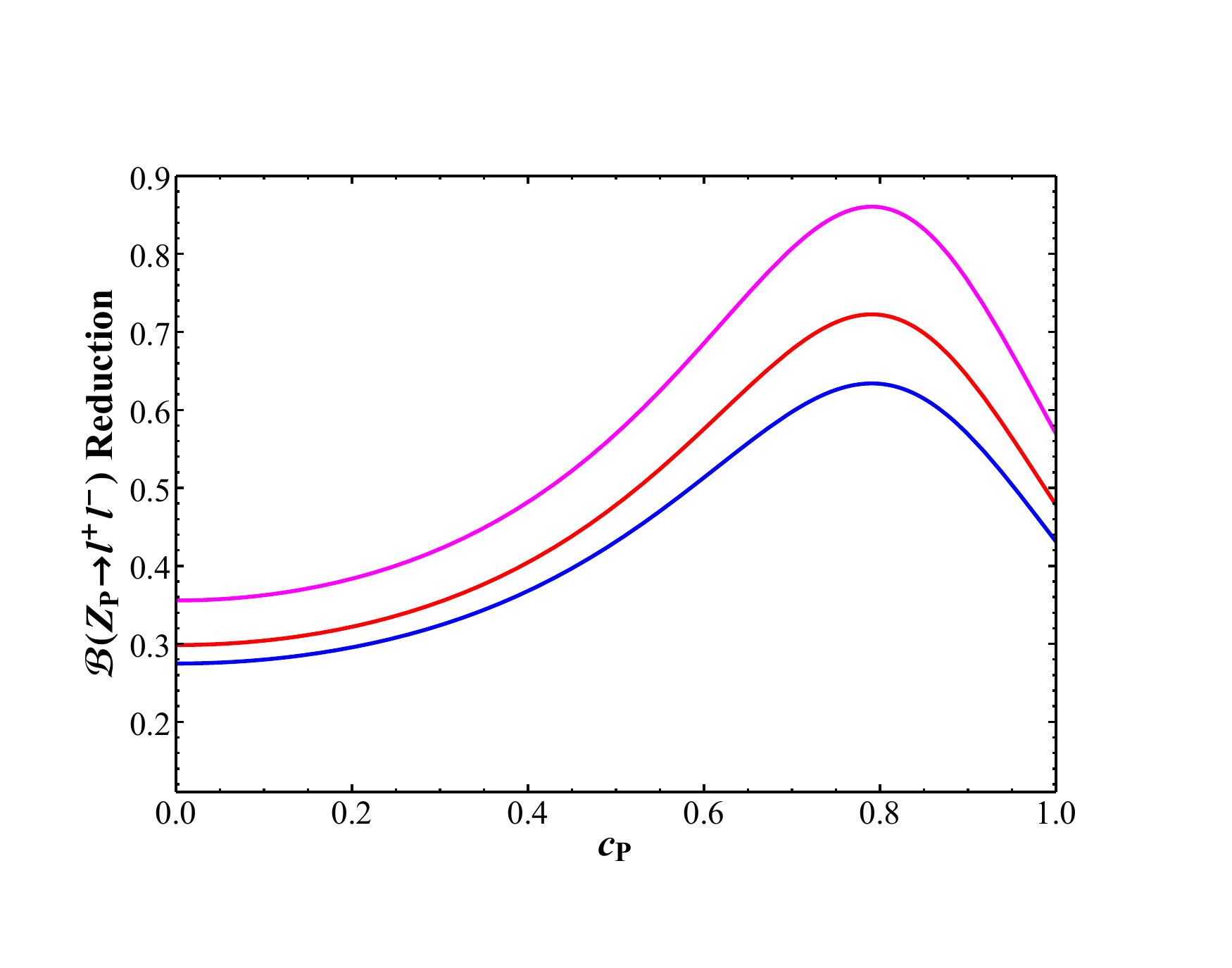}}
\vspace*{-1.0cm}
\caption{(Top) The proportional suppression of the branching ratio of $Z_P$ to leptons, compared to the scenario where $Z_P$ only decays to SM states, as a function of $c_P$, assuming portal matter states have the same mass $m_P$, assuming $m_P/m_{Z_P}=0.2$ (blue), 0.3 (red), and 0.4 (magenta).
(Bottom) Same as above, but assuming that only one of the two lighter portal matter states in each sector, rather than both, are kinematically accessible for this decay.}
\label{fig5}
\end{figure}

Examining Figure \ref{fig4}, we note that for realistic $O(1)$ values of $g_4/c_P$ (or, equivalently, $O(10 \; \textrm{TeV})$ values of $v_P$), the present constraint roughly requires $m_{Z_P} \gsim 5 \; \textrm{TeV}$; if we assume that $g_4/c_P = g/c_w$, where $g$ is the $SU(2)_L$ coupling constant and $c_w$ is the cosine of the Weinberg angle, the limit is, for example, $m_{Z_P} \geq 5.15 \; \textrm{TeV}$. A null result at the HL-LHC meanwhile suggests that these limits could be increased by approximately 1 TeV: If $g_4/c_P=g/c_w$, the expected limit from a null result of HL-LHC data would be $m_{Z_P} \geq 6.2 \; \textrm{TeV}$. However, we note that similar to the setup in \textbf{II}, this circumstance can be somewhat modified in regions of parameter space in which decays to exotic fermion species are kinematically allowed. In particular, constraints may be somewhat relaxed by assuming that the portal matter fields, $P_{1,2}^{u,d,e,\nu}$ are sufficiently light to allow for $Z_P$ to decay into them (which, assuming $g_4/c_P \sim O(1)$, would require at worst a mild $O(10^{-1})$ tuning of the Yukawa coupling parameters $y_{P1}$ and $y_{P2}$). While the decay into the top partner can reduce the branching fraction of the $Z_P$ to dilepton channels by approximately $10\%$, the large number of portal matter states with roughly degenerate masses (if we assume that decays to all of these portal matter states are kinematically accessible, for example, $Z_P$ would have 8 different new particles, some color triplets, to which it can decay) can provide as much as an order of magnitude reduction in the cross section observable in dilepton final state searches. We depict the proportional suppression factor of the branching ratio $\mathcal{B}(Z_{P}\rightarrow l^+ l^-)$ assuming that some or all portal matter states are kinematically accessible in Figure \ref{fig5}, where we note that, as Eq.(\ref{ZPPortalWidth}) suggests, the suppression factor is a function of $c_P$.

In Figure \ref{fig5}, we have specifically considered two sets of simple benchmarks for the possibility that the portal matter states are kinematically accessible in $Z_P$ decays -- either that all portal matter states have the same mass (which, we note, would not be an unreasonable approximation if $y_{P1} \approx y_{P2}$, in which case all portal matter states would have degenerate masses up to radiative corrections), or that only half of the portal matter states are light enough for $Z_P$ to decay into them (as we might expect if, for example, $y_{P1}< y_{P2}$). The suppression factors in Figure \ref{fig5}, in turn, correspond to as much as $O(1)$ increases in the maximum allowed $g_4/c_P$ values (or equivalently, $O(1)$ decreases in the minimum $v_P$ values) depicted in Figure \ref{fig4}. We note, however, that even under the best of circumstances (for example, if the suppression factor is $\sim 0.2$, very near the lowest we can achieve), the $g_4/c_P$ values are only increased by a factor of $\sim 2$. A cursory inspection of Figure \ref{fig4} indicates that this amelioration is enough to perhaps allow $m_{Z_P} \sim 4 \; \textrm{TeV}$ while maintaining a reasonable $O(1)$ value of $g_4/c_P$ (at least based on 13 TeV data), but certainly cannot enable much smaller $m_{Z_P}$ values. Of course, the potential ameliorating effects of these exotic decay channels on dilepton searches may be counteracted by new distinctive signals due to the exotic matter production. In particular, the production of particle-antiparticle pairs of portal matter or top partner quarks via a resonant $Z_P$ might have a significant effect on the overall production cross section of these states, especially in the case of leptonic portal matter, which otherwise may only be produced via electroweak and dark photon interactions. Null results for portal matter and top partner searches, then, will likely apply a non-trivial constraint on the possible mass of the $Z_P$ as well, however treating this situation quantitatively is beyond the scope of this paper.

\section{Discussion and Conclusions}\label{ConclusionSection}

The distinctive requirements for fermionic portal matter fields in vector portal DM models provide an intriguing framework for UV extensions of these models. In particular, constraints from SM anomalies and lifetimes of new SM-charged particles suggest that these fields are vector-like, albeit with collider signatures that are quite different from those seen with more ``conventional'' vector-like fermions. Here, we have seen that highly complex additional physics, not a priori related to DM, can in fact be incorporated into a portal matter model by following simple rules: Extending the SM gauge group to a semisimple one guarantees that KM with a $U(1)_D$ embedded in a separate dark group will be finite and calculable, at which point the only requirements are to ensure that the complete theory is anomaly-free and contains only the SM and heavy vector-like particles.

To demonstrate the potential of following this recipe, we have developed a model with Pati-Salam symmetry extended by $SU(4)_F \times U(1)_F = 4_F 1_F$, in which the $4_F 1_F$ symmetry contains both a dark photon gauge group $U(1)_D$ and an $SU(3)$ flavor symmetry. The Pati-Salam symmetry is assumed to be broken at a high scale $\sim 10^{13} \; \textrm{GeV}$, at which point KM between $U(1)_F$ and $U(1)_Y$ occurs. Scalars in the adjoint representation of $SU(4)_F$ then break the $4_F 1_F$ down to $1_{F'} 1_F$, completely breaking the embedded $SU(3)_F$ flavor symmetry, at multi-TeV scales to avoid flavor constraints and generate the fermion mass hierarchy observed in the SM. The $1_{F'} 1_F$ symmetry is then broken down to $U(1)_D$ at a scale of $\sim 1-10 \; \textrm{TeV}$, and $U(1)_D$, the gauge symmetry corresponding to the DM vector portal, is finally broken by small vev terms of $O(0.1-1 \; \textrm{GeV})$. The presence of diverse scales in the symmetry breaking in turn translates to diverse scales of the new exotic particles proposed. However, we find that among the new fermions, only certain portal matter fields, which we have labelled $P_{1,2}^{u,d,e,\nu}$, and a vector-like new partner to the top quark, which we call $T$, can be expected to be light enough to be observed at present collider experiments, having masses at the scale of $\sim$ a few TeV. Among the new gauge bosons we have anticipated, the only ones which we find to have potentially observable effects are the dark photon of mass $O(1 \; \textrm{GeV})$ $A_D$, a TeV-scale boson with flavor-universal couplings $Z_P$, and a TeV-scale flavor-changing neutral boson $Z_F$ associated with a combination of generators of the flavor group $SU(3)_F$.

With the model set up, we proceeded to explore the phenomenological implications of the experimentally observable (that is, TeV-scale or less) sector of the theory, in particular focusing on the phenomenology of quarks and their portal matter partners, which we expect to have more phenomenologically visible signals than the corresponding physics in the lepton sector. We found that present constraints from flavor-changing neutral currents arising from the dark photon, in particular from $K \rightarrow \pi A_D$ transitions, placed harsh limits on the form of the vev parameters which might break $U(1)_D$, and in turn can substantially influence the branching fractions of portal matter fields to SM particles. We next noted that the vector-like top partner $T$ and the new vector boson $Z_P$ contributed non-trivially to $\Delta B = 2$ flavor-changing processes, the $Z_P$ at tree-level and the $T$ at the one-loop level. These effects were comparable in magnitude and heavily dependent on the value of the model parameter $y_S v_S$, a dimensionful quantity which set the scale of both $T$ and $Z_F$'s mass. However, even  when combined the sum of these contributions still provided only a limited constraint on our parameter space based on present measurements, although Belle II and LHCb data may constrain these more stringently in the near future \cite{Charles:2020dfl}.

Beyond the rich collider phenomenology already explored in \textbf{I} and \textbf{II}, much of which is replicated in our model, we find additional model-building flexibility arising from the potential for the portal matter fields $P_{2}^u$ to decay into $T$ (or vice versa, depending on the particles' relative masses), which can provide distinctive collider signatures appearing in neither the treatments discussed in \textbf{I} and \textbf{II} nor more conventional models of vector-like quarks \cite{VLFs,Vatsyayan:2020jan,DeSimone:2012fs}.

Looking forward, we note that the present model represents a single specific realization of a much broader recipe for developing models with non-minimal portal matter sectors. There are a number of different avenues through which this particular effort can be further explored, for example by incorporating an explanation for the small masses and near-maximal mixing observed in the neutrino sector, addressing the scalar sector rigorously, considering scenarios in which the Pati-Salam symmetry (or some components of it) are broken at a similar or lower scale to the one at which the $SU(4)_F \times U(1)_F$ symmetries are broken, or explicitly incorporating either scalar or fermionic DM in the construction. Perhaps a broader conclusion to be drawn from this work, however, is the potential that portal matter model building possesses to elaborate on other extensions of the SM. In particular, we have seen that incorporating a local $SU(3)$ flavor symmetry with a model of portal matter leads to rich phenomenological signatures, some of which do not appear in more conventional portal matter models or models with local flavor symmetries individually, such as a potential correlation between portal matter lifetime, branching fractions, and the allowed dark photon masses, or decays of portal matter to more conventional vector-like quarks. Presumably, similar paths can be taken to explore potential links between portal matter models and other popular extensions of the SM, predicting entirely different unique signatures that may depart radically from those anticipated by conventional new physics searches.

% DISCUSS FUTURE EXTENSIONS OF THIS FRAMEWORK: ALTERNATE $SU(3)_F$ FLAVOR CONSTRUCTIONS, INCLUSION OF NEUTRINOS, EXPLORATION OF THE SCALAR SECTOR, AND EXPLICIT INCLUSION OF DARK MATTER.

%------------------------------------ ACKNOWLEDGEMENTS ---------------------------------------%
\section*{Acknowledgements}
GNW would like to thank T.D. Reuter for discussions relating to this work. This work was supported by
the Department of Energy, Contract DE-AC02-76SF00515.

%------------------------------------ APPENDICES ---------------------------------------%
\appendix
\section{$SU(4)$ Generator Matrices}\label{appendix:SU4Generators}

Here we list the generators of the $SU(4)$ algebra, $t^i$. The first eight generators correspond to the embedded group $SU(3)_F$ in $SU(4)_F$, and are given by
\begin{align}
    t^1 = \frac{1}{2}\begin{pmatrix}
    0 & 1 & 0 & 0\\
    1 & 0 & 0 & 0\\
    0 & 0 & 0 & 0\\
    0 & 0 & 0 & 0
    \end{pmatrix},\;\; & t^2 = \frac{1}{2} \begin{pmatrix}
    0 & -i & 0 & 0\\
    i & 0 & 0 & 0\\
    0 & 0 & 0 & 0\\
    0 & 0 & 0 & 0
    \end{pmatrix}, \nonumber\\
    t^4 = \frac{1}{2}\begin{pmatrix}
    0 & 0 & 1 & 0\\
    0 & 0 & 0 & 0\\
    1 & 0 & 0 & 0\\
    0 & 0 & 0 & 0
    \end{pmatrix},\;\; & t^5 = \frac{1}{2} \begin{pmatrix}
    0 & 0 & -i & 0\\
    0 & 0 & 0 & 0\\
    i & 0 & 0 & 0\\
    0 & 0 & 0 & 0
    \end{pmatrix},\\
    t^6 = \frac{1}{2}\begin{pmatrix}
    0 & 0 & 0 & 0\\
    0 & 0 & 1 & 0\\
    0 & 1 & 0 & 0\\
    0 & 0 & 0 & 0
    \end{pmatrix},\;\; & t^7 = \frac{1}{2} \begin{pmatrix}
    0 & 0 & 0 & 0\\
    0 & 0 & -i & 0\\
    0 & i & 0 & 0\\
    0 & 0 & 0 & 0
    \end{pmatrix}, \nonumber\\
    t^3 = \frac{1}{2}\begin{pmatrix}
    1 & 0 & 0 & 0\\
    0 & -1 & 0 & 0\\
    0 & 0 & 0 & 0\\
    0 & 0 & 0 & 0
    \end{pmatrix},\;\; & t^8 = \frac{1}{2\sqrt{3}} \begin{pmatrix}
    1 & 0 & 0 & 0\\
    0 & 1 & 0 & 0\\
    0 & 0 & -2 & 0\\
    0 & 0 & 0 & 0
    \end{pmatrix}. \nonumber
\end{align}
The next 6 generators are then
\begin{align}
    t^9 = \frac{1}{2}\begin{pmatrix}
    0 & 0 & 0 & 1\\
    0 & 0 & 0 & 0\\
    0 & 0 & 0 & 0\\
    1 & 0 & 0 & 0
    \end{pmatrix},\;\; & t^{10} = \frac{1}{2} \begin{pmatrix}
    0 & 0 & 0 & -i\\
    0 & 0 & 0 & 0\\
    0 & 0 & 0 & 0\\
    i & 0 & 0 & 0
    \end{pmatrix}, \nonumber\\
    t^{11} = \frac{1}{2}\begin{pmatrix}
    0 & 0 & 0 & 0\\
    0 & 0 & 0 & 1\\
    0 & 0 & 0 & 0\\
    0 & 1 & 0 & 0
    \end{pmatrix},\;\; & t^{12} = \frac{1}{2} \begin{pmatrix}
    0 & 0 & 0 & 0\\
    0 & 0 & 0 & -i\\
    0 & 0 & 0 & 0\\
    0 & i & 0 & 0
    \end{pmatrix},\\
    t^{13} = \frac{1}{2}\begin{pmatrix}
    0 & 0 & 0 & 0\\
    0 & 0 & 0 & 0\\
    0 & 0 & 0 & 1\\
    0 & 0 & 1 & 0
    \end{pmatrix},\;\; & t^{14} = \frac{1}{2} \begin{pmatrix}
    0 & 0 & 0 & 0\\
    0 & 0 & 0 & 0\\
    0 & 0 & 0 & -i\\
    0 & 0 & i & 0
    \end{pmatrix}.
\end{align}
Finally, $t^{15}$ is the generator corresponding to the embedded group $U(1)_F'$ in $SU(4)_F$, which mixes with $U(1)_F$ to form the dark charge group. It is given by
\begin{align}
    t^{15} = \frac{1}{2 \sqrt{6}} \begin{pmatrix}
    1 & 0 & 0 & 0\\
    0 & 1 & 0 & 0\\
    0 & 0 & 1 & 0\\
    0 & 0 & 0 & -3
    \end{pmatrix}.
\end{align}

\end{document}